\let\latexaddtocontents\addtocontents
\documentclass[aps,pra,reprint,superscriptaddress]{revtex4-2}
\let\addtocontents\latexaddtocontents
\pdfoutput=1
\usepackage{easy-todo}
\usepackage{color}
\usepackage{graphicx}
\usepackage{float}
\usepackage{amsmath}
\usepackage{amsfonts}
\usepackage{amssymb}
\usepackage{microtype}
\usepackage{verbatim}
\usepackage{bbold}

\usepackage{hyperref}
\usepackage{cleveref}
\usepackage{dsfont}
\usepackage[normalem]{ulem} 
\allowdisplaybreaks
\usepackage{multirow}
\usepackage{bbm}

\usepackage{braket}

\usepackage{array}   
\newcolumntype{L}{>{$}l<{$}} 

\hypersetup{
  colorlinks   = true, 
  urlcolor     = blue, 
  linkcolor    = blue, 
  citecolor   = blue 
}
\graphicspath{{./Figures/}}

\usepackage[smallerops]{newtxmath}

\usepackage{array}
\newcolumntype{C}[1]{>{\centering\let\newline\\\arraybackslash\hspace{0pt}}m{#1}}

\DeclareMathOperator{\tr}{tr}

\begin{document}

\author{Jiangtian Yao}
\email{jyao@pks.mpg.de}
\affiliation{Max Planck Institute for the Physics of Complex Systems, 01187 Dresden, Germany}

\author{Pieter W. Claeys}

\affiliation{Max Planck Institute for the Physics of Complex Systems, 01187 Dresden, Germany}

\title{Temporal Entanglement Barriers in Dual-Unitary Clifford Circuits with Measurements}

\begin{abstract}
We study temporal entanglement in dual-unitary Clifford circuits with probabilistic measurements preserving spatial unitarity. We exactly characterize the temporal entanglement barrier in the measurement-free regime, exhibiting ballistic growth and decay and a volume-law peak. In the presence of measurements, we relate the temporal entanglement to the scrambling properties of the circuit. For ``good scramblers'' measurements do not qualitatively change the temporal entanglement profile but only result in a reduced entanglement velocity, whereas for ``poor scramblers'' the initial ballistic growth of temporal entanglement with bath size is modified to diffusive. This difference is understood through a mapping of the underlying operator dynamics to a biased and an unbiased persistent random walk respectively. In the latter case measurements additionally modify the ballistic decay to the perfect dephaser limit, with vanishing temporal entanglement, to an exponential decay, which we describe through a spatial transfer matrix method. This spatial dynamics is shown to be described by a non-Hermitian hopping model, exhibiting a PT-breaking transition at a critical measurement rate $p=1/2$. In all cases the peak value of the temporal entanglement barrier exhibits volume-law scaling for all measurement rates.
\end{abstract}

\maketitle
\section{Introduction}
Quantum circuit models recently emerged as a field of rapidly growing interest due to both experimental progress on noisy intermediate-scale quantum (NISQ) devices and newly-developed theoretical treatments. Experimentally, immense progress was made on realizing novel many-body quantum phases on quantum processors, such as topologically ordered states and time-crystalline eigenstate order \cite{bharti_noisy_2022, satzinger_realizing_2021, mi_time_2022}. Numerically, tensor-network based methods find natural applications in representing and simulating quantum circuits with built-in local structures, such as the brickwork or the staircase circuit geometries \cite{smith_simulating_2019, liu_methods_2022, jobst_finite_2022, lin_real_2021}. Restricted classes of quantum circuits were additionally found to admit exact solvability, which allows for benchmarking of numerical and experimental results. One such example is the class of dual-unitary circuits, possessing unitarity along both the temporal and the spatial directions. Such circuits can act as minimal models for capturing a wide range of phenomena in many-body quantum dynamics \cite{bertini_exact_2019, bertini_entanglement_2019, gopalakrishnan_unitary_2019, reid_entanglement_2021, bertini_operator_2020, claeys_maximum_2020, piroli_exact_2020}. On the one hand, unitarity in both space and time yields analytical solvability of such models; on the other, the constraint is loose enough to allow for generic behaviors ranging from integrable to chaotic dynamics \cite{claeys_ergodic_2021, bertini_entanglement_2019, flack_statistics_2020, bertini_operator_2020}. Besides these advances in 
methodology, quantum circuits present a natural setup for studying and observing new intriguing physical phenomena, with measurement-induced phase transitions as one of the paradigmatic examples \cite{li_measurement_2019, li_quantum_2018, chen_emergent_2020, bao_theory_2020, choi_quantum_2020,  zabalo_critical_2020, gullans_dynamical_2020,skinner_measurement_2019, jian_measurement_2020, cao_entanglement_2019, alberton_entanglement_2021, lunt_measurement_2020}. These are new classes of nonequilibrium quantum phase transitions that manifest themselves in the entanglement scaling of the quantum systems of interest. Except for certain limiting cases, the universality classes of such transitions do not match any known classes, and immense effort is devoted to analytically characterizing the nature of such transitions. 

For circuits with generic choice of gates, either originating from Trotterized Hamiltonian dynamics or directly representing Floquet unitary dynamics, exact results are generally out of reach, and matrix product state (MPS) evolution presents a natural choice of numerical method. In conventional approaches, one starts with the wave function represented as an MPS and updates it along the temporal direction by successively applying the appropriate unitary evolution operators \cite{paeckel_time_2019, cazalilla_time_2002, white_real_2004, daley_time_2004, verstraete_matrix_2004, vidal_efficient_2003}. The numerical simulability of the system dynamics is determined by the scaling of the required bond dimension for storing the MPS wave function, which is physically governed by the growth of spatial entanglement \cite{calabrese_evolution_2005, schuch_entropy_2008, osborne_efficient_2006}.

An alternative approach dubbed the ``folding algorithm" was proposed in Ref. \cite{banuls_matrix_2009}, where one updates the time-evolved density matrix as a so-called ``folded" MPS. In the folded representation, the density matrix $\rho$ is expressed as a wave function using the operator-state mapping: $\rho=|\Psi\rangle\langle\Psi| \rightarrow |\rho\rangle=|\Psi\rangle\otimes|\Psi\rangle^*$. The dynamics of this state follows as $|\rho(t)\rangle=(U\otimes U^*)|\Psi_0\rangle\otimes|\Psi_0\rangle^*$. 
The trace operation, naturally appearing when calculating expectation values of observables, can be written as an inner product
\begin{equation}
\begin{split}
\tr(\rho(t)) \equiv\braket{\tr |\rho(t)}= \bra{\tr} (U\otimes U^*)|\Psi\rangle\otimes|\Psi\rangle^*,
\end{split}
\end{equation}
where the first equation defines the inner product between a state vector and the ``trace vector" $\bra{\tr}$.

The expectation value of a local observable $O$ can be expressed in the folded representation as
\begin{equation}
\begin{split}
    \langle O (t)  \rangle& = \braket{\tr|O|\rho(t)}=\langle\tr|O (U\otimes U^*)|\Psi\rangle\otimes|\Psi\rangle^* \\
   & = \quad\vcenter{\hbox{\includegraphics[width=0.5\linewidth]{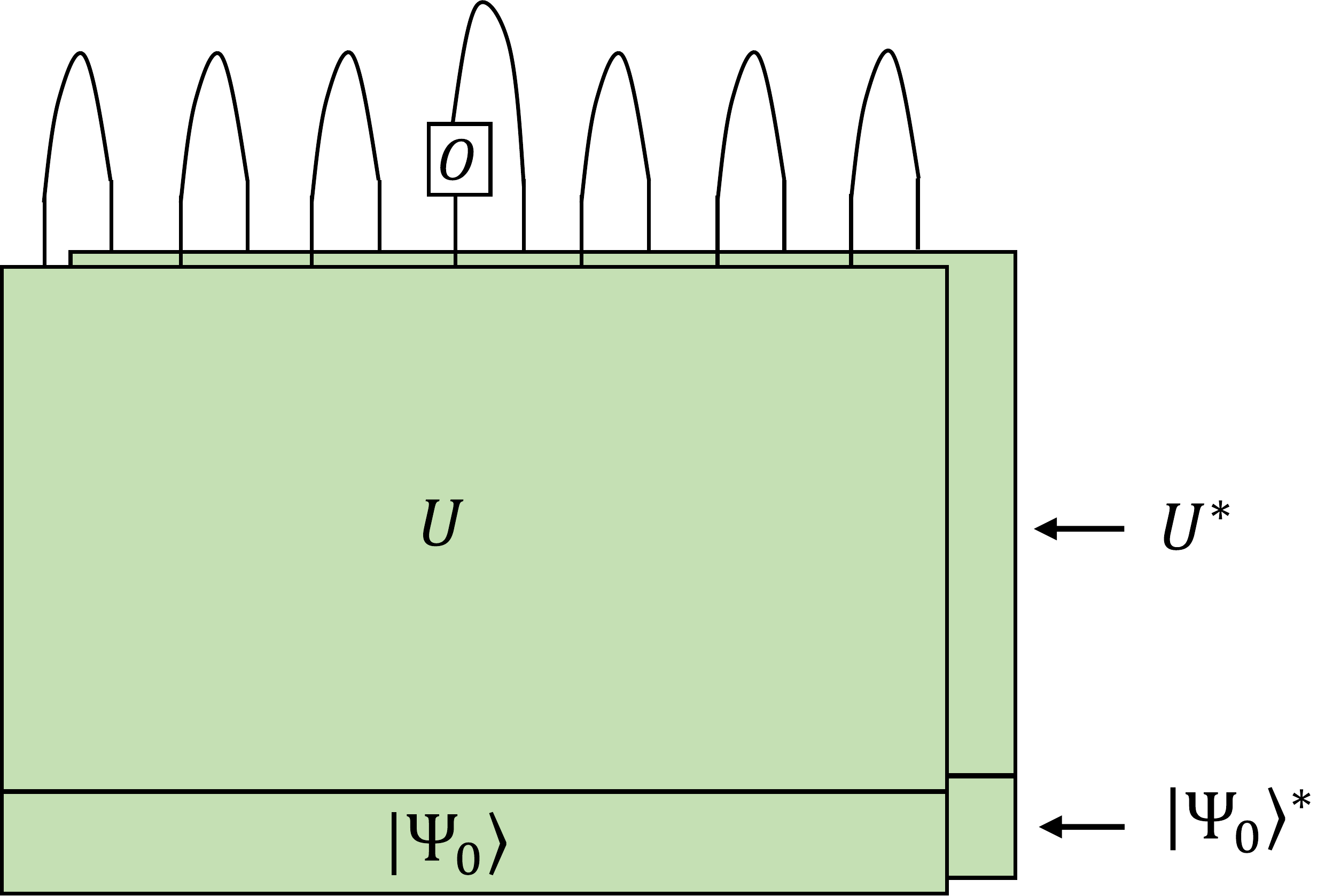}}}
\end{split}
\end{equation}
In the final equality we graphically represent the equation in the tensor network language (see e.g. Ref.~\cite{orus_practical_2014}), making explicit the folding.

For a local observable $O$ supported on e.g. a single site, one could treat the spatial slice where $O$ acts non-trivially separately from the regions to its left and right, where it acts trivially. To do so, $\langle O(t) \rangle$ can be re-expressed as  
\begin{equation}
    \begin{split}
        \langle O (t) \rangle&=\langle I_{\text{left}} | \mathcal{T}_O|I_{\text{right}}\rangle\\ &=\quad\vcenter{\hbox{\includegraphics[width=0.45\linewidth]{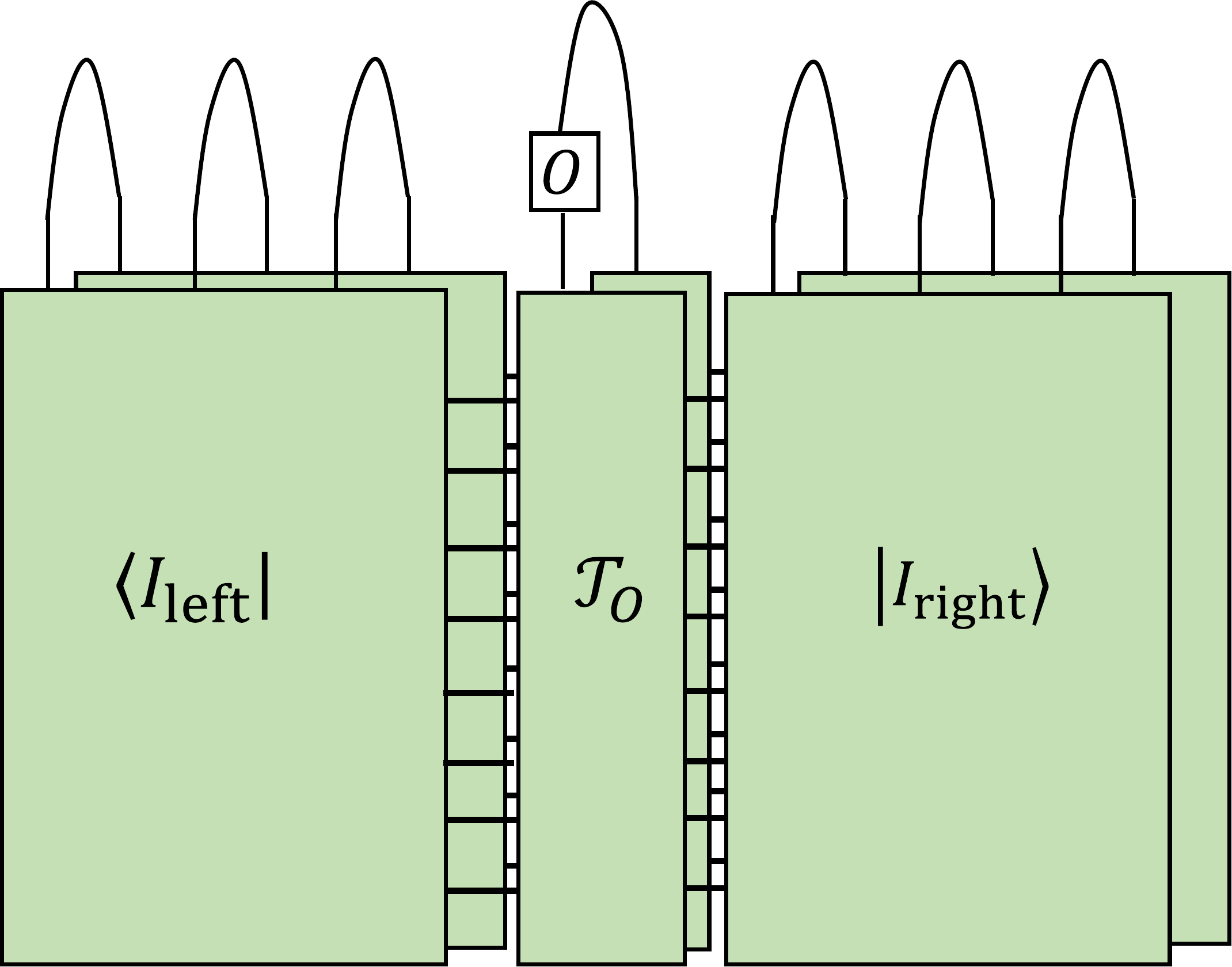}}}
    \end{split}
\end{equation}

For a spatially homogeneous evolution it is possible to identify a spatial transfer matrix such that the regions to the right and to the left can be written as powers of this matrix. Using $\langle I_{\text{left}}|$ as an example, in the folding algorithm, one starts with an arbitrary temporal MPS, $\langle \Phi| \otimes\langle \Phi|^* $, fixing the left boundary and successively applies to it the spatial transfer matrix $\mathcal{T}$. The thermodynamic limit of infinite system size can be taken by projecting $\langle \Phi|\otimes\langle \Phi|^*$ onto $\langle L|$, the dominant left eigenvector of $\mathcal{T}$:
\begin{equation}
    \begin{split}
        \langle I_{\text{left}}|_{L\rightarrow\infty}&=\lim_{L\rightarrow\infty}\left( \left(\langle \Phi|\otimes\langle \Phi|^*\right)\mathcal{T}^{L}\right)\\&=\lim_{L\rightarrow\infty}\quad\left(\vcenter{\hbox{\includegraphics[width=0.5\linewidth]{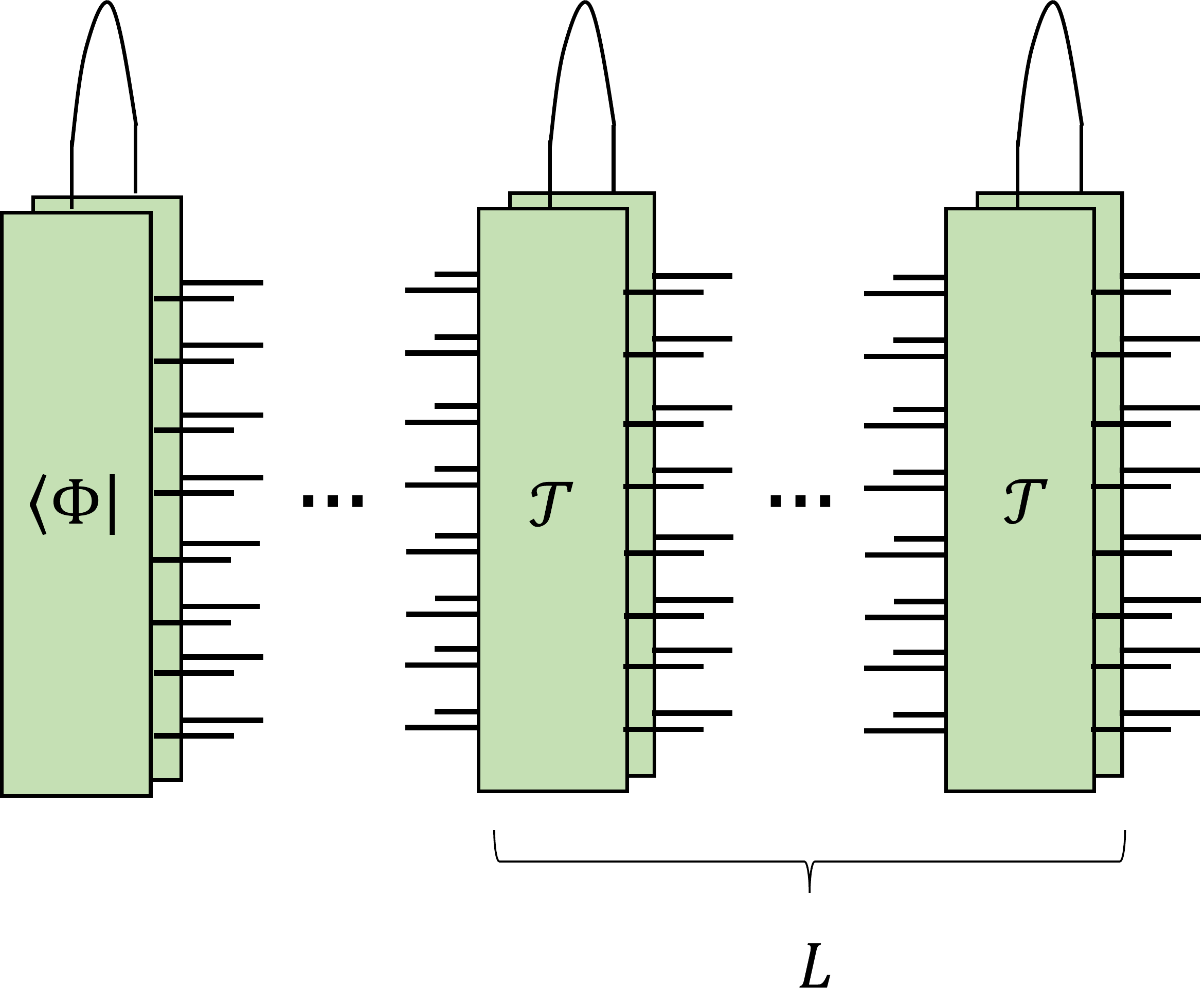}}}\right)\\&=\quad\vcenter{\hbox{\includegraphics[width=0.12\linewidth]{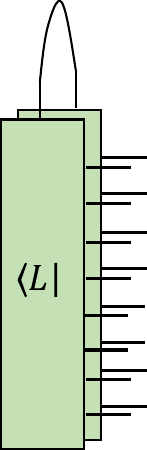}}}\quad=\langle L|
    \end{split}
\end{equation}

Analogously, $| I_{\text{right}}\rangle_{L\rightarrow\infty}= \lim_{L\rightarrow\infty}\left(\mathcal{T}^{L}(|\Phi\rangle \otimes |\Phi\rangle^*)  \right)=|R\rangle $ is the dominant right eigenvector of $\mathcal{T}$. 

Once these dominant eigenvectors are obtained, the value of $\langle O (t) \rangle$ in the thermodynamic limit can be computed as:
\begin{equation}
    \begin{split}
        \lim_{L\rightarrow\infty}\langle O(t)\rangle=\quad\vcenter{\hbox{\includegraphics[width=0.3\linewidth]{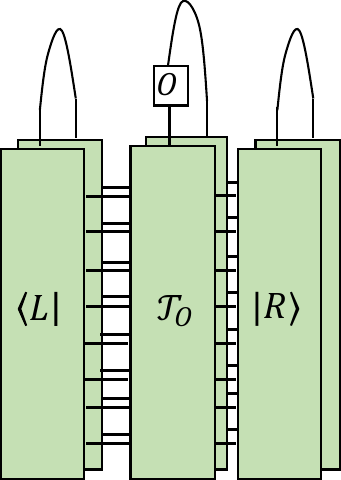}}}
    \end{split}
\end{equation}

In the context of quenched dynamics, the folding algorithm typically allows for dynamical studies that can reach longer times than conventional methods. The numerical complexity is determined by the scaling of maximal entanglement of the temporal MPS's $\langle I_{\text{left}}|$ and $| I_{\text{right}}\rangle$, dubbed the ``temporal entanglement" \cite{muellerhermes_tensor_2012, hastings_connecting_2015}.

Using $\langle I_{\text{left}}|$ as example, the temporal entanglement $S_T$ is defined as 
\begin{equation}
    \begin{split}
        S_T=\max_{t_i=t_1\ldots t_{T-1}} S_T^{(t_i)} \equiv\max_{t_i=t_1\ldots t_{T-1}}S_{\text{vN}}(\rho^{\langle I_{\text{left}}|}_{t_i})
    \end{split}
\end{equation}
where $S_{\text{vN}}$ denotes the von Neumann entanglement entropy and $\rho^{\langle I_{\text{left}}|}_{t_i}$ is the reduced density matrix of $\langle I_{\text{left}}|$ with respect to the temporal bipartition at time $t_i$:
\begin{equation}
    \begin{split}
        \rho^{\langle I_{\text{left}}|}_{t_i}=\quad \vcenter{\hbox{\includegraphics[width=0.5\linewidth]{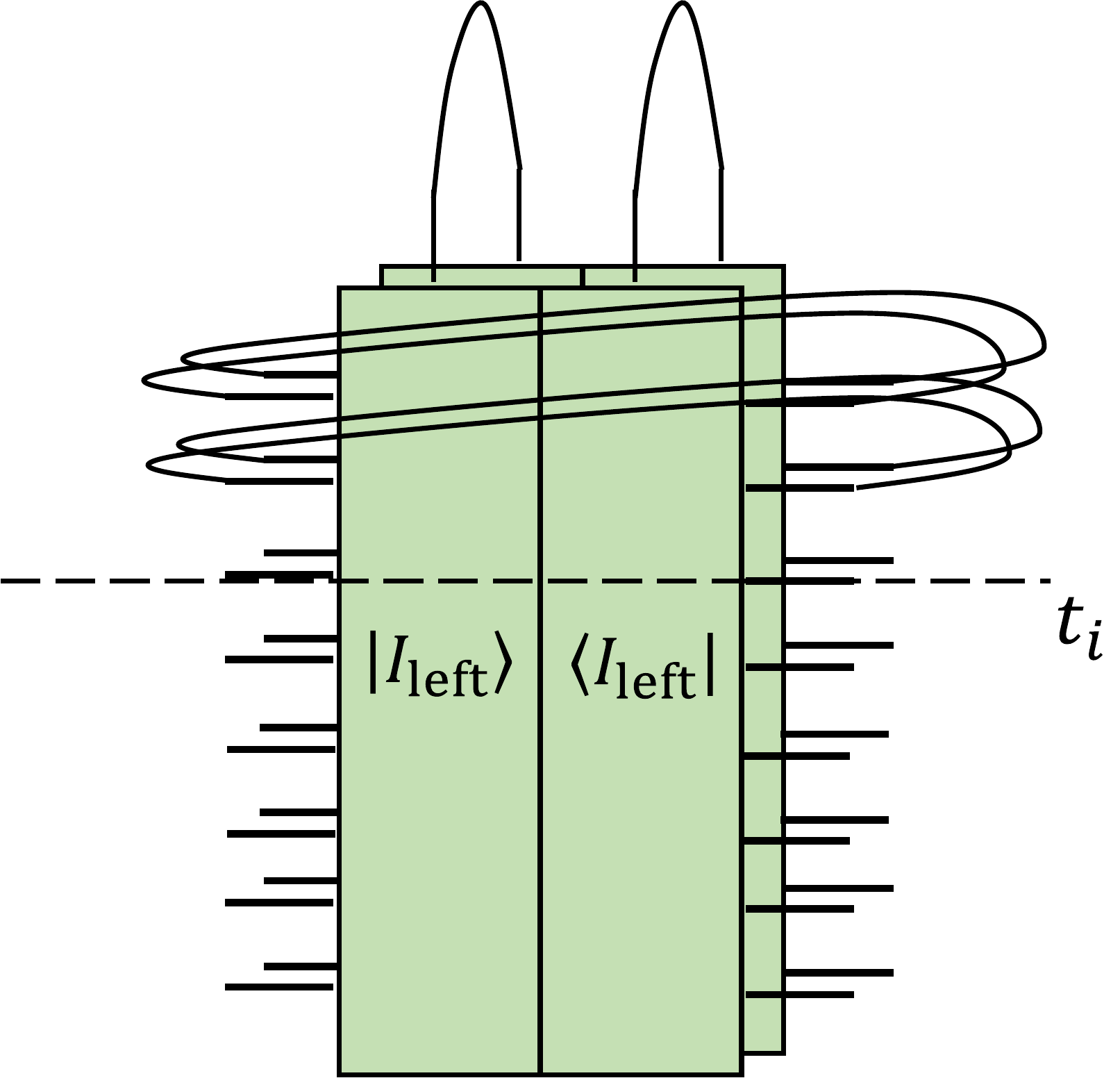}}}
    \end{split}
\end{equation}

Analogous to the Feynman-Vernon influence functional \cite{feynman_theory_1963}, the vectorized tensors $\langle I_{\text{left}}|$ and $|I_{\text{right}}\rangle$ are dubbed ``influence matrices" and interpreted as effective baths for the ``impurity" $\mathcal{T}_O$ in Ref. \cite{lerose_influence_2021}. Temporal entanglement therefore characterizes the memory effects, or non-Markovianity, of the effective bath.

Within the context of open quantum systems coupled to non-Markovian environments, an object analogous to the influence matrix was proposed and dubbed the ``process tensor" \cite{pollock_non_2018, strathearn_efficient_2018, gribben_using_2022}. The process tensor captures the effects of a non-Markovian environment and proves useful for studying the dynamics of open quantum system. In the context of the present work, however, the division between the subsystem and the environment is arbitrary due to the translational invariance of the system. The focus is therefore on the influence matrix, which encodes the dynamics of the closed quantum system, possibly subjected to measurements. 

Various works were recently conducted on different aspects of the influence matrix and temporal entanglement. Areas of interest include the behavior of temporal entanglement for exactly solvable dynamics and dynamics close to integrability \cite{ klobas_exact_2021,lerose_scaling_2021, giudice_temporal_2022}, using the influence matrix for treating quantum impurity problems \cite{thoenniss_efficient_2023, kloss_equilibrium_2023, thoenniss_nonequilibrium_2023}, as well as using temporal entanglement as a way of characterizing generic quantum many-body dynamics \cite{foligno_temporal_2023}. 

Due to the action of tracing out these bath degrees of freedom at the end of the time evolution, or the ``temporal boundary", temporal entanglement generally displays behaviors different from its spatial counterpart. For instance, a system with spatial entanglement scaling as volume-law with system size might have temporal entanglement that scales as area-law with the total evolution time, and vice versa. A system which cannot be efficiently simulated along the temporal direction could admit a chance of being efficiently simulated along the spatial direction \cite{hastings_connecting_2015,tirrito_characterizing_2022,carignano_temporal_2023}. Observations as such motivate, from a numerical methodology point of view, study on the temporal entanglement scaling of (1+1) D quantum systems. 

From a phenomenology perspective, temporal entanglement attracts interest in its own right, since it serves as a potential diagnostic for the nature of the quantum many-body dynamics. Depending on whether the system is free, interacting integrable, or chaotic, its temporal entanglement has been shown to display different scaling behaviors. With certain free systems admitting area-law temporal entanglement (TE) \cite{sonner_influence_2021}, certain interacting integrable systems admitting log-law TE \cite{giudice_temporal_2022}, and generic chaotic systems admitting volume-law TE \cite{foligno_temporal_2023}.

Despite growing interest in TE and its implication on the dynamics, analytical treatments of the TE at finite bath size -- particularly its scaling with bath size -- remains lacking. Such properties are relevant since 1) TE is known to not increase monotonously with bath size, but rather assumes a ``barrier-like" shape \cite{lerose_overcoming_2023}; it is the peak of the barrier, rather than the infinite-bath limit value of TE, that ultimately determines the numerical simulability of the dynamics; 2) without proper knowledge of the scaling of TE with bath size, or shape of the temporal entanglement barrier (TEB), convergence of the influence matrix (IM) to its thermodynamic-limit value may be difficult to determine; 3) the shape of the TEB carries additional information about the many-body dynamics that is not accessible from just the thermodynamic-limit value.
\begin{figure}
    \centering
    \includegraphics[width=0.7\columnwidth]{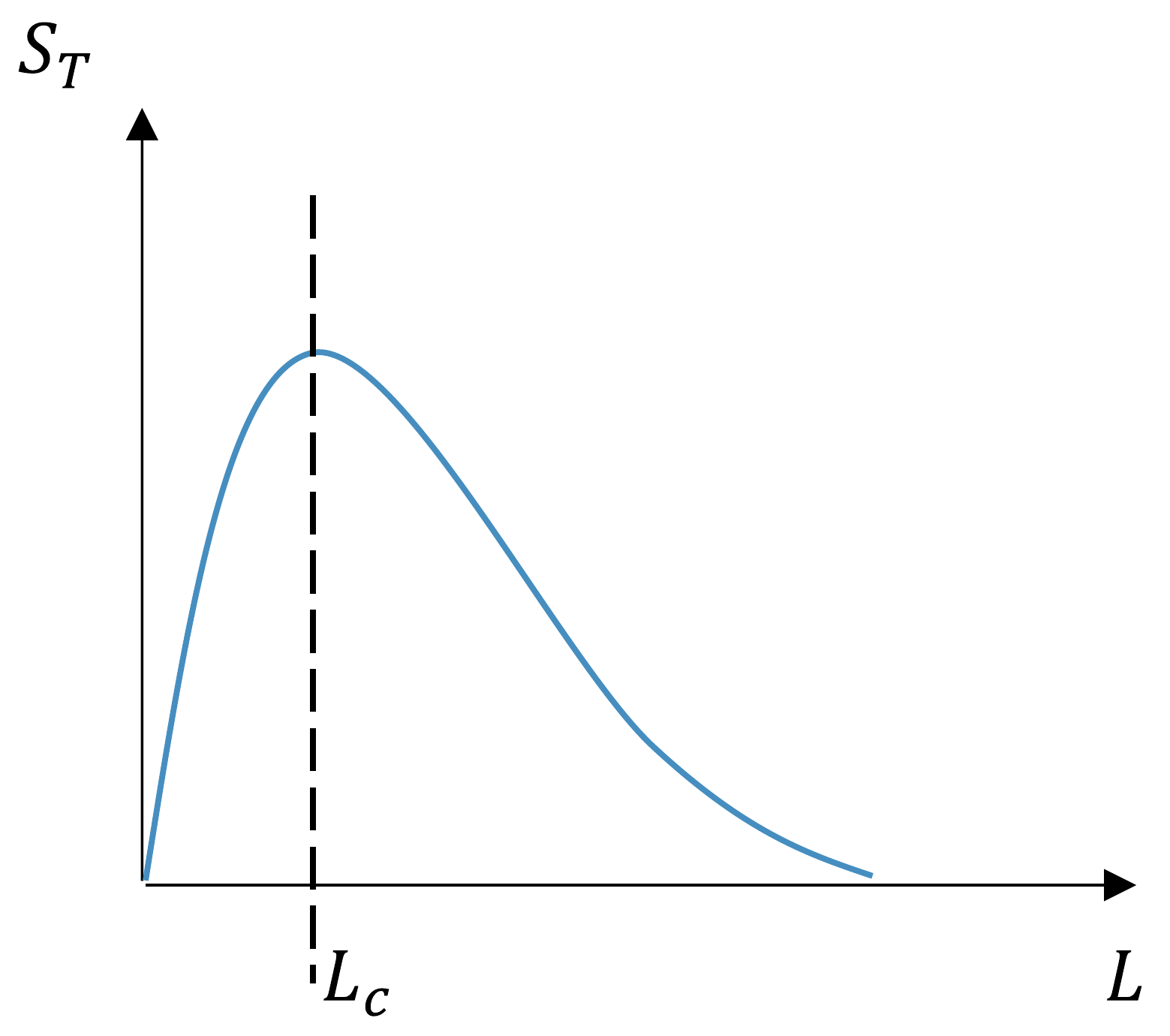}
    \caption{Schematic illustration of the temporal entanglement $S_T$ at different bath sizes $L$. $S_T$ does not increase monotonously with $L$ but rather assumes a barrier-like shape. The bath size corresponding to the peak of the barrier is dubbed the critical bath size $L_c$.}
    \label{fig:schematic}
\end{figure}
The typical behavior of temporal entanglement with bath size is illustrated in Fig.~\ref{fig:schematic}, peaking at a critical bath size $L_c$. 
Furthermore, only systems with unitary dynamics have been studied so far, and TE in non-unitary systems, particularly monitored quantum circuits, have not yet been treated. Given the rapidly growing relevance of non-unitary quantum dynamics, it is natural to aim at developing a treatment for TE in non-unitary systems. 

The present work fills the aforementioned two gaps: numerical and analytical characterizations are provided on the scaling of temporal entanglement with both total evolution time and bath size in quantum circuits with and without probabilistic measurements. The study is restricted to the class of dual-unitary Clifford circuits, with only measurements that preserve the spatial unitarity, in order to admit analytic results and allow for numerical simulations for large system sizes. However, most results on the dynamics without measurements directly extend to generic, i.e. non-Clifford, dual-unitary circuits. Furthermore, even for this restricted choice of gates and measurements qualitatively different behaviors of the TE can be observed.

\subsection{Outline of the Paper}

This paper is structured as follows. Sec. \ref{sec:setup} introduces the structure of the quantum circuits under study. Sec. \ref{sec:DU_no_meas} presents numerical and analytical results on temporal entanglement (TE) in circuits without measurements. Sec. \ref{sec:DU_with_meas} presents numerical results on monitored ``poor scrambler" and ``good scrambler" classes of circuits, where qualitatively different behaviors are observed. Secs.\ref{sec:analytical-growth}, \ref{sec:analytical_decay} and \ref{sec:peak} then respectively present analytical results on the growth, decay, and peak value of the temporal entanglement barriers in the two classes of circuits. Sec. \ref{sec:nonH} presents numerical and analytical results on a non-Hermitian phase transition as the measurement rate is varied. Conclusions are presented in Sec.~\ref{sec:conclude}.

\subsection{Summary of Key Results}
\begin{table*}
    \centering
    \begin{tabular}{|C{0.25\textwidth}|C{0.1\textwidth}|C{0.1\textwidth}|C{0.1\textwidth}|C{0.1\textwidth}|C{0.1\textwidth}|C{0.1\textwidth}|}
        \hline
         &   $S_T$ growth &  $S_T$ decay & $S_{T}$ peak & $\lim_{L\rightarrow \infty} S_T$   & $L_c$ \\
         \hline
        Dual-Unitary &  $2L$ & $T-L$ & $2T/3$ & 0 & $T/3$\\
        \hline
        Monitored Good Scrambler & $2v_{E}L$  & $T-v_{E}L $  & 
        $2T/3 $ & 0& $T/(3v_{E})$ \\
        \hline
        Monitored Poor Scrambler & $\propto \sqrt{L}$  & $\propto \exp{(-L/\xi)} $  & 
        $T/3 $ & 0 & $\propto T^2$ \\
        \hline
    \end{tabular}
    \caption{Summary of key results. $S_T$ denotes temporal entanglement, $L$ denotes the bath size, and $T$ denotes the total evolution time. The entanglement velocity $v_{E}$ sets the scale of the growth, satisfying $v_E=1$ for dual-unitary dynamics, and $\xi$ denotes a characteristic decay scale, both depending on the measurement rate. The temporal entanglement reaches its peak value at the critical bath size $L_c$. }
    \label{table:key_res}
\end{table*}

Table \ref{table:key_res} summarizes the key findings of this paper. The shape of the temporal entanglement barrier (TEB) is characterized in dual-unitary Clifford circuits with and without measurements, dubbed ``monitored" and ``dual-unitary" respectively. In monitored ``good scrambler" circuits, introducing measurements modifies the speed at which TE grows and decays with system size but otherwise leaves the qualitative features intact. In monitored ``poor scrambler" circuits, introducing measurements changes the initial growth from linear to diffusive and the later decay from linear to exponential. The steady-state value of $S_T$ in the thermodynamic limit of an infinite bath size $L$ remains zero in all three cases, and the peak $S_T$ value in monitored poor scrambler circuits is half that of the dual-unitary and monitored good scrambler circuits. The critical bath size $L_c$ at which the peak value of $S_T$ is reached scales quadratically with $T$ in the monitored case, to be contrasted with the linear scaling with $T$ in the dual-unitary and monitored good scrambler circuits.

\section{Quantum Circuit Setup}\label{sec:setup}

\subsection{Brickwork Circuit Geometry}

The quantum circuit of interest has the so-called ``brickwork" geometry. 
The unitary evolution operators consists of alternating odd and even layers, with each layer consisting of tensor products of two-site unitary gates acting on odd and even bonds respectively:
\begin{equation}\label{eqn:unitary_operator}
\begin{split}
U(T)&=\left(\prod_{i \:\text{odd}}U_{i,\, i+1}\prod_{i\: \text{even}}U_{i,\, i+1}\right)^{T/2}\\
&\text{}\\
&=\quad\vcenter{\hbox{\includegraphics[width=0.55\linewidth]{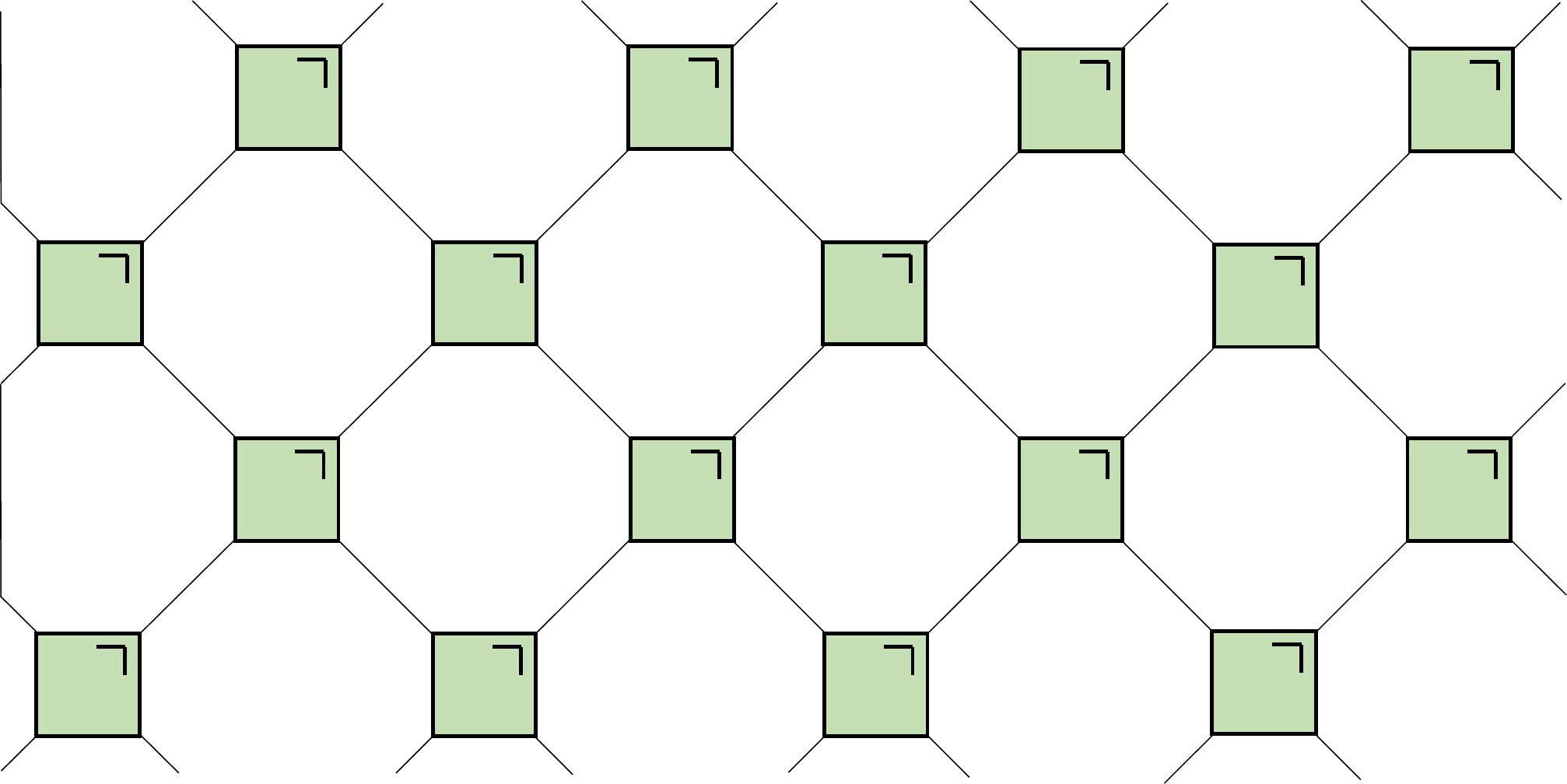}}} \quad\ldots
\end{split}
\end{equation}
where $T$ is the total number of update steps for the Floquet evolution, which is henceforth referred to as the total evolution time. The building blocks of the full evolution operator are given by unitary matrices (gates) graphically represented as 
\begin{equation}
    \begin{split}
        U_{i,\, i+1}=\vcenter{\hbox{\includegraphics[width=0.12\linewidth]{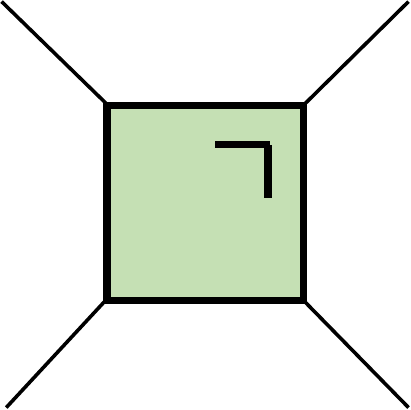}}}\,.
    \end{split}
\end{equation}
The brickwork geometry originates naturally from Trotterized Hamiltonian dynamics, where one alternates between switching on local interactions on all even bonds and all odd bonds, as is done in the Time-Evolving Block Decimation (TEBD) algorithm \cite{verstraete_matrix_2004, vidal_efficient_2004}.   

For simplicity and analytic tractability, the initial state is chosen to be short-range entangled and takes the form:
\begin{equation}\label{eqn:init_pairs}
    \begin{split}
        |\Psi_{0}\rangle&=|\psi\rangle\otimes|\psi\rangle\otimes\ldots\otimes|\psi\rangle \\&=\quad\vcenter{\hbox{\includegraphics[width=0.45\linewidth]{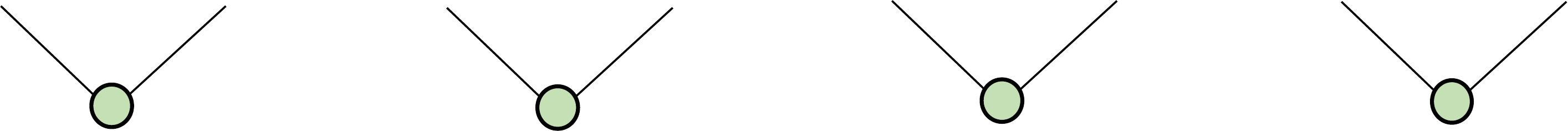}}}\,,
    \end{split}
\end{equation}
where each two-site pair is denoted $|\psi\rangle$:
\begin{equation}\label{eqn:single_pair}
    \begin{split}
        |\psi\rangle=\sum_{a,\, b}\psi_{ab}\,|a\rangle\otimes|b\rangle, \qquad
        \psi_{ab}=\;\vcenter{\hbox{\includegraphics[width=0.12\linewidth]{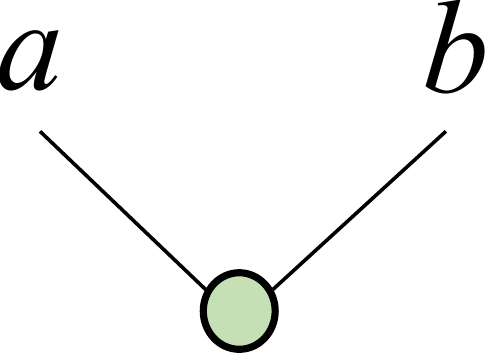}}}\,.
    \end{split}
\end{equation}
Taken together, the time-evolved state in the folded picture under the brickwork circuit can be graphically denoted as
\begin{equation}
    \begin{split}
       &(U\otimes U^*)|\Psi_0\rangle\otimes|\Psi_0\rangle^* = \\
       &\qquad\vcenter{\hbox{\includegraphics[width=0.55\linewidth]{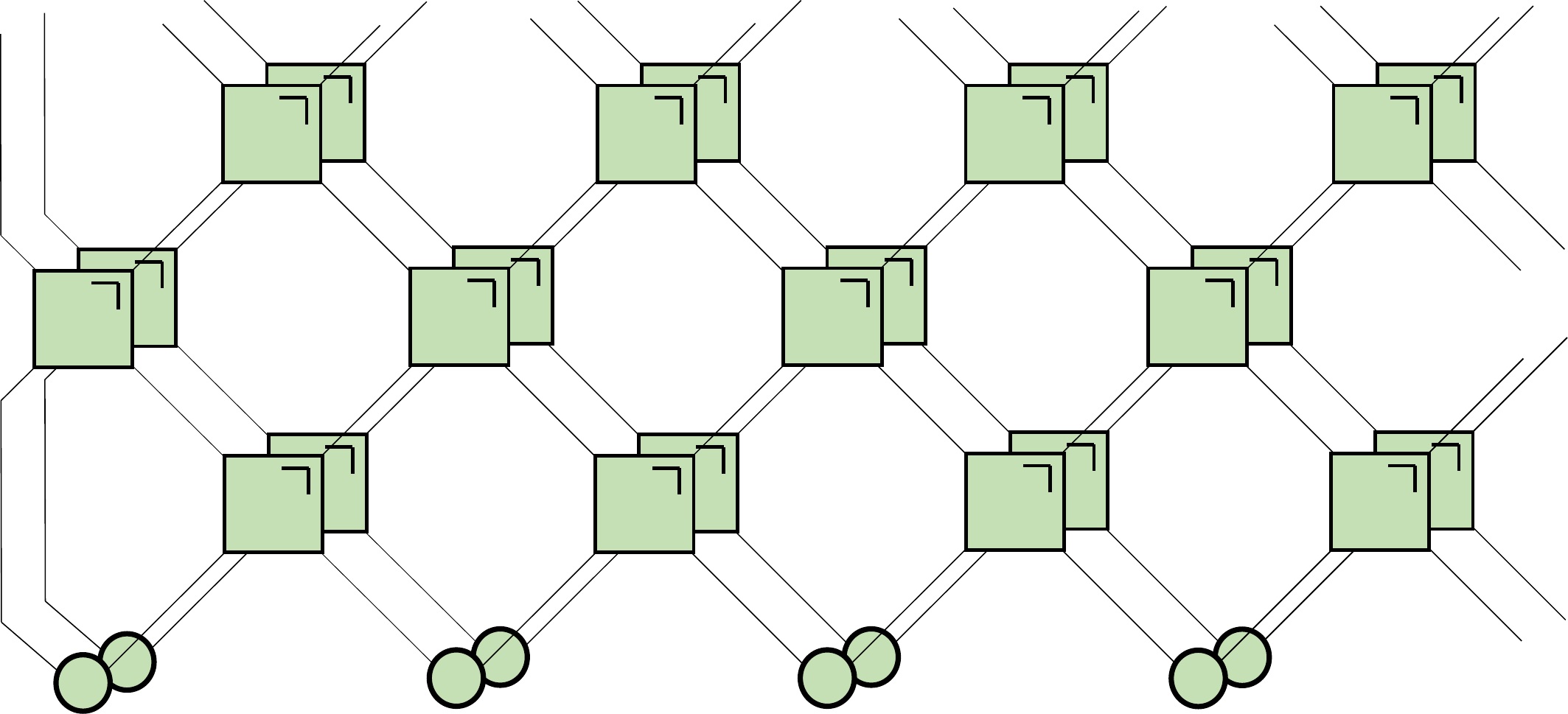}}}
    \end{split}
\end{equation}
For convenience, we use the same graphical notation for $U$ and $U^*$, with the implicit convention that all circuits in the top (bottom) layer correspond to $U$ ($U^*$).
The trace operation is then applied at the end of the time evolution: 
\begin{equation}\label{eq:Ileft_circ}
    \begin{split}
        \langle I_{\text{left}}|&=\langle \tr|(U\otimes U^*)|\Psi_0\rangle\otimes|\Psi_0\rangle^* \\
        &=\vcenter{\hbox{\includegraphics[width=0.55\linewidth]{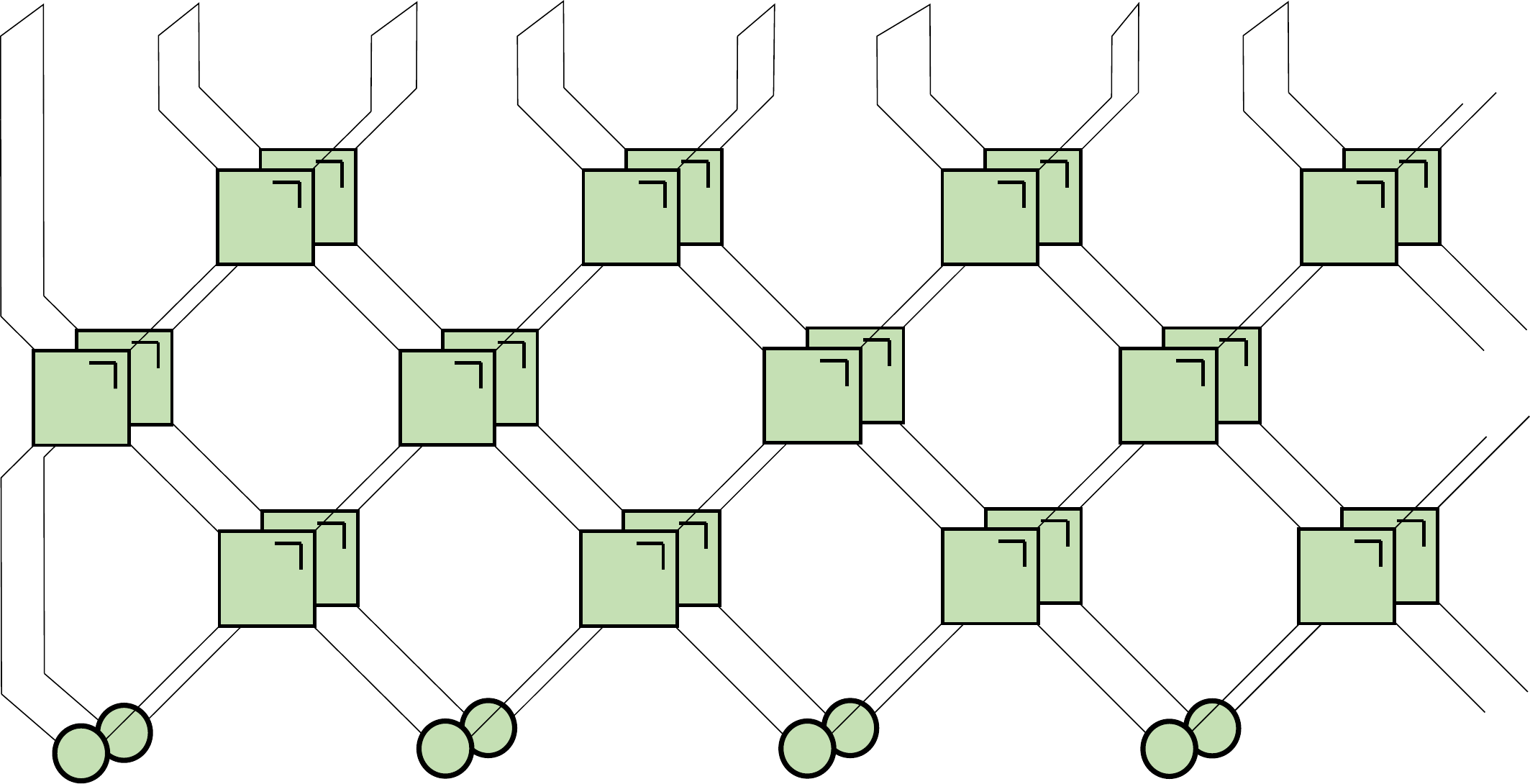}}}
    \end{split}
\end{equation}
Following Ref. \cite{lerose_influence_2021}, the layer containing $U$ is dubbed the ``forward time contour", and the layer containing $U^*$ is dubbed the ``backward time contour". 
We can identify a spatial transfer matrix $\mathcal{T}$ as
\begin{equation}\label{eqn:spatial_transfer_matrix}
    \mathcal{T}= \quad\vcenter{\hbox{\includegraphics[width=0.2\linewidth]{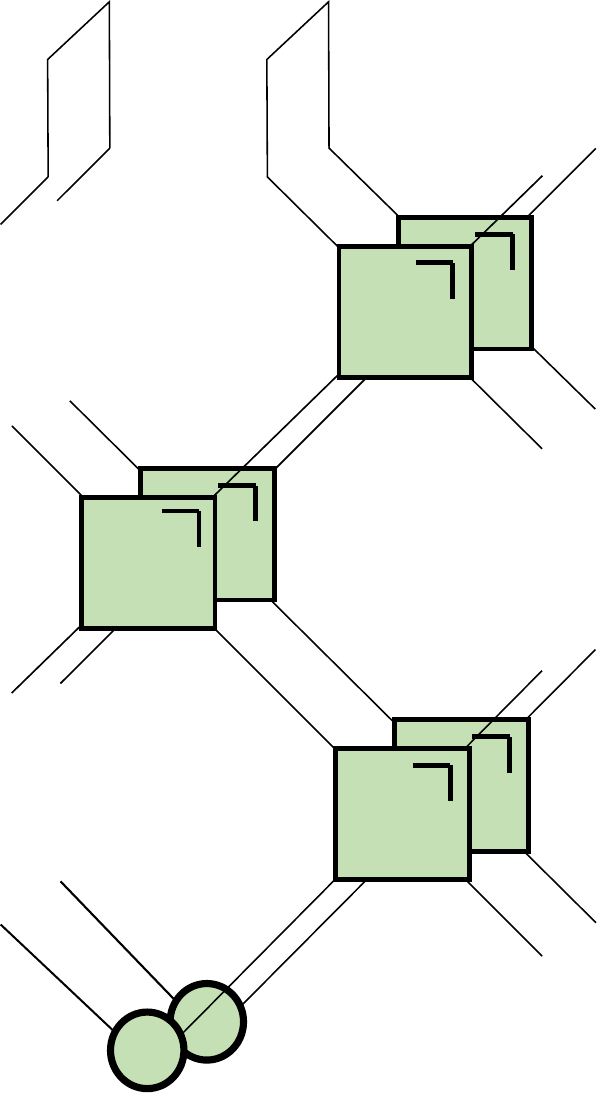}}}
\end{equation}

\subsection{Dual-Unitary and Clifford Gates}\label{sec:DU_clifford}

In this work we will restrict ourselves to dual-unitary Clifford gates. The two-site unitary gates appearing in the brickwork circuit can generally be any element of $U(4)$. Any choice of local unitary gates leads to global unitary time evolution, and we will refer to the unitarity as temporal unitarity. Graphically, this property reads:
\begin{equation}
    \text{Temporal Unitarity} \quad\Leftrightarrow \quad \vcenter{\hbox{\includegraphics[width=0.12\linewidth]{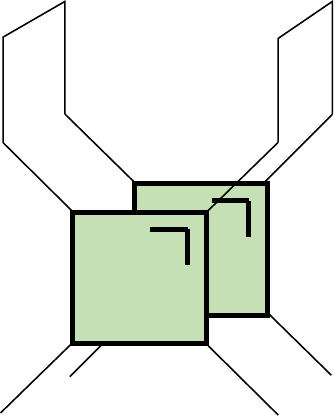}}}\quad=\quad \vcenter{\hbox{\includegraphics[width=0.12\linewidth]{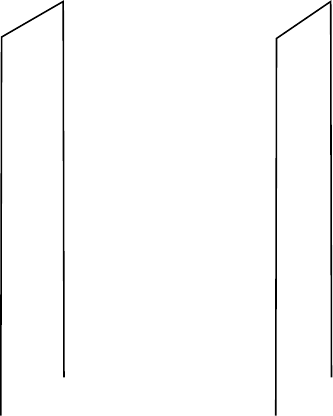}}}
\end{equation}
Spatial unitarity can be analogously defined as:
\begin{equation}
    \text{Spatial Unitarity} \quad\Leftrightarrow \quad \vcenter{\hbox{\includegraphics[width=0.14\linewidth]{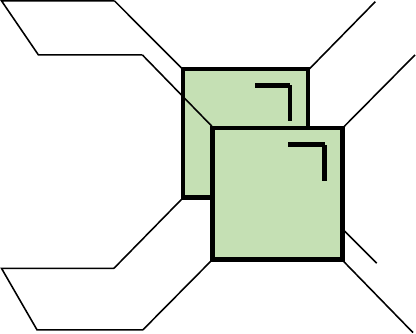}}}\quad=\quad \vcenter{\hbox{\includegraphics[width=0.14\linewidth]{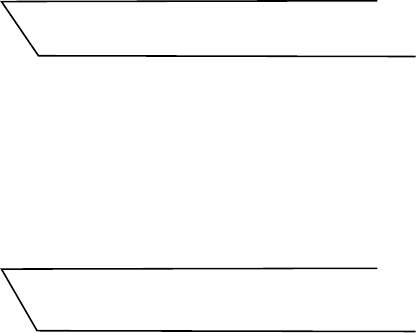}}}
\end{equation}
A gate with temporal unitarity generically does not possess spatial unitarity. In the case where the gate possesses both, it is referred to as being ``dual-unitary"~\cite{gopalakrishnan_unitary_2019,bertini_exact_2019}.

A general parametrization for two-site dual-unitary gates on qubits is given by \cite{bertini_exact_2019} as
\begin{equation}
    U=e^{i\phi}(u_+\otimes u_-)V[J](v_-\otimes v_+)\,,
\end{equation}
where $\phi,\, J\in \mathbb{R}$, $u_{\pm}, \, v_{\pm}\in SU(2) $, and the entangling gate is defined as
\begin{equation}\label{eqn:trotter_xxz}
    V[J]=\exp\left[ -i\left( \frac{\pi}{4}\sigma_x\otimes\sigma_x+\frac{\pi}{4}\sigma_y\otimes\sigma_y+J\sigma_z\otimes\sigma_z   \right)   \right]\,,
\end{equation}
also known as the Trotterized XXZ gate. A brickwork circuit consisting of dual-unitary gates is itself dual-unitary. 

The Clifford property refers to the fact that the gate can be generated from a specific set of gates, namely
\begin{align}
    U_{\text{Clifford}}\in \langle \{H, S, \text{CNOT}\}\rangle\,
\end{align}
where, in the standard computational basis,
\begin{equation}
    \begin{split}
        H=\frac{1}{\sqrt{2}}\begin{bmatrix}
            1 & 1 \\
            1 & -1
        \end{bmatrix}, \,\, S=\begin{bmatrix}
            1 & 0 \\
            0 & i
        \end{bmatrix},  \,\,
        \text{CNOT}= \begin{bmatrix}
            1&0&0&0\\
            0&1&0&0\\
            0&0&0&1\\
            0&0&1&0
        \end{bmatrix}.
    \end{split}
\end{equation}
Up to single-site Clifford gates, there are two classes of two-site gates that are both dual-unitary and Clifford: the SWAP class and the iSWAP class. Brickwork circuits consisting of dual-unitary Clifford gates were previously studied in Ref.~\cite{sommers_crystalline_2023}, where such nonrandom quantum circuits are dubbed ``crystalline quantum circuits". 

The SWAP and the iSWAP gates corresponds to the Heisenberg and the XX points in the dual-unitary parameterization, where
\begin{equation}
    U_{\text{SWAP}}=V[J=\frac{\pi}{4}], \qquad U_{\text{iSWAP}}=V[J=0],
\end{equation}
with $V[J]$ defined in Eq. \eqref{eqn:trotter_xxz}. 
As pointed out in Ref.~\cite{sommers_crystalline_2023}, there exist two types of gates  within the iSWAP class displaying qualitatively distinct behaviors, dubbed ``poor scramblers" and ``good scramblers" respectively. In the present work, we study one representative gate of each type, both of which are closely related to the circuit representation of the self-dual kicked Ising (SDKI) model. The generic kicked Ising model is described by the Floquet unitary operator:
\begin{equation}
    U_{KI}=e^{-ib\sum_j \sigma^{x}_j}e^{-i\sum_j J\sigma^{z}_{j}\sigma^{z}_{j+1}+h_j \sigma^{z}   },
\end{equation}
The model can be decomposed in two-site gates~\cite{gopalakrishnan_unitary_2019} and is self-dual for parameter values $|J|=|b|=\pi/4$, where it is dubbed the SDKI model. This model is known to be free and integrable at zero longitudinal field strength $h_j=0$ and non-integrable for any longitudinal field strength $h_j\neq 0$

In this work, we consider uniform longitudinal field strength $h_j=h$, for which the SDKI model admits Clifford-circuit decomposition at the points $h=0$ and $h=\pi/4$. We henceforth refer to the free integrable point $h=0$ as SDKI-f and the non-integrable point $h=\pi/4$ as SDKI-S, giving rise to poor and good scramblers respectively. 
The two-site SDKI-f gate and SDKI-S gate are defined through the gate decompositions
\begin{align}
    U_{\text{SDKI-f}}&=CZ(H\otimes H)\text{CZ}, \label{eqn:SDKI-f}\\
    U_{\text{SDKI-S}}&=(S\otimes I)\,U_{\text{SDKI-f}}\,(S\otimes I), \label{eqn:SDKI-s}
\end{align}
where CZ denotes the controlled $Z$ gate, 
\begin{align}
        \text{CZ}= \begin{bmatrix}
            1&0&0&0\\
            0&1&0&0\\
            0&0&1&0\\
            0&0&0&-1
        \end{bmatrix}.
\end{align}


The representative from the poor scrambler type is chosen to be the SDKI-f gate. As for the good scrambler type, however, a gate different from the SDKI-S is chosen, which we dub as the SDKI-r gate, given by:
\begin{equation}\label{eqn:SDKI-r}
        U_{\text{SDKI-r}}=(v_1\otimes I)\,U_{\text{SDKI-f}}\,(v_2\otimes I),
\end{equation}
where $v_1,\, v_2$ are single-site gates chosen independently and uniformly-randomly from the group of single-site Clifford gates. The SDKI-S and the SDKI-r gates display qualitatively similar behaviors, but the SDKI-r gate has the advantage that averaging over single-site Clifford gates yields analytical tractability and allows us to make exact predictions for the operator dynamics. 

The kicked Ising model (KIM) is extensively studied in early works on dual-unitarity \cite{akila_particle_2016, gutkin_exact_2020}, (temporal) entanglement \cite{bertini_entanglement_2019,lerose_influence_2021, lerose_overcoming_2023, giudice_temporal_2022}, emergent quantum state designs \cite{ho_exact_2022, wilming_high_2022}, and measurement-based quantum computing~\cite{stephen2022universal}. The Clifford points, albeit being singular points in the continuous parameter space, admit efficient numerical simulability \cite{aaronson_improved_2004}. This motivates using the Clifford SDKI circuits alongside the SWAP circuit as minimal models for investigating the temporal entanglement profile. As pointed out in Ref. \cite{ippoliti_fractal_2022}, the set of two-site dual-unitary Clifford gates makes up 50\% of total two-site Clifford gates. The other two classes are the CNOT class and the identity class \cite{crooks_gates_2020}. As such, the gate choice of being dual-unitary and Clifford is arguably not overly-constrained. While circuits consisting of CNOT gates are not dual-unitary, they are extensively studied in the contexts of the Floquet quantum East model \cite{bertini_localised_2023, bertini_exact_2023} as well as realizing generalized dual-unitary circuits~\cite{yu_hierarchical_2023}. Various recent works studied the entanglement membrane of such circuits~\cite{rampp_entanglement_2023, foligno_quantum_2023}, with Ref.~\cite{sommers_zero-temperature_2024} focusing on generalized dual-unitary Clifford circuits.
 

\subsection{Space-time Rotation, Significance of the Trace Operation and the Perfect-Dephaser Limit}

The move from spatial entanglement to temporal entanglement fits within the larger frame of space-time rotation, exchanging the role of discrete time and discrete space in lattice circuit models~\cite{ippoliti_postselection_2021, lu_spacetime_2021,garratt_local_2021,ippoliti_fractal_2022,kos_correlations_2021}. 
After space-time rotation, a two-site unitary gate $U$ with matrix elements $U_{ab,cd}$ becomes a gate $\tilde{U}$ with matrix elements $\Tilde{U}_{bd,ac} = U_{ab,cd}$:
\begin{equation}
    \begin{split}
        U_{ab,cd}=\; \vcenter{\hbox{\includegraphics[width=0.15\linewidth]{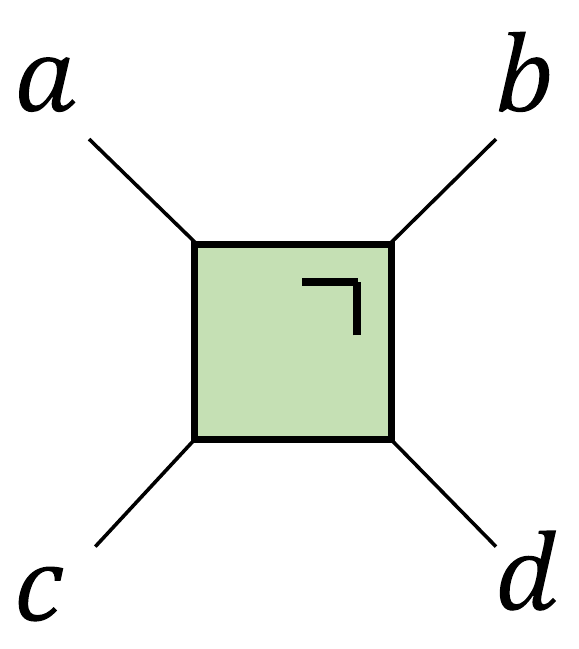}}}\; \xrightarrow{\text{space-time rotation}}\; \Tilde{U}_{bd,ac}=\vcenter{\hbox{\includegraphics[width=0.15\linewidth]{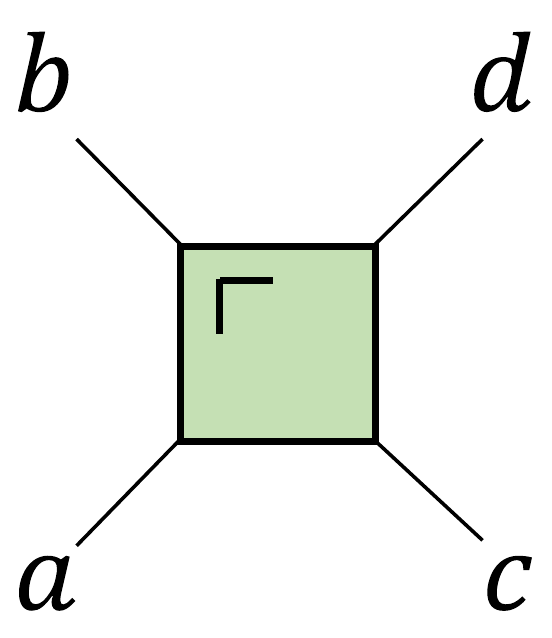}}}
    \end{split}
\end{equation}
Dual-unitary gates and dual-unitary circuits remain unitary after space-time rotation by construction.

The choice of initial state can possibly undermine the dual-unitarity of the circuit. For short-range entangled states as introduced in Eq. \eqref{eqn:init_pairs} and \eqref{eqn:single_pair}, unitarity along the spatial direction results in the condition $\sum_j \psi_{i,j}\psi^*_{k,j}=\delta_{ik}$, leading to so-called ``solvable" initial states satisfying~\cite{piroli_exact_2020,foligno_temporal_2023}:
\begin{equation}
    \begin{split}
    \vcenter{\hbox{\includegraphics[width=0.2\linewidth]{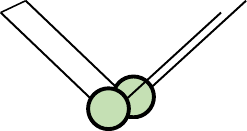}}}\quad =\quad \vcenter{\hbox{\includegraphics[width=0.2\linewidth]{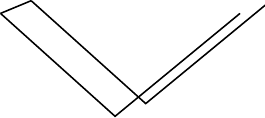}}}
    \end{split}
\end{equation}
By construction, such states possess spatial unitarity. When contracted to a dual-unitary circuit, the contracted circuit remains unitary after space-time rotation. 

Beyond short-range entangled states, solvability for generalized MPS initial states is discussed in Ref. \cite{piroli_exact_2020}. For simplicity and in order to preserve the Clifford property, we choose $|\psi\rangle=(|00\rangle+|11\rangle)/\sqrt{2}\equiv |B_I\rangle$, such that:
\begin{equation}
    \begin{split}
        \vcenter{\hbox{\includegraphics[width=0.18\linewidth]{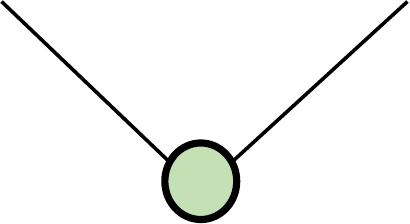}}}\quad =\quad \vcenter{\hbox{\includegraphics[width=0.15\linewidth]{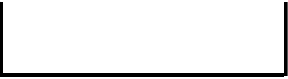}}}
    \end{split}
\end{equation}

Given the open boundary condition defined by the brickwork circuit structure in Eq. \eqref{eqn:unitary_operator}, the temporal MPS $\langle\Phi|$ at the spatial boundary consists of Bell-pair states connecting the site at time $t_i$ to the site at time $t_{i+1}$:
\begin{equation}\label{eqn:phi1}
\begin{split}
    \langle \phi|\equiv\quad \vcenter{\hbox{\includegraphics[width=0.1\linewidth]{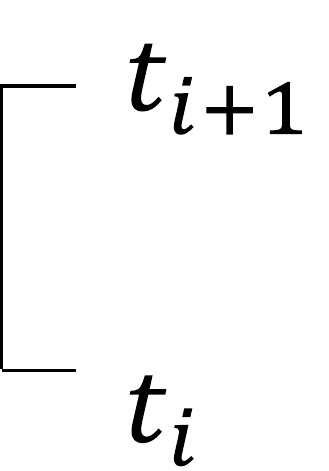}}}\;,\quad
    \langle \Phi|=\langle \phi|\otimes \langle \phi|\otimes \ldots \otimes \langle \phi|
    =\; \vcenter{\hbox{\includegraphics[width=0.08\linewidth]{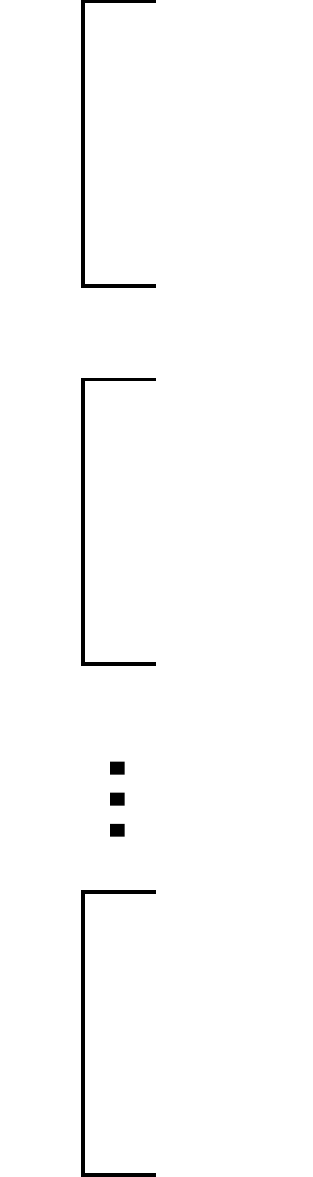}}}
    \end{split}
\end{equation}
It will prove to be convenient to additionally define $\langle\theta|$ as a Bell pair connecting two equal-time sites at $t_i$ on the forward and backward time contours:
\begin{equation}
    \langle\theta|\equiv\quad\vcenter{\hbox{\includegraphics[width=0.12\linewidth]{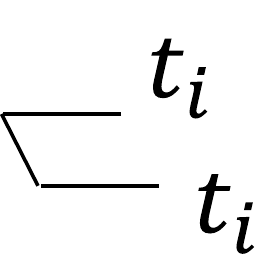}}}
\end{equation}

After space-time rotation, each pair of trace operations becomes a projector $|\theta\rangle\langle\theta|$:

\begin{equation}
    \begin{split}
        \vcenter{\hbox{\includegraphics[width=0.2\linewidth]{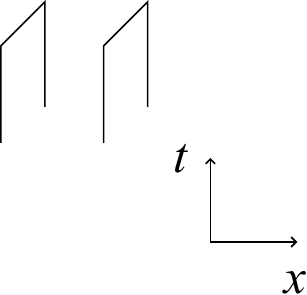}}}\quad \xrightarrow[\text{trace becomes projector}]{\text{space-time rotation}}\quad \vcenter{\hbox{\includegraphics[width=0.2\linewidth]{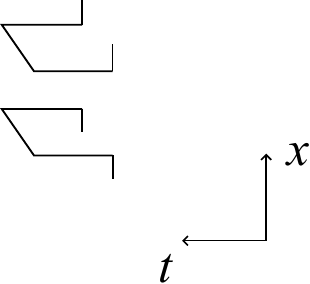}}}
    \end{split}
\end{equation}
Such projector is henceforth dubbed the ``trace projector". 

Although dual-unitary gates and solvable initial conditions preserve spatial unitarity, the trace projector does not, rendering $\mathcal{T}$ non-unitary. Therefore, the behavior of temporal entanglement under update by $\mathcal{T}$ is generally different from the behavior of spatial entanglement under update by $U\otimes U^*$. 

Ref. \cite{lerose_influence_2021} pointed out that for dual-unitary circuits with solvable initial states the influence matrix $\langle I_{\text{left}}|$ always reduces to the so-called ``perfect-dephaser" form for $L\ge T$, where $T$ is the total number of time steps of the Floquet evolution:
\begin{align}\label{eqn:perfect_dephaser}
       & \langle I_{\text{PD}}|\equiv\langle \theta|\otimes\langle \theta |\otimes\ldots\otimes\langle\theta|
        =\; \vcenter{\hbox{\includegraphics[width=0.1\linewidth]{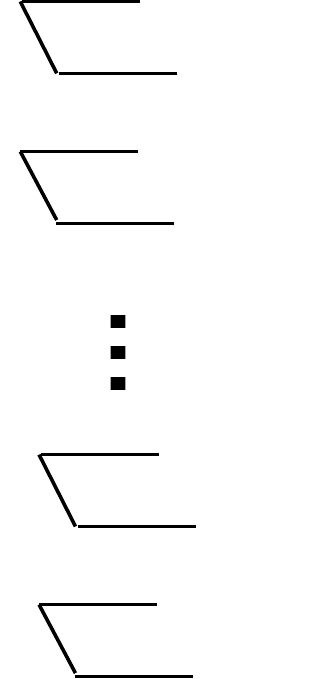}}}
\end{align}
satisfying
\begin{align}
 \langle I_{\text{PD}}|\mathcal{T}=\langle I_{\text{PD}}|\,.
\end{align}

The perfect-dephaser influence matrix $\langle I_{\text{PD}}|$ has zero temporal entanglement, since temporal bipartitions would not cut across any Bell pairs. Therefore, dual-unitary circuit with solvable initial states always reaches zero temporal entanglement at system sizes $L\ge T$.  


\section{Dual-Unitary Clifford Circuits without Measurements}\label{sec:DU_no_meas}
This Section presents numerical and analytical results on temporal entanglement in dual-unitary Clifford circuits. The numerical results are obtained using the stabilizer formalism for simulating Clifford circuits \cite{aaronson_improved_2004, li_measurement_2019}. 
\subsection{Numerical Results}\label{sec:num_no_meas}

\begin{figure}[tb!]
\includegraphics[width=\columnwidth]{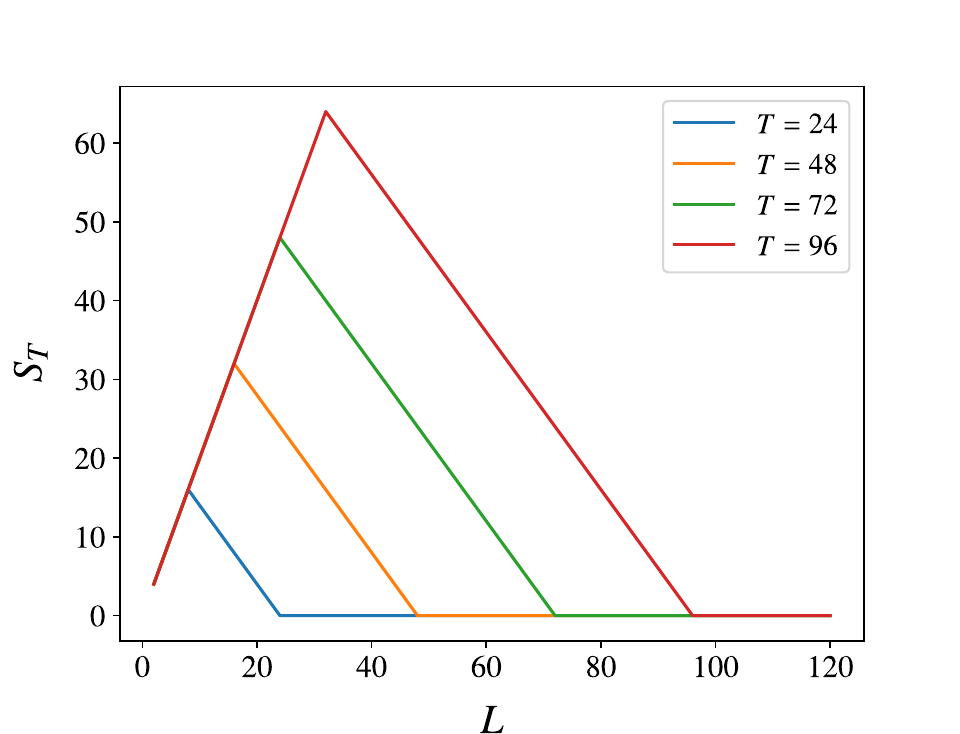}
\caption{Temporal entanglement $S_T$ plotted against bath size $L$, with total evolution times $T=24, 48, 72, 96$. Results are identical for all dual-unitary Clifford circuits considered in this work.}
\label{fig:TE_no_meas}
\end{figure}
Fig. \ref{fig:TE_no_meas} shows the temporal entanglement as a function of the bath size $L$ at various total evolution times $T$. The gates are chosen to be SDKI-f for concreteness, with identical results for the other gates discussed in this work. We can clearly identify three distinct regimes: an initial linear growth of temporal entanglement is followed by a linear decay, before saturating at a zero value. The overall function is piecewise linear and given by:
\begin{align}\label{eqn:num_res_no-meas}
S_T(L) = \begin{cases}
2L &\text{\; for\; }L<\frac{T}{3}\\
T-L &\text{\; for\; } \frac{T}{3}\le L<T\\
0 &\text{\; for \; } L\ge T
\end{cases}
\end{align}

Let us comment on some qualitative features. First, at $L=T/3$ and $L=T$, the TE changes non-analytically. This is because entanglement in Clifford circuits comes in the form of Bell pairs, and the number of Bell pairs across a given bipartition can only increase or decrease as integers. Second, temporal entanglement always decays to exactly zero for finite $L\ge T$. In this regime we recover the perfect dephaser limit [Eq. \eqref{eqn:perfect_dephaser}], with $\langle I_{\text{left}}|=\langle I_{\text{PD}}|$ for  $L\ge T$. Third, the peak TE scales linearly with the total evolution time $T$. Therefore, in the alternative order of limits where first $T\rightarrow\infty$ then $L\rightarrow\infty$ is taken, the TE would exhibit volume-law growth with $T$.

\subsection{Analytical Derivation Through Diagrammatic Contractions}\label{sec:analytical_no_meas}
These different regimes and the corresponding TE can be analytically obtained using standard graphical manipulations. For stabilizer states, all orders of R\'enyi entropies are identical and equal the von Neumann entropy. We can hence focus on the second R\'enyi entropy, since it requires the smallest tensor power of $\langle I_{\text{left}}|$:
\begin{align}\label{eq:renyi}
         S_T^{(t_i)}&=S_{\text{vN}}\left( \rho_{t_i}^{\langle I_{\text{left}}|} \right) = -\log_2 \frac{\tr\left( \left(\rho_{t_i}^{\langle I_{\text{left}}|}\right)^2 \right)}{\left(\tr \rho^{\langle I_{\text{left}}|}\right)^2} \nonumber\\
         &\equiv-\log_2 D_{t_i}^{\langle I_{\text{left}}|},
\end{align}
where $D_{t_i}^{\langle I_{\text{left}}|}$ is the purity with respect to bipartition at $t_i$. 
Using the operator-state mapping, one may write $\rho^{\bra{I_{\text{left}}}}$ as
\begin{equation}
    \begin{split}
        \rho^{\langle I_{\text{left}}|}=|I_{\text{left}}\rangle\langle I_{\text{left}}|\rightarrow |\rho^{\langle I_{\text{left}}|}\rangle\equiv |I_{\text{left}}\rangle\otimes |I_{\text{left}}\rangle^*\,.
    \end{split}
\end{equation}

Graphically representing this expression, each $|I_{\text{left}}\rangle$ contains a layer of $U$ and $U^*$ such that $\left(\rho^{\langle I_{\text{left}}|}\right)^{\otimes 2}$ contains 4 layers of $U$ and $U^*$ each, and we can write:
\begin{equation}
    \begin{split}
        \left(\rho^{\langle I_{\text{left}}|}\right)^{\otimes 2}&=\quad\vcenter{\hbox{\includegraphics[width=0.5\linewidth]{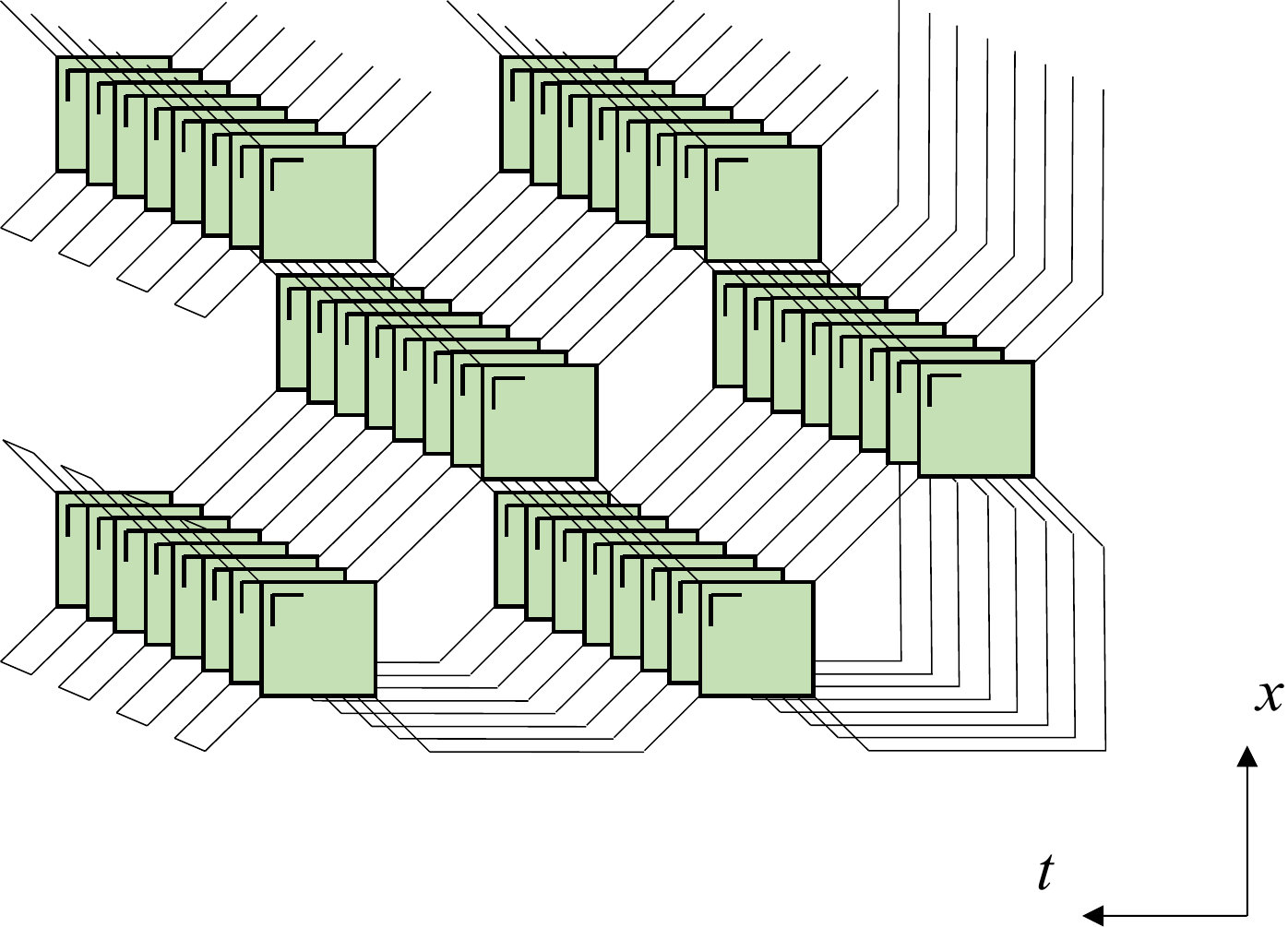}}}
    \end{split}
\end{equation}
Note that the graphical representation circuit is rotated w.r.t. its original representation [Eq.~\eqref{eq:Ileft_circ}], corresponding to a space-time rotation.

It is convenient to define a new merged representation, where:
\begin{equation}
    \begin{split}
        \vcenter{\hbox{\includegraphics[width=0.12\linewidth]{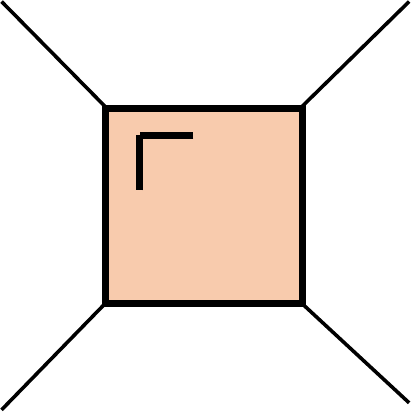}}}\quad\equiv \quad \vcenter{\hbox{\includegraphics[width=0.28\linewidth]{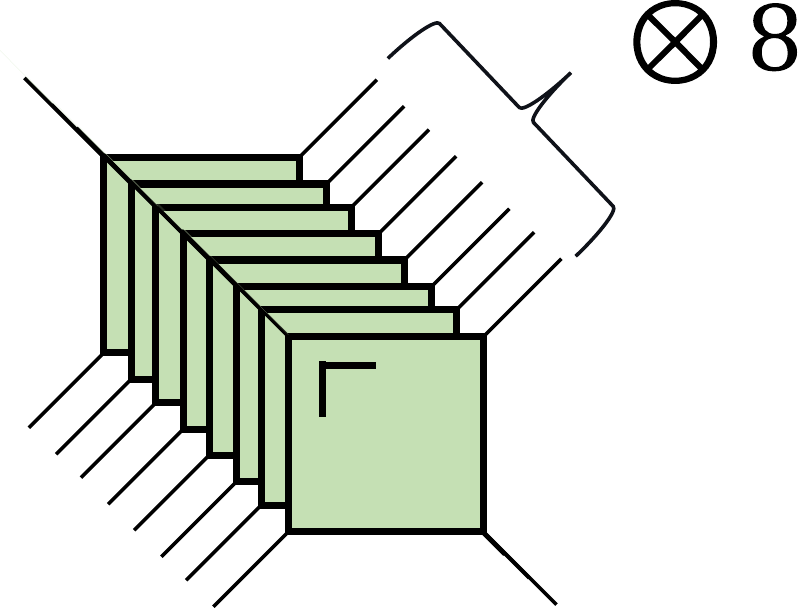}}}
    \end{split}
\end{equation}
These gates are operators acting on pairs of 8 copies of the local Hilbert space. The contraction order associated with the trace projector is denoted by a triangle:
\begin{equation}\label{eqn:boundary_triangle}
    \begin{split}
        \vcenter{\hbox{\includegraphics[width=0.15\linewidth]{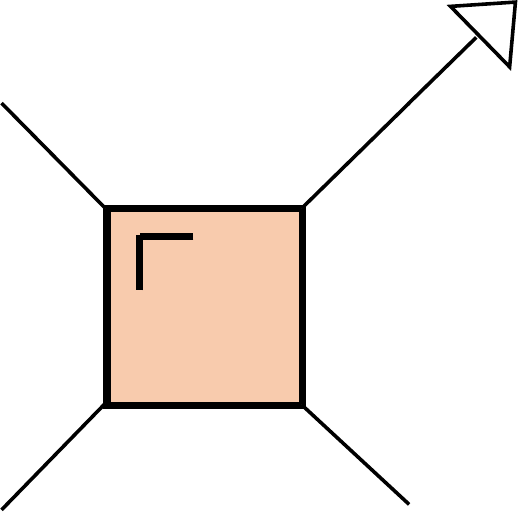}}}\;\equiv \quad \vcenter{\hbox{\includegraphics[width=0.2\linewidth]{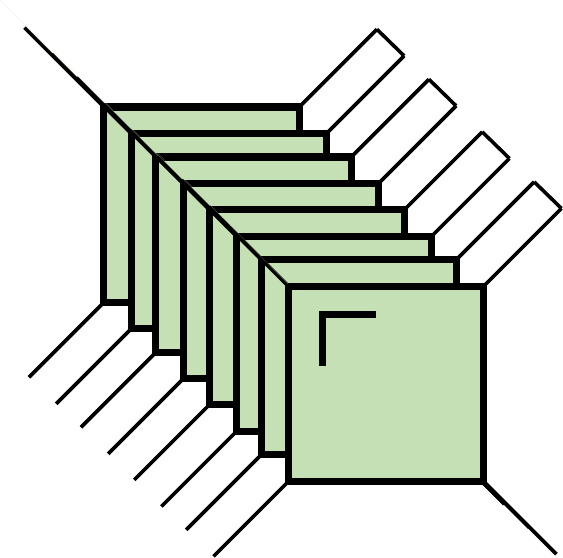}}}\quad,\quad \vcenter{\hbox{\includegraphics[width=0.05\linewidth]{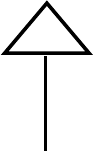}}}\;\equiv \; \vcenter{\hbox{\includegraphics[width=0.2\linewidth]{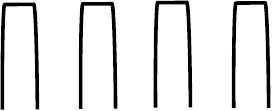}}}
    \end{split}
\end{equation}
corresponding to a state in 8 copies of the local Hilbert space. All necessary contractions for our calculation can be similarly represented, where two additional contraction orders appear on the two sides of the bipartition:
\begin{equation}\label{eqn:boundary_circle_square}
    \begin{split}
        \vcenter{\hbox{\includegraphics[width=0.15\linewidth]{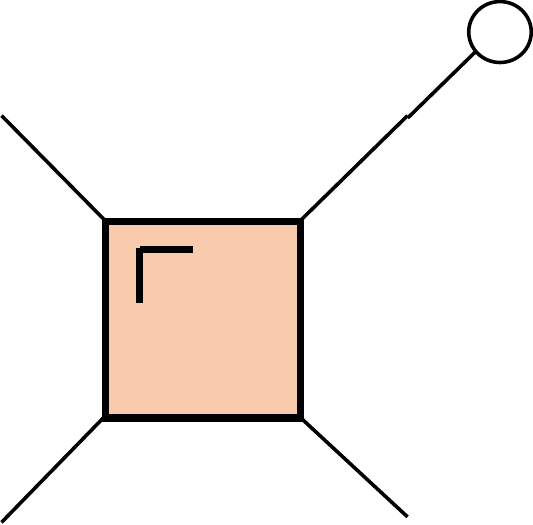}}}\quad&\equiv\quad\vcenter{\hbox{\includegraphics[width=0.2\linewidth]{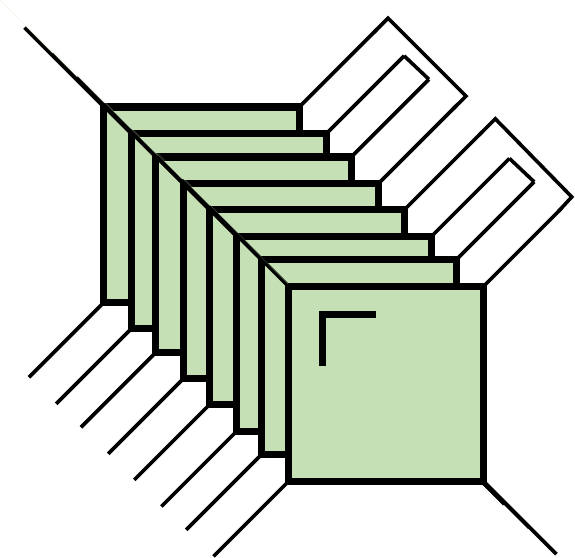}}}
        \quad,\quad \vcenter{\hbox{\includegraphics[width=0.04\linewidth]{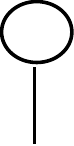}}}\; \equiv \; \vcenter{\hbox{\includegraphics[width=0.2\linewidth]{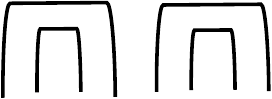}}}\\
        \vcenter{\hbox{\includegraphics[width=0.15\linewidth]{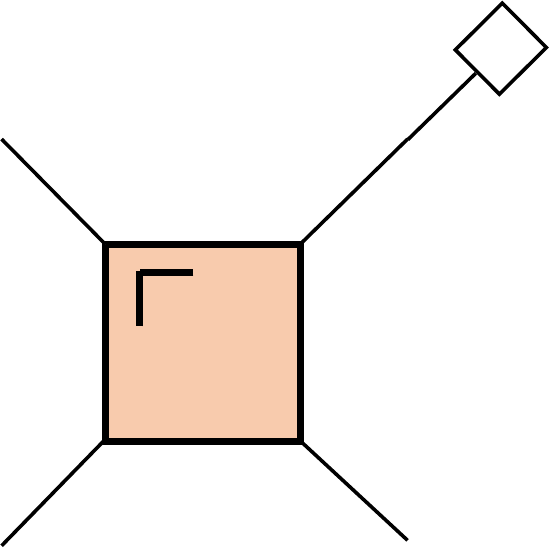}}}\quad&\equiv\quad\vcenter{\hbox{\includegraphics[width=0.2\linewidth]{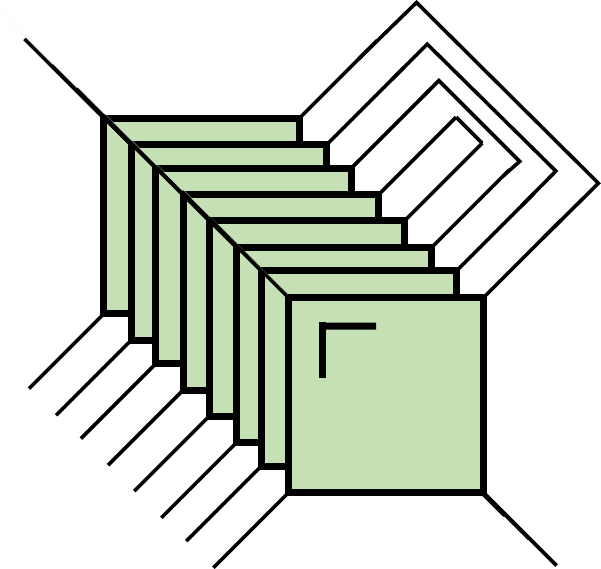}}}
        \quad,\quad \vcenter{\hbox{\includegraphics[width=0.03\linewidth]{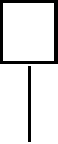}}}\;\equiv \; \vcenter{\hbox{\includegraphics[width=0.2\linewidth]{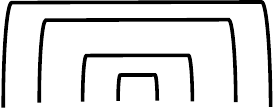}}}
    \end{split}
\end{equation}
Temporal and spatial unitarity lead to a set of graphical identities allowing specific boundary vectors to propagate through the system:
\begin{equation}\label{eq:graphical_U_DU}
    \begin{split}
    \vcenter{\hbox{\includegraphics[width=0.14\linewidth]{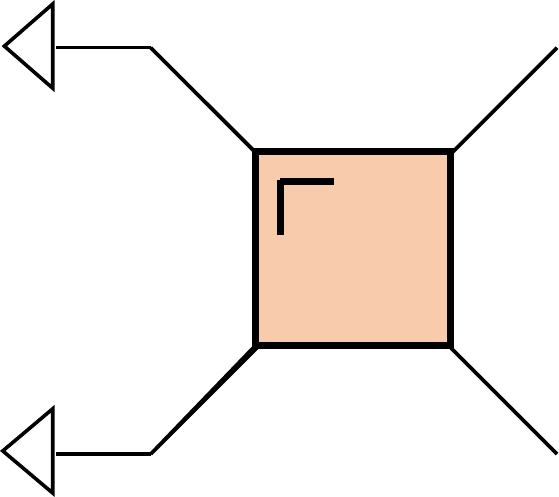}}}\;=\;\vcenter{\hbox{\includegraphics[width=0.14\linewidth]{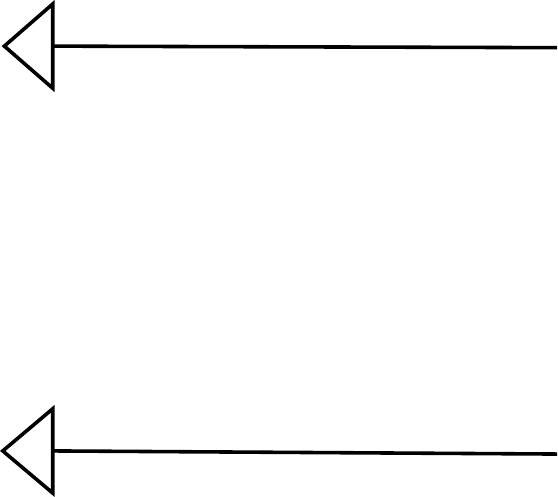}}}\;,\quad \vcenter{\hbox{\includegraphics[width=0.12\linewidth]{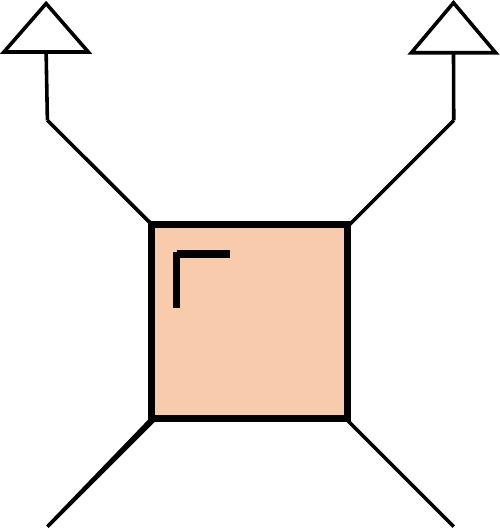}}}\;=\;\vcenter{\hbox{\includegraphics[width=0.12\linewidth]{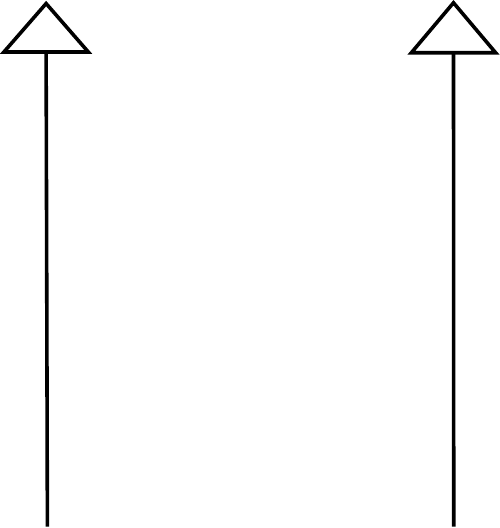}}}
    \end{split}
\end{equation}
and similarly for the circles and the squares in Eq.~\eqref{eqn:boundary_circle_square}.

Let us illustrate how these contractions appear in a simple example. We consider a circuit with $L=4$, $T=8$ and calculate the TE for a bipartition across $t_i=t_1$. The R\'enyi entropy \eqref{eq:renyi} requires evaluating the following two diagrams:
\begin{equation}\label{eqn:purity_contraction}
    \begin{split}
        \tr\left( \left(\rho_{t_i}^{\langle I_{\text{left}}|}\right)^2 \right) &=\;\vcenter{\hbox{\includegraphics[width=0.65\linewidth]{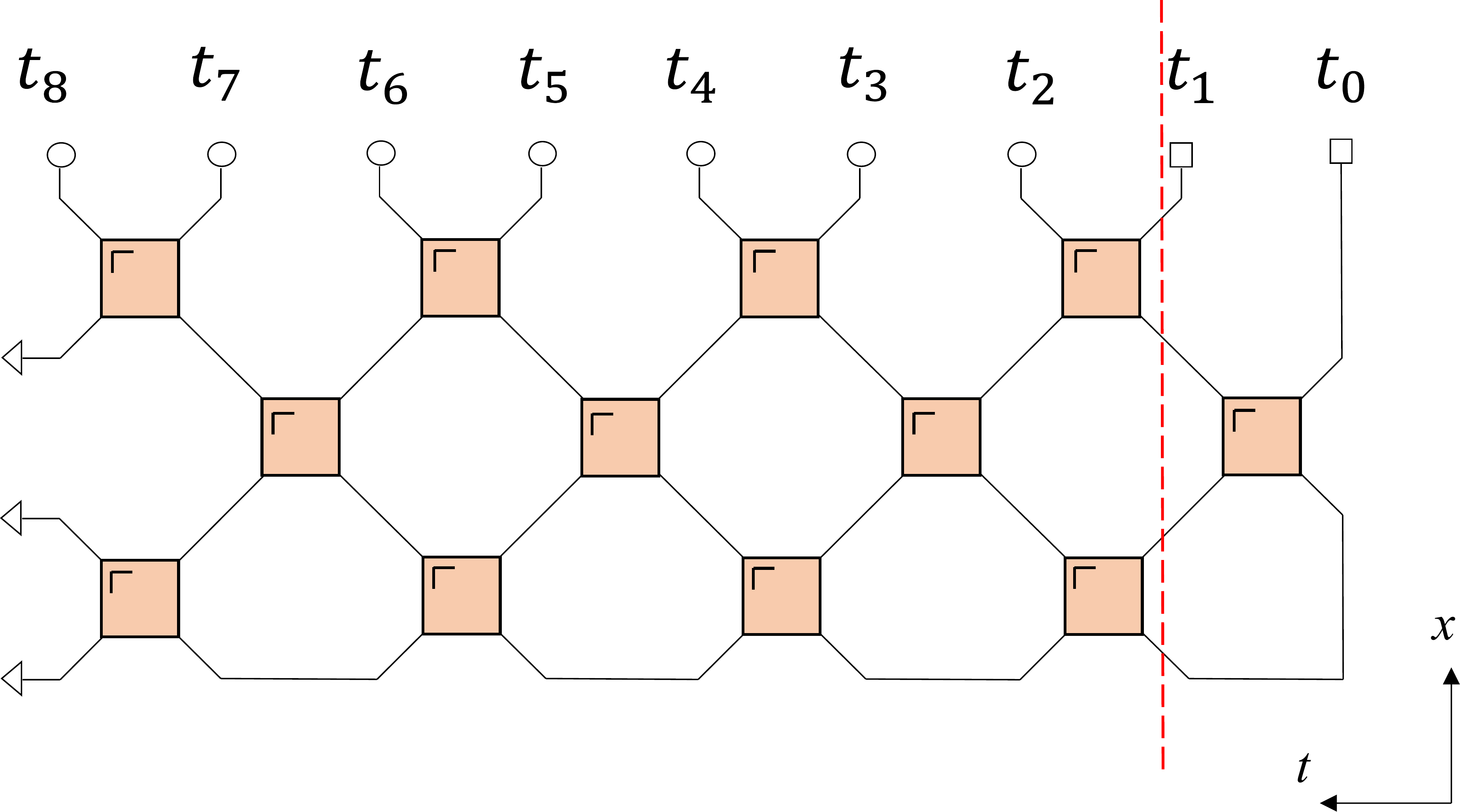}}}\\
         \left(\tr \rho^{\langle I_{\text{left}}|}\right)^2 &=\;\vcenter{\hbox{\includegraphics[width=0.65\linewidth]{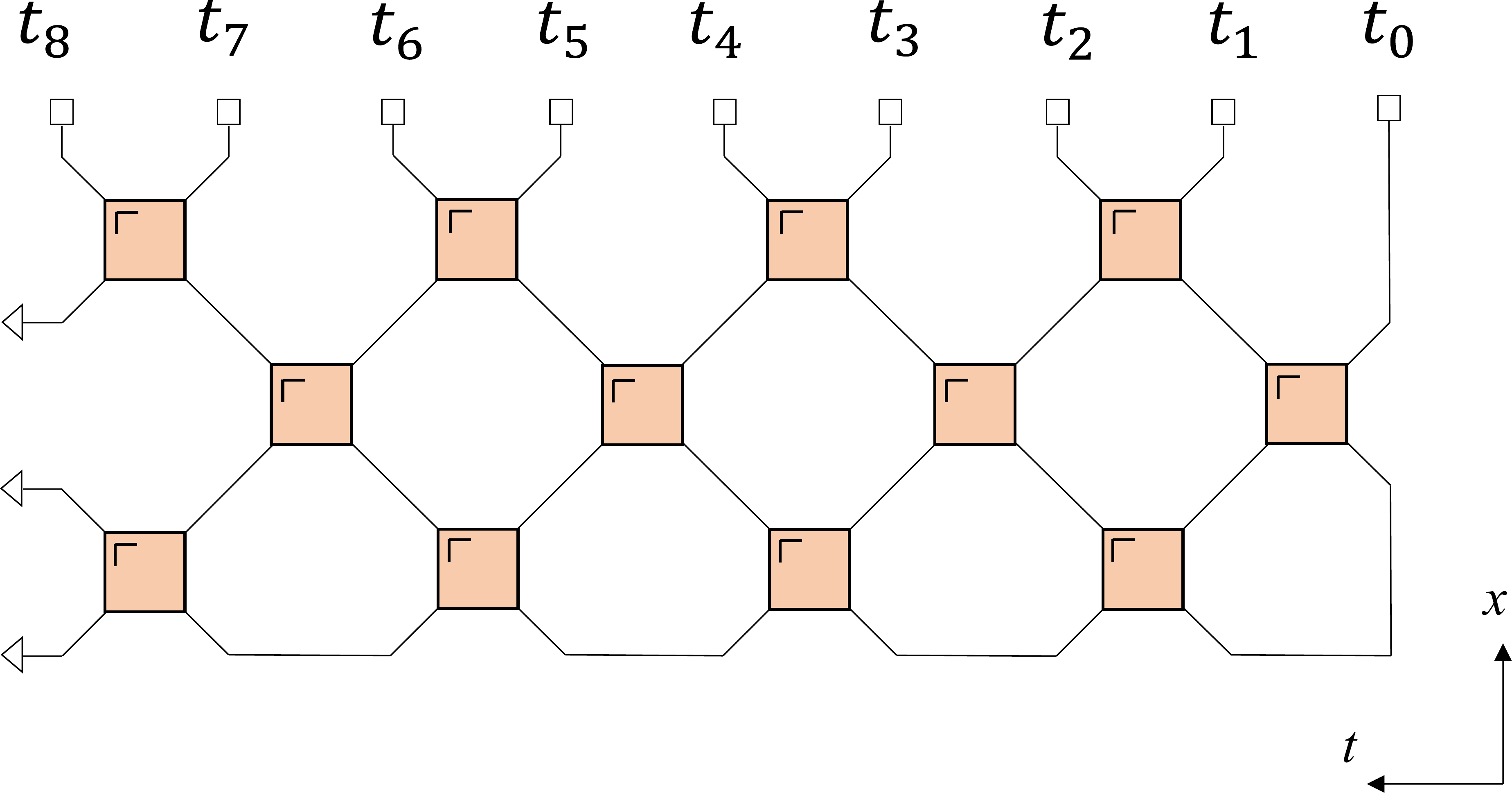}}}
    \end{split}
\end{equation}
Using the graphical identities~\eqref{eq:graphical_U_DU}, contracting the diagram requires evaluating the overlap between the different boundary vectors representing different contraction orders. These overlaps correspond to counting the number of loops, with each contracted loop contributing a factor of $q=2$ to the overall purity calculation, where $q$ is the local Hilbert space dimension:
\begin{equation}
    \vcenter{\hbox{\includegraphics[width=0.1\linewidth]{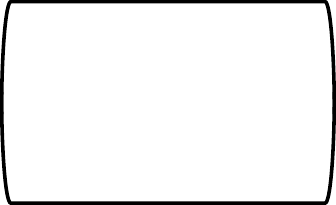}}}\quad=2,
\end{equation}
leading to, e.g.,
\begin{equation}
    \begin{split}
        \vcenter{\hbox{\includegraphics[width=0.1\linewidth]{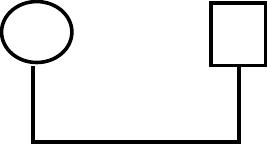}}}\quad = \quad\vcenter{\hbox{\includegraphics[width=0.1\linewidth]{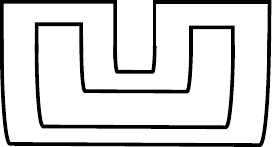}}}\quad=2^2=4.
    \end{split}
\end{equation}
The required overlaps follow as
\begin{equation}
    \begin{split}
        \vcenter{\hbox{\includegraphics[width=0.1\linewidth]{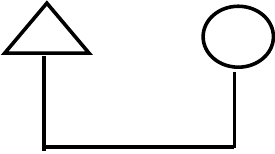}}}\quad&=\quad\vcenter{\hbox{\includegraphics[width=0.1\linewidth]{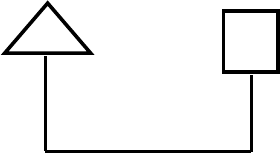}}}\quad=\;2^2,\\
        \vcenter{\hbox{\includegraphics[width=0.1\linewidth]{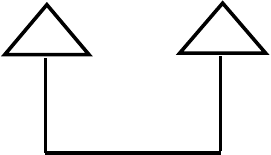}}}\quad&=\quad
        \vcenter{\hbox{\includegraphics[width=0.1\linewidth]{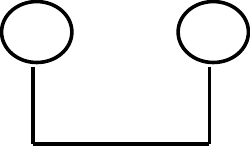}}}\quad=\quad
        \vcenter{\hbox{\includegraphics[width=0.1\linewidth]{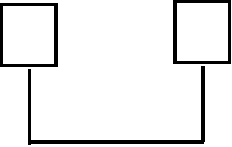}}}\quad=\;2^4.
    \end{split}
\end{equation}
Using spatial unitarity to contract vertically yields:
\begin{equation}
    \begin{split}
        \tr\left( \left(\rho_{t_i}^{\langle I_{\text{left}}|}\right)^2 \right) &=\;\vcenter{\hbox{\includegraphics[width=0.65\linewidth]{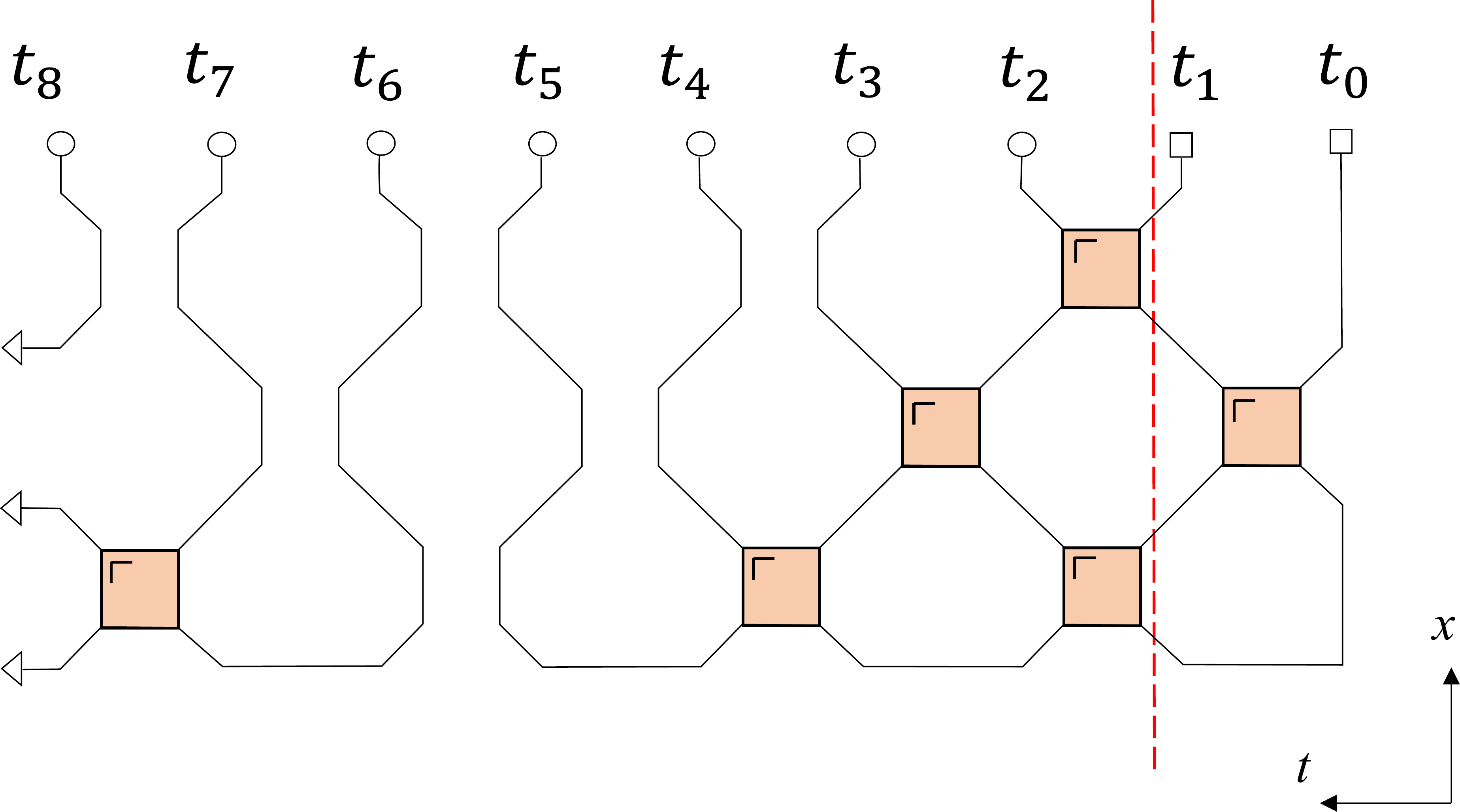}}}
    \end{split}
\end{equation}
This diagram can now be further simplified using temporal unitarity to contract horizontally, resulting in:
\begin{equation}
    \begin{split}
        \tr\left( \left(\rho_{t_i}^{\langle I_{\text{left}}|}\right)^2 \right) &=\;\vcenter{\hbox{\includegraphics[width=0.65\linewidth]{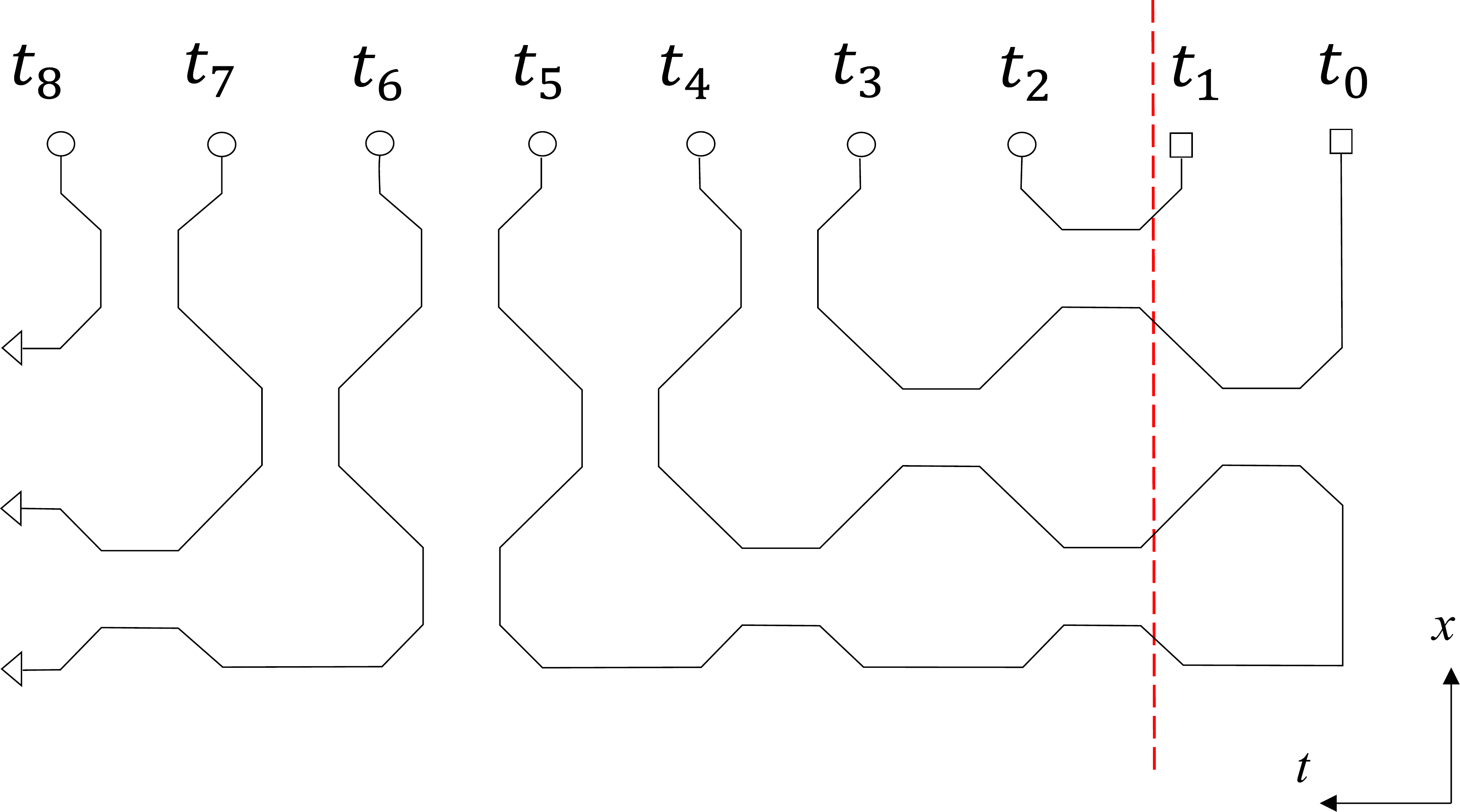}}}\\
        &=\; \left(\vcenter{\hbox{\includegraphics[width=0.1\linewidth]{triangle_circle.pdf}}}\right)^3\left(\vcenter{\hbox{\includegraphics[width=0.1\linewidth]{circle1.pdf}}}\right)^1\left(\vcenter{\hbox{\includegraphics[width=0.1\linewidth]{circle_square_bell.pdf}}}\right)^2=2^{14}
    \end{split}
\end{equation}
An analogous calculation holds for the denominator,
\begin{equation}
    \begin{split}
         \left(\tr \rho^{\langle I_{\text{left}}|}\right)^2 
        =\; \left(\vcenter{\hbox{\includegraphics[width=0.1\linewidth]{triangle_square.pdf}}}\right)^3\left(\vcenter{\hbox{\includegraphics[width=0.1\linewidth]{square1.pdf}}}\right)^3=2^{18},
    \end{split}
\end{equation}
resulting in $D^{\langle I_{\text{left}}|}_{t_i}=2^{-4}$ and $S_2\left(\rho^{\langle I_{\text{left}}|}_{t_i}\right)=2$. 

Upon normalization, only connections between a circle and a square contributes to the purity, and each such connection contributes a factor of $2^{-2}$. Therefore,
\begin{equation}
    S_T^{(t_i)}=2\times\text{Number of}\; \left(  \vcenter{\hbox{\includegraphics[width=0.1\linewidth]{circle_square_bell.pdf}}}\right)\;\equiv 2\cdot n_{\text{cs}},
\end{equation}
where $n_{\text{cs}}$ is the number of circle-square pairs. This result admits a direct interpretation: physically, each Bell pair crossing the boundary of the bipartition contributes one unit of $S_T^{(t_i)}$. Since each $\langle I_{\text{left}}|$ contains both the forward and the backward time contours, each connection between a circle and a square represents two Bell pairs crossing the boundary and contributes two units of $S_T^{(t_i)}$.

After dropping trivial pairings, different classes of diagrams can occur depending on the values of $L$, $T$, and $t_i$. Each parameter regime is now discussed separately. 


\subsubsection{Regime 1: $0<L<T/3$}

\begin{figure}[tb!]
\includegraphics[width=\columnwidth]{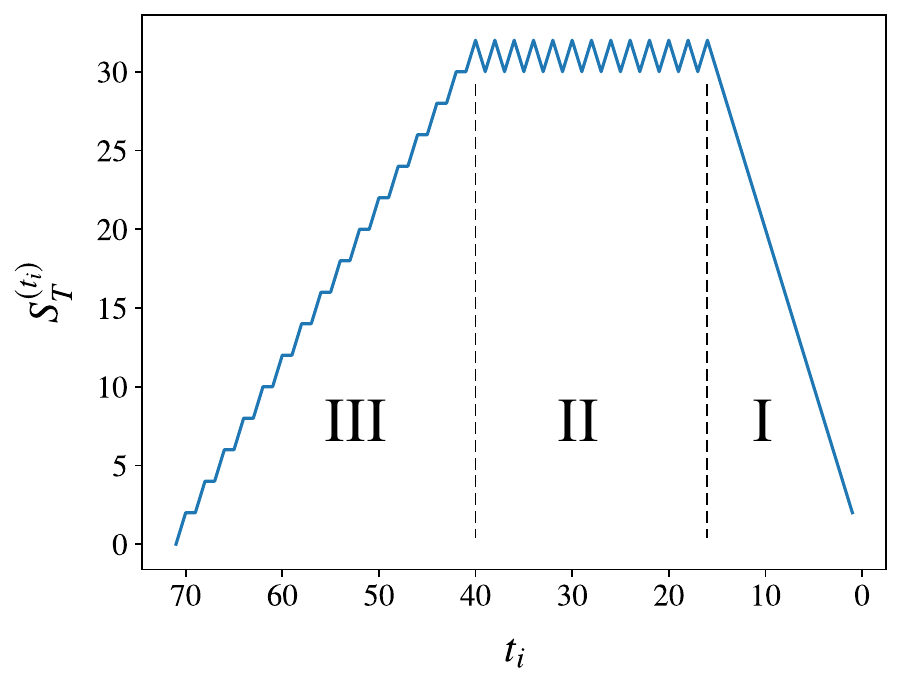}
\caption{Temporal entanglement plotted against temporal bipartition point $t_i$, with total evolution time $T=72$ and bath size $L=16$. There are 3 bipartition intervals, each showing different phenomenologies. Interval I is $0<t_i<L$, Interval II is $L\le t_i<T-2L$, and Interval III is $T-2L\le t_i<T$.}
\label{fig:profile_T72L16SDKI}
\end{figure}

We will first consider the regime where the bath size $L$ is small compared to the total time evolution $T$, more specifically with $0 < L < T/3$. The temporal entanglement profile for this regime is shown in Fig. \ref{fig:profile_T72L16SDKI}. The three intervals of bipartition location $t_i$ are now analyzed separately. To avoid even-odd parity effects, $t_i$ is chosen to be always odd. 

The first bipartition interval is $0<t_i<L$. There, the contracted diagram is of the shape:
\begin{equation}\label{eq:diag_class1}
    \begin{split}
        \vcenter{\hbox{\includegraphics[width=0.4\linewidth]{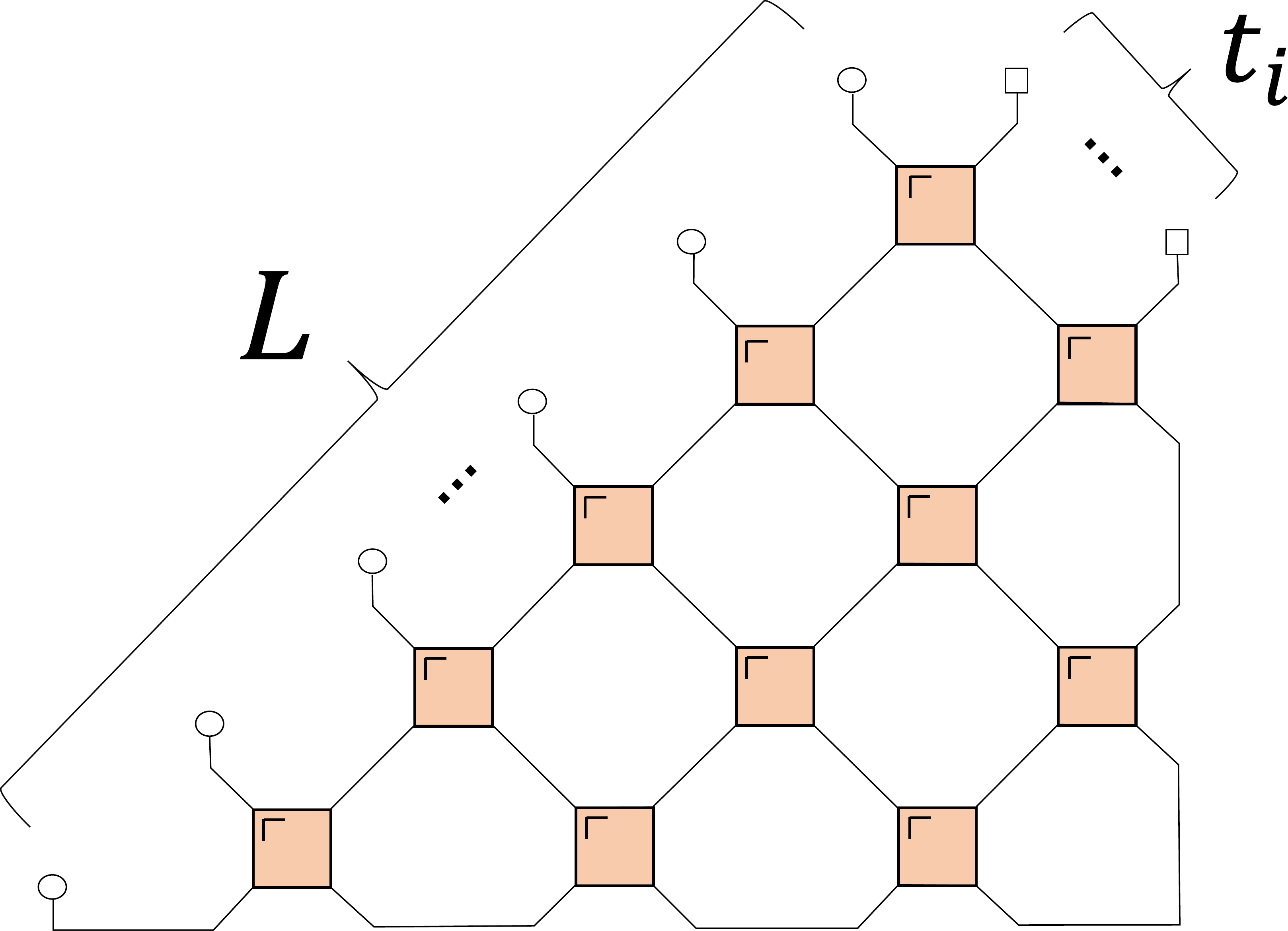}}}\;=\;\vcenter{\hbox{\includegraphics[width=0.4\linewidth]{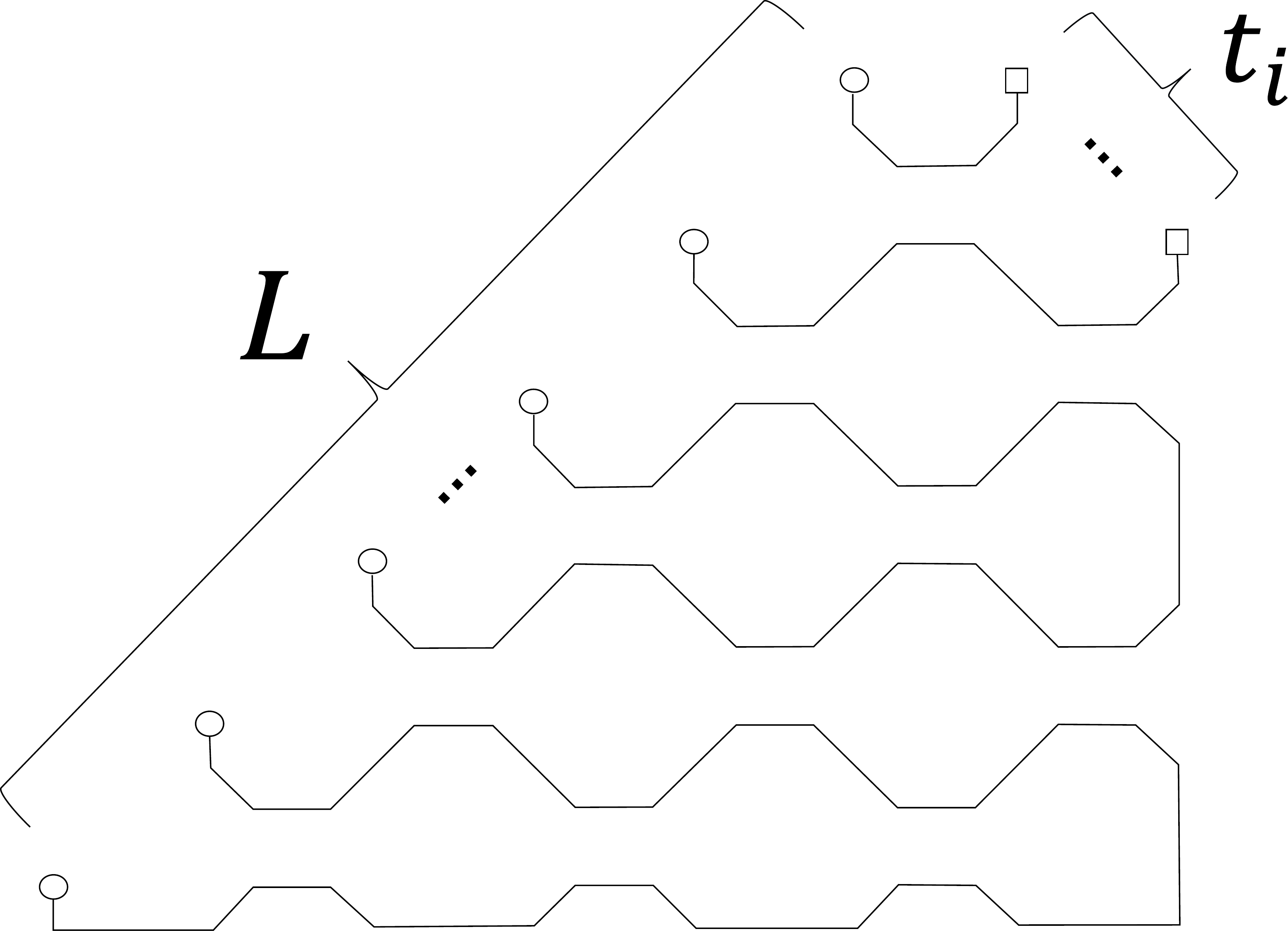}}}
    \end{split}
\end{equation}

All circle contractions can be propagated from the left, leading to a number of circle-square pair $n_{\text{cs}}=t_i$.
In this interval, all legs between $t=0$ and $t=t_i$ are paired up with legs from the other bipartition. Therefore, the number of Bell pairs crossing the bipartition increases linearly with $t_i$, with slope 1 per time contour. This result indicates a maximal TE bounded only by the size of the bipartition: due to the small bath size in this regime all information that initially ``leaks" into the bath will strongly influence future dynamics.

The second bipartition interval is $L\le t_i<T-2L$. There, the contracted diagram is of the shape:
\begin{equation}
    \begin{split}
        \vcenter{\hbox{\includegraphics[width=0.45\linewidth]{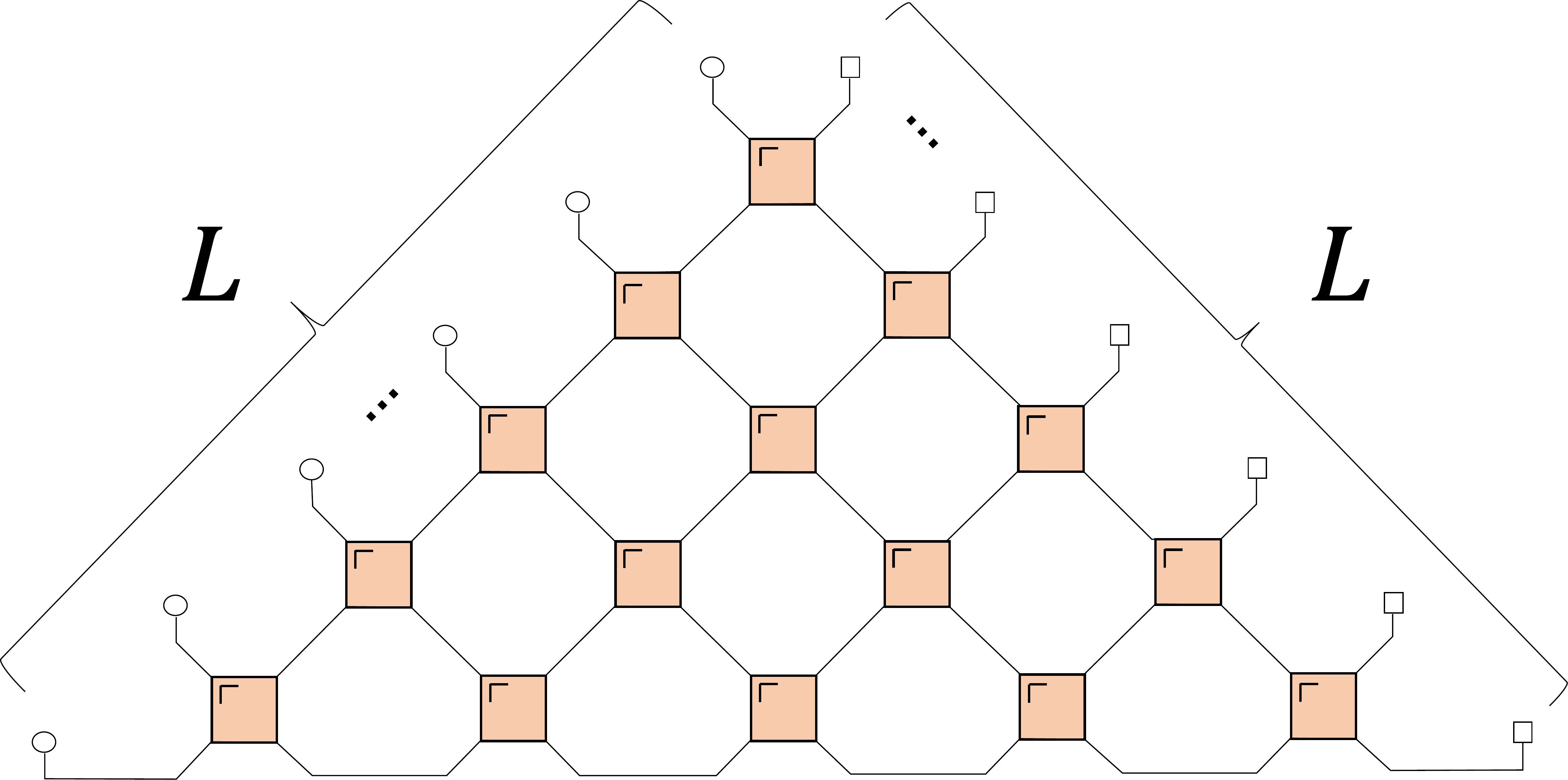}}}\;=\;\vcenter{\hbox{\includegraphics[width=0.45\linewidth]{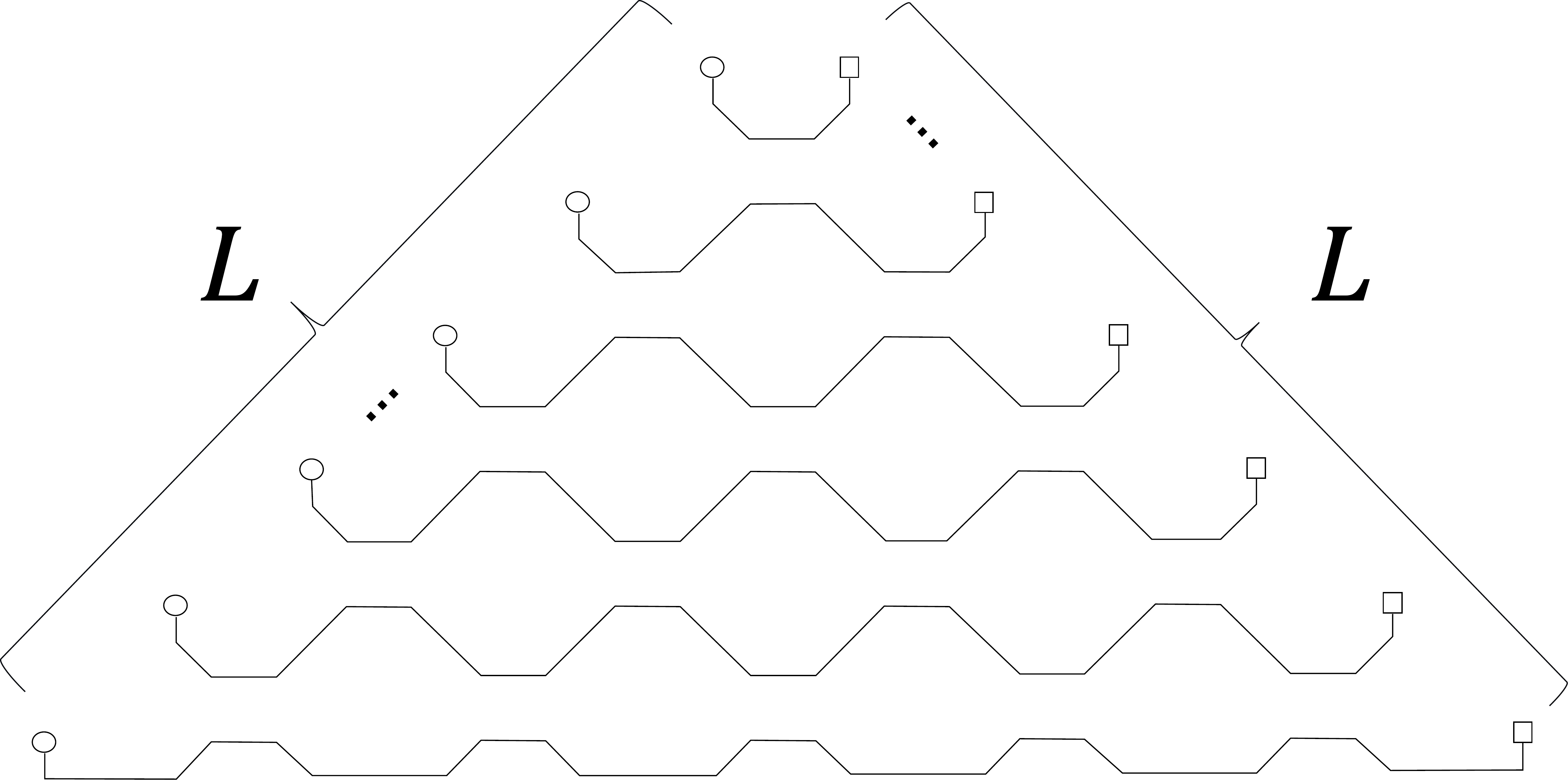}}}
    \end{split}
\end{equation}
Here the contractions can be propagated either from the left or to the right. 
In this interval, $n_{\text{cs}}=L$ independent of $t_i$, and the number of contributing Bell pairs is limited by $L$ and therefore insensitive to the precise location of the bipartition.  The TE has effectively saturated at a maximal value bounded by the bath size, behaving strongly non-Markovian.

The last bipartition interval is $T-2L<t_i<T$. Here, the contracted diagram is of the shape:
\begin{equation}\label{eq:diag_class3}
    \begin{split}
        \vcenter{\hbox{\includegraphics[width=0.45\linewidth]{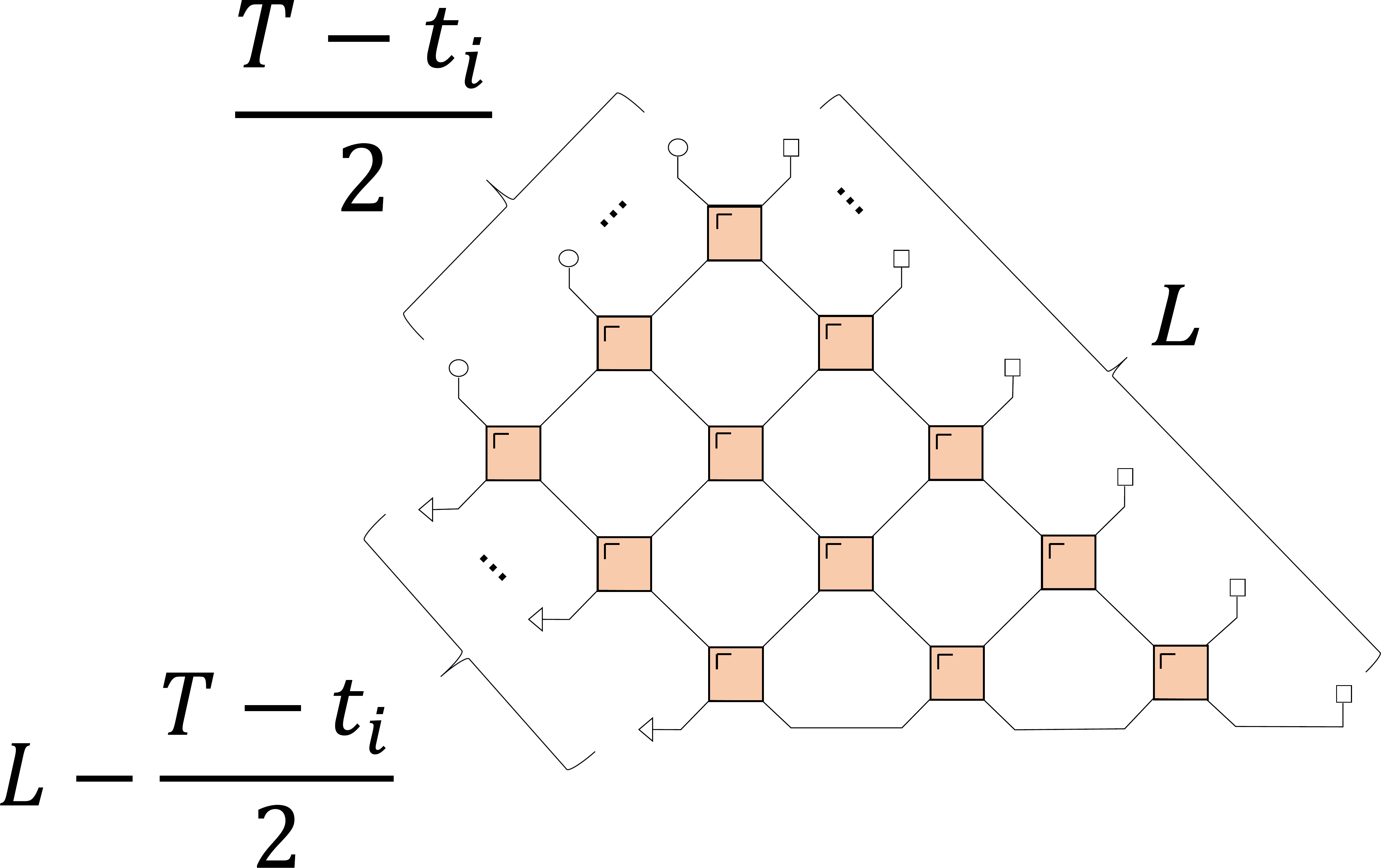}}}\;=\;\vcenter{\hbox{\includegraphics[width=0.45\linewidth]{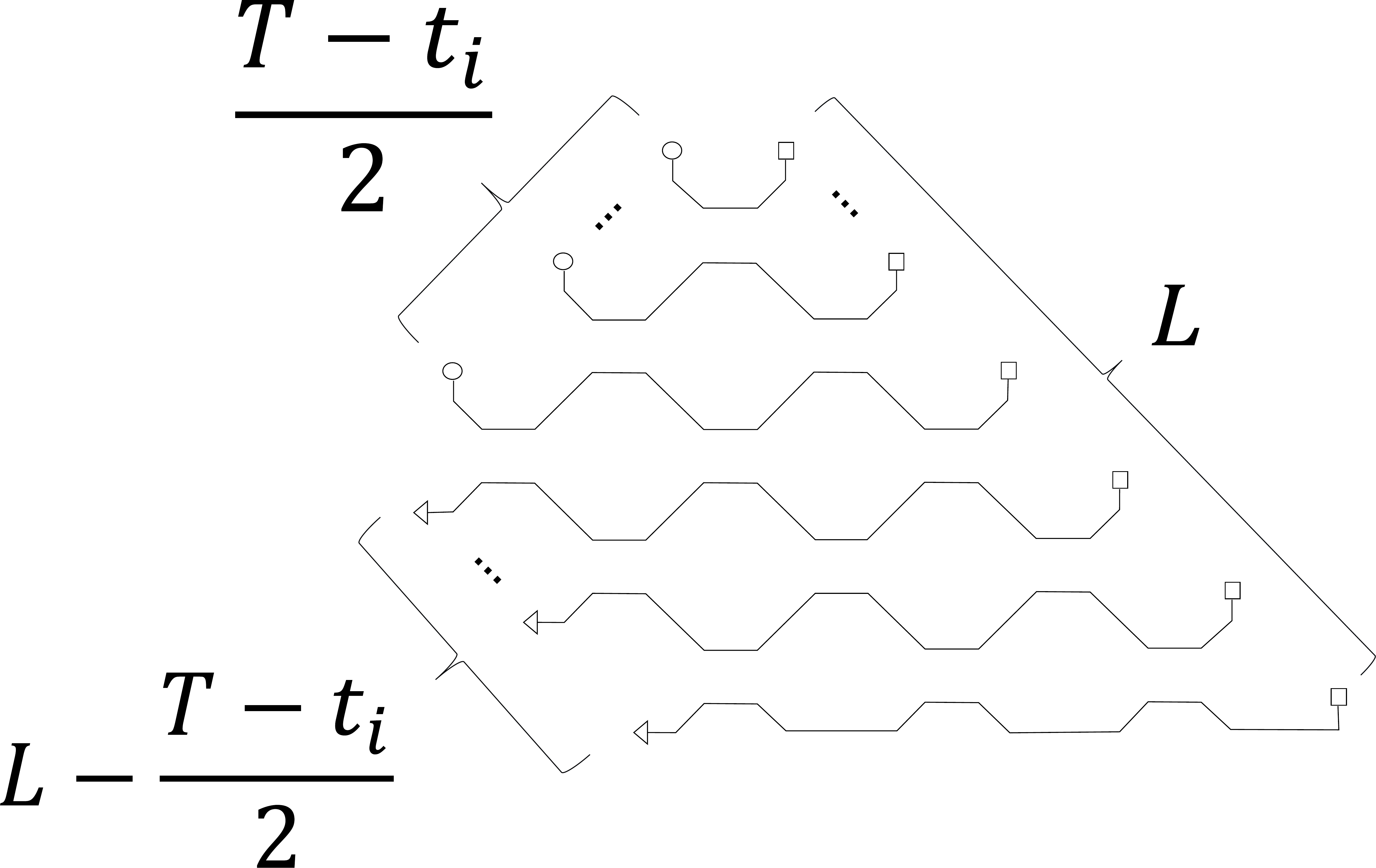}}}
    \end{split}
\end{equation}

In this interval $n_{\text{cs}}=(T-t_i)/2$, the number of Bell pairs annihilated by the trace projectors increases linearly with $t_i$, with slope $1/2$. The slope accounts for the fact that for every increment of $t_i$ by 2, two additional Bell pairs per time contour cross the bipartition, similar to the diagrams of Eq.~\eqref{eq:diag_class1}. However, due to the trace operator, one contributing Bell pair from each time contour is annihilated, and some information that enters the bath is no longer accessible. The different boundaries in time (initial state vs. trace operator) hence introduce an asymmetry between the short-time and late-time bipartitions.

The analysis for the three intervals of this regime matches the profile shown in Fig. \ref{fig:profile_T72L16SDKI}. The choice of bipartition $t_i$ that maximizes the entanglement entropy is then anywhere within the interval $L\le t_i\le T-2L$. The corresponding temporal entanglement follows as $S_T=2\cdot \text{max}_{t_i}  (n_{\text{cs}})=2L$.

\subsubsection{Regime 2: $T/3\le L< T/2$}

\begin{figure}[tb!]
\includegraphics[width=\columnwidth]{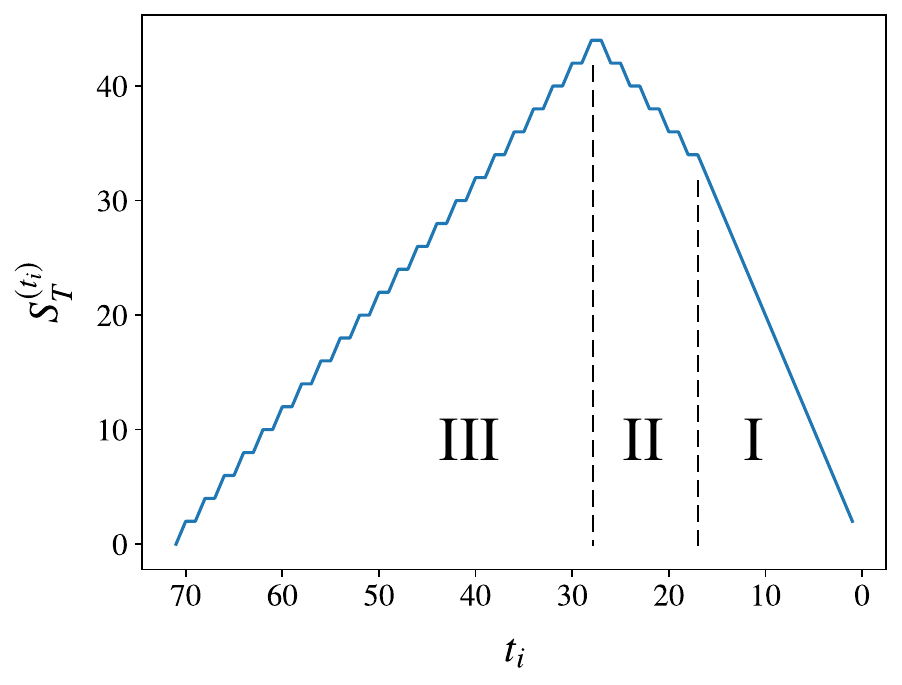}
\caption{Temporal entanglement plotted against temporal bipartition point $t_i$, with total evolution time $T=72$ and bath size $L=28$. There are 3 bipartition intervals, each showing different phenomenologies. Interval I is $0<t_i<T-2L$, Interval II is $T-2L\le t_i<L$, and Interval III is $L\le t_i<T$.}
\label{fig:profile_T72L28SDKI}
\end{figure}

Next, we consider the regime where the size of the bath is larger than $T/3$ but smaller than half the total time evolution $T/2$, such that Bell pairs propagating ballistically through the bath can hit the boundary and return, and memory effects are hence expected to play a role here. The temporal entanglement profile for this regime is shown in Fig. \ref{fig:profile_T72L28SDKI}. The three intervals of the bipartition location $t_i$ are again analyzed separately. To avoid even-odd site effects, $t_i$ is again chosen to be always odd. 

The first bipartition interval is $0< t_i<T-2L$. There, the contracted diagram have the shape of Eq.~\eqref{eq:diag_class1}, and $n_{\text{cs}}=t_i$. The phenomenology is the same as the one for the first interval in Regime 1. 

The second bipartition interval is $T-2L\le t_i<L$. There, the contracted diagram is of the shape:
\begin{equation}\label{eq:diag_class4a}
    \begin{split}
        \vcenter{\hbox{\includegraphics[width=0.5\linewidth]{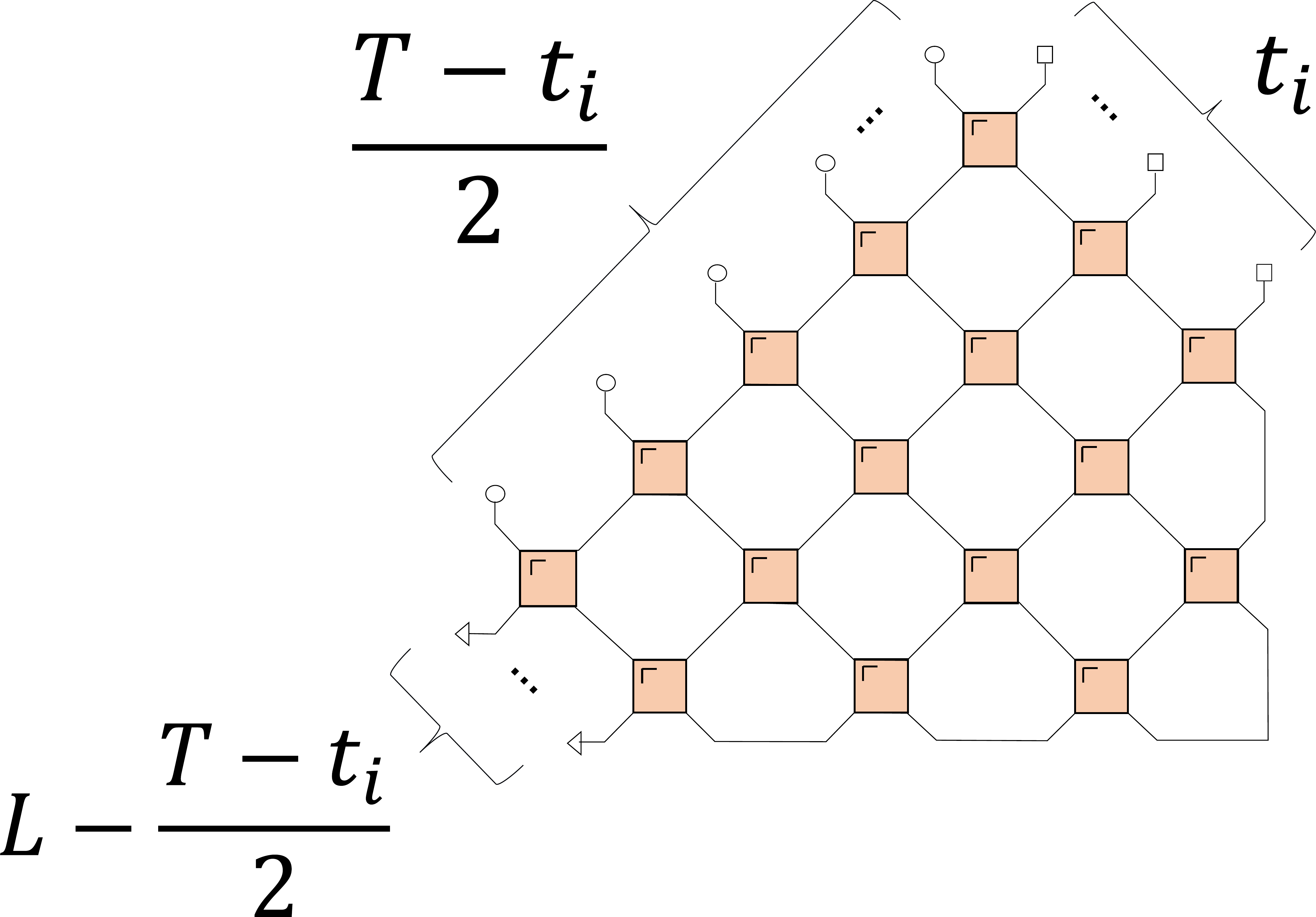}}}
    \end{split}
\end{equation}
where
\begin{equation}
    \begin{split}
     L<\frac{T}{2}\;\Rightarrow\; \#\left( \;\vcenter{\hbox{\includegraphics[width=0.05\linewidth]{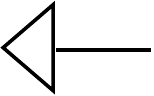}}} \;\right)-\#\left(\;\vcenter{\hbox{\includegraphics[width=0.02\linewidth]{square_single.pdf}}}\;\right) = L-\frac{T-t_i}{2}-t_i<0.
    \end{split}
\end{equation}

While all results so far held for general dual-unitary circuits, the diagram \eqref{eq:diag_class4a} cannot be further simplified using dual-unitarity alone. However, the SWAP gate and the SDKI-f gate both possess the additional symmetries of being self-dual and real:
\begin{equation}
    \begin{split}
        \vcenter{\hbox{\includegraphics[width=0.1\linewidth]{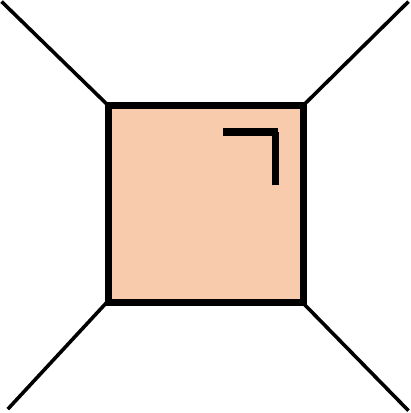}}}\quad=\quad\vcenter{\hbox{\includegraphics[width=0.1\linewidth]{merged_gate1.pdf}}}, \qquad 
        \vcenter{\hbox{\includegraphics[width=0.1\linewidth]{merged_gate1_original.pdf}}}\quad=\quad\left(\;\vcenter{\hbox{\includegraphics[width=0.1\linewidth]{merged_gate1_original.pdf}}}\;\right)^*\,.
    \end{split}
\end{equation}
Therefore, if the circuit consists solely of SWAP or solely of SDKI-f gates, exactly the Clifford gates under consideration, the following identity holds:
\begin{equation}\label{eqn:corner}
    \begin{split}
        \vcenter{\hbox{\includegraphics[width=0.18\linewidth]{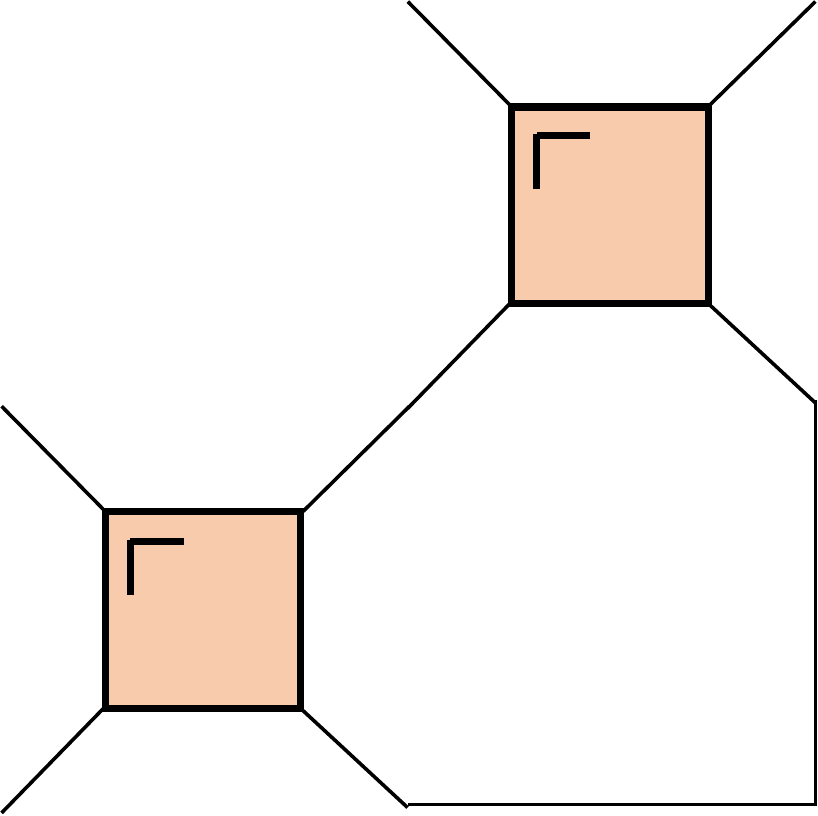}}}\;=\;\vcenter{\hbox{\includegraphics[width=0.18\linewidth]{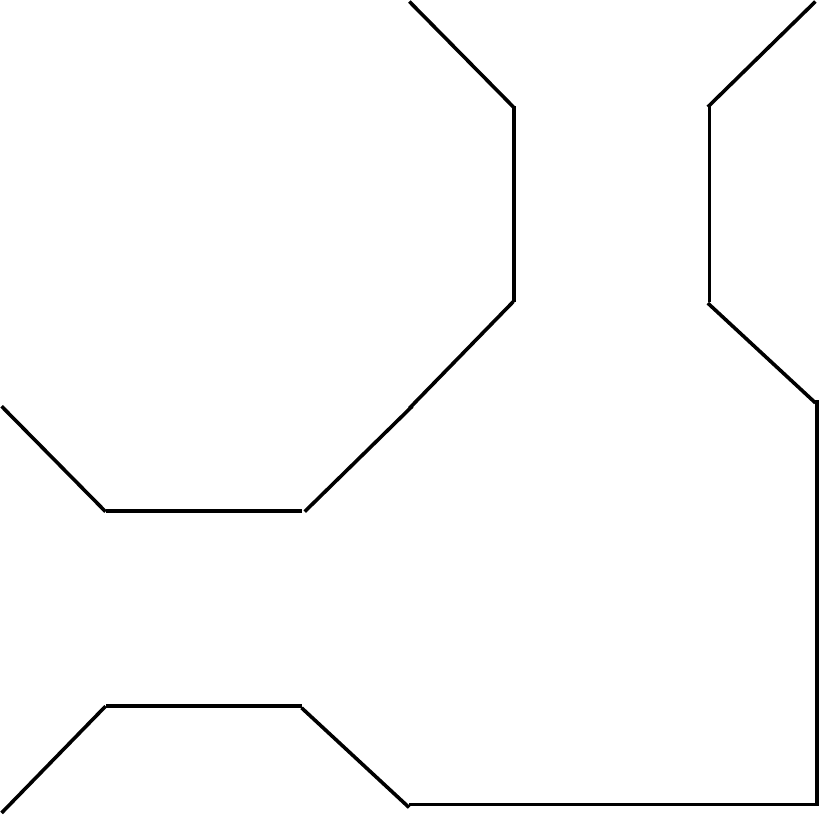}}}
    \end{split}
\end{equation}
Using Eq.~\eqref{eqn:corner} the diagram \eqref{eq:diag_class4a} can be further simplified, and $n_{\text{cs}}=t_i/2 -L+T/2$:
\begin{equation}
    \begin{split}
        \vcenter{\hbox{\includegraphics[width=0.45\linewidth]{topo4.pdf}}}\;=\vcenter{\hbox{\includegraphics[width=0.45\linewidth]{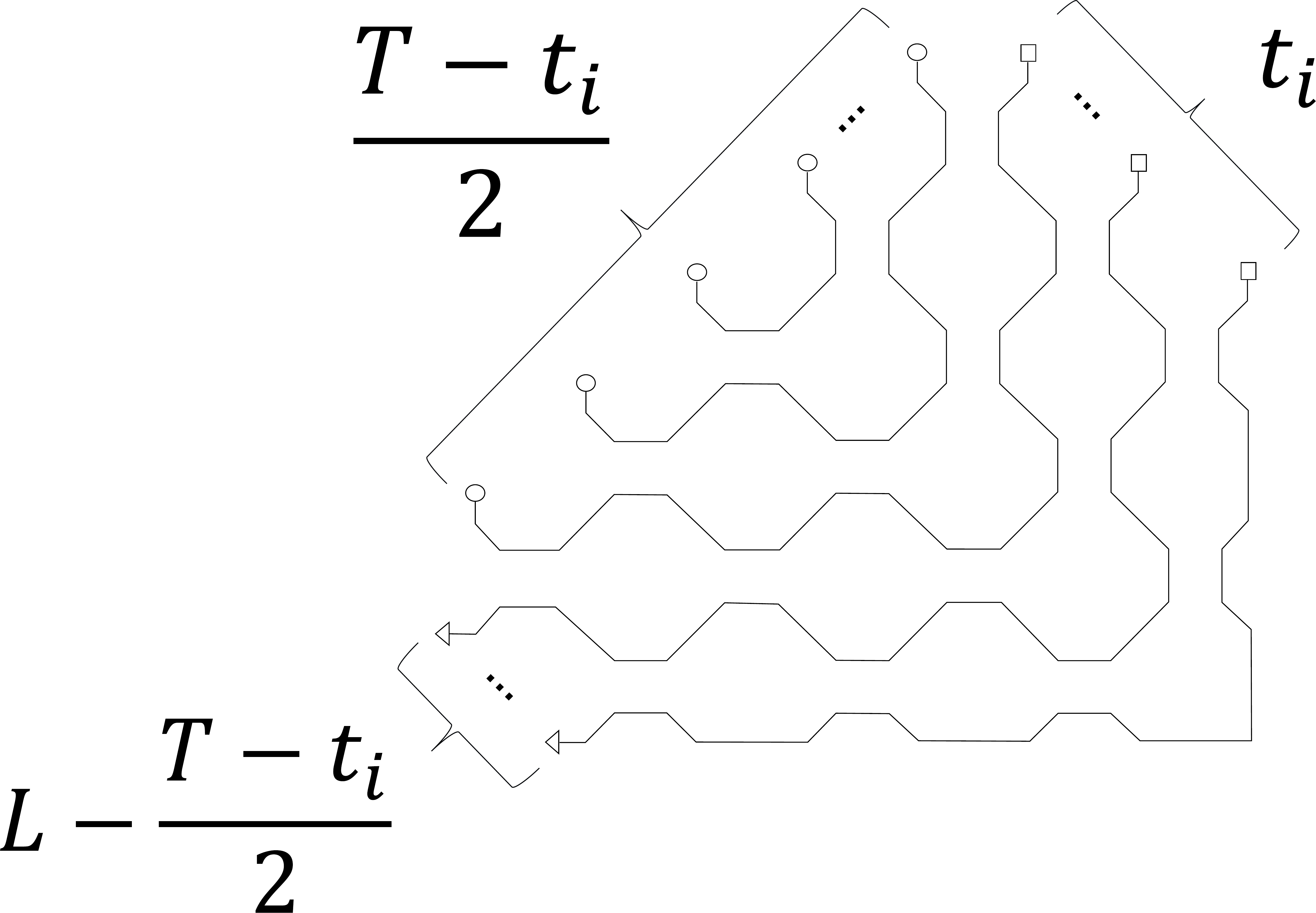}}}
    \end{split}
\end{equation}
In this interval, effects from both temporal boundaries need to be taken into account: the number of contributing Bell pairs increases linearly with $t_i$ with slope 1 per time contour, while the number of Bell pairs annihilated by the trace projectors also increases linearly with $t_i$ with slope $1/2$ per time contour. Therefore, the net increase in number of contributing Bell pairs is linear in $t_i$ with slope $1/2$.  
In the generic case where Eq. \eqref{eqn:corner} does not hold, the temporal entanglement profiles of these diagrams are different, as is discussed in Appendix \ref{app:generic_DU_class_4}. 

The third bipartition interval is $L\le t_i< T$. There, the contracted diagram has the shape of Eq.~\eqref{eq:diag_class3}, and $n_{\text{cs}}=(T-t_i)/2$. The phenomenology is the same as the one for the third interval in Regime 1. 

The analysis for the three intervals of Regime 2 matches the profile shown in Fig. \ref{fig:profile_T72L28SDKI}. The $t_i$ for maximal entanglement entropy is at $t_i=L$. The corresponding temporal entanglement is $S_T=2\cdot \text{max}_{t_i}(n_{\text{cs}})=T-L$.

\subsubsection{Regime 3: $T/2\le L<T$}

\begin{figure}[tb!]
\includegraphics[width=\columnwidth]{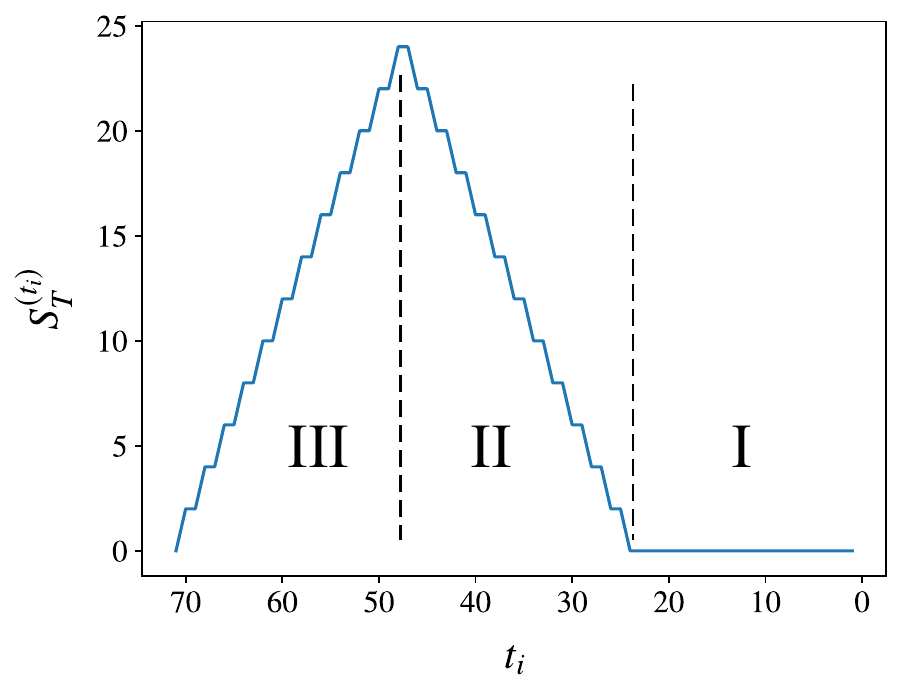}
\caption{Temporal entanglement plotted against temporal bipartition point $t_i$, with total evolution time $T=72$ and bath size $L=48$. There are 3 bipartition intervals, each showing different phenomenologies. Interval I is $0<t_i<2L-T$, Interval II is $2L-T\le t_i<L$, and Interval III is $L\le t_i<T$.}
\label{fig:profile_T72L48SDKI}
\end{figure}

Let us now consider the limit where the bath size is on the same order as the total evolution time, but constrained to $L \geq T/2$ such that ballistically propagating Bell pairs can not traverse the length of the bath twice and the right boundary is hence expected to not play a role. The temporal entanglement profile for Regime 3 is shown in Fig. \ref{fig:profile_T72L48SDKI}. There are again three intervals of bipartition location $t_i$, with $t_i$ again chosen to be always odd. 

The first bipartition interval is $0< t_i<2L-T$. There, the contracted diagram is of the shape:
\begin{equation}\label{eq:diag_class4b}
    \begin{split}
        \vcenter{\hbox{\includegraphics[width=0.5\linewidth]{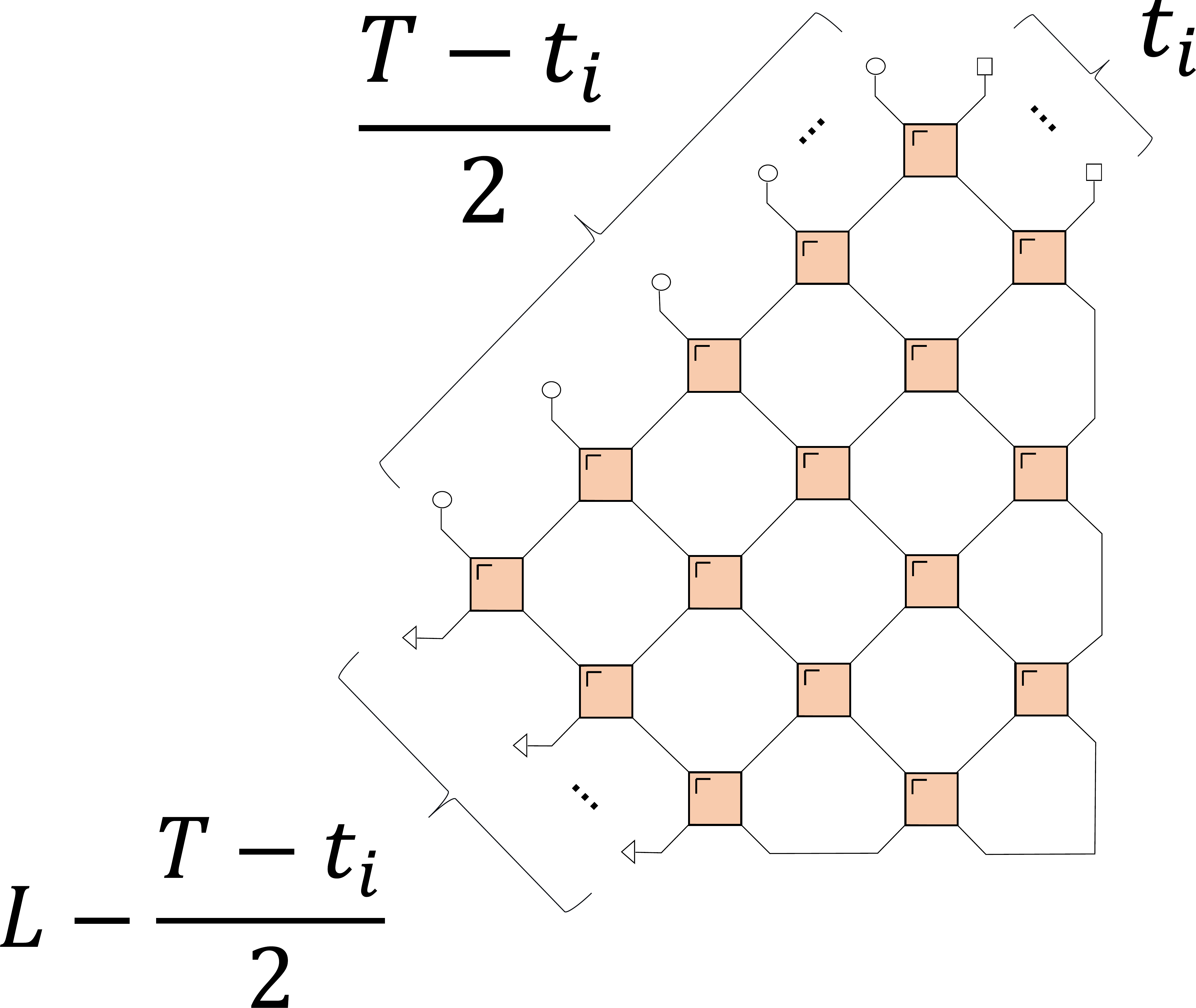}}}
    \end{split}
\end{equation}
where
\begin{equation}
    \begin{split}
     &L>\frac{T}{2},\, t_i<2L-T\;\\&\Rightarrow\; \#\left( \;\vcenter{\hbox{\includegraphics[width=0.05\linewidth]{triangle_single.pdf}}} \;\right)-\#\left(\;\vcenter{\hbox{\includegraphics[width=0.02\linewidth]{square_single.pdf}}}\;\right) = L-\frac{T-t_i}{2}-t_i>0.
    \end{split}
\end{equation}

Similar to the diagrams of Eq.~\eqref{eq:diag_class4a}, these cannot be further simplified using dual-unitarity alone. For all-SWAP or all-SDKI-f circuits, Eq. \eqref{eqn:corner} again holds, and $n_{\text{cs}}=0$ independent of $t_i$:
\begin{equation}
    \begin{split}
        \vcenter{\hbox{\includegraphics[width=0.45\linewidth]{topo4b.pdf}}}\;=\;\vcenter{\hbox{\includegraphics[width=0.45\linewidth]{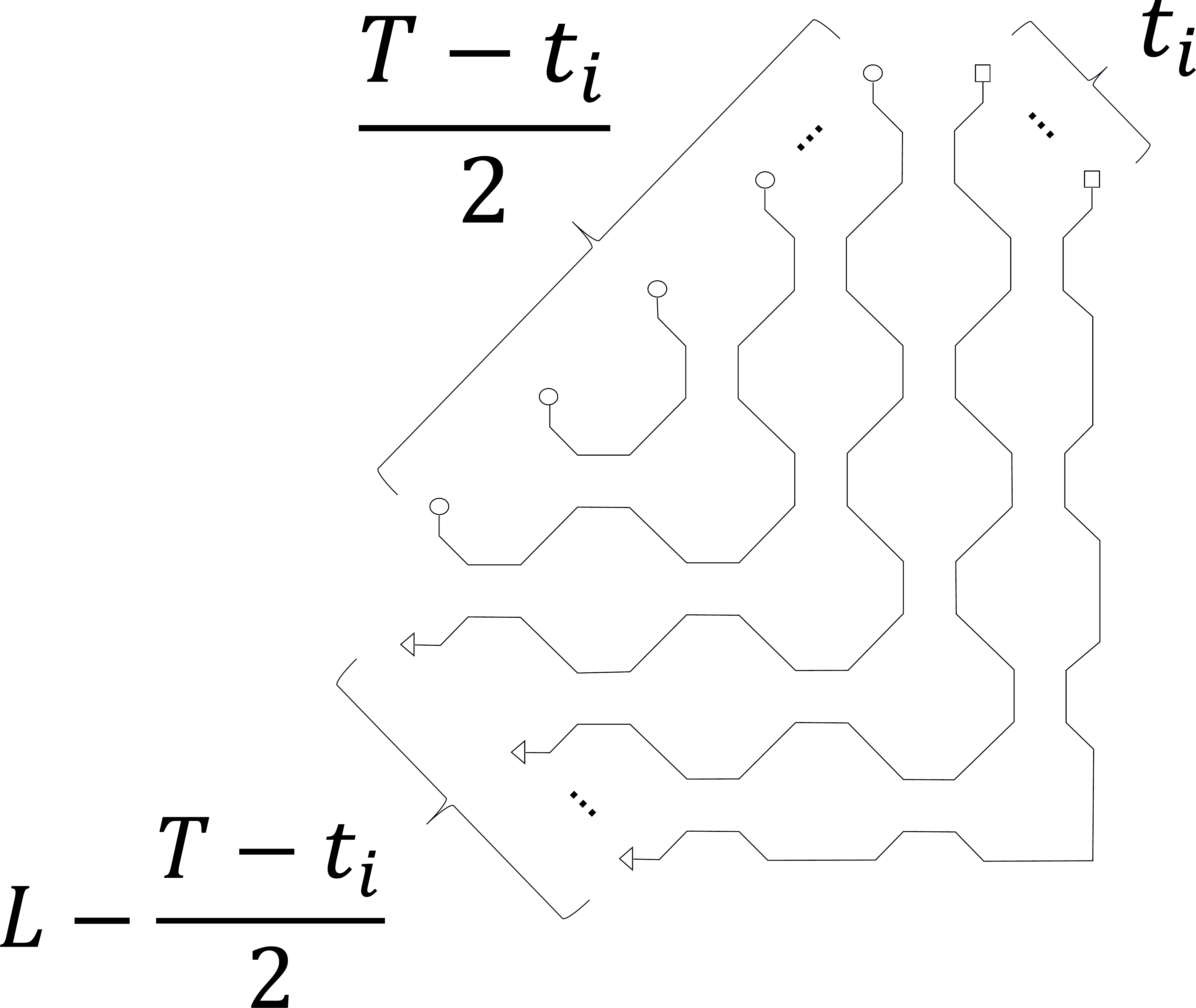}}}
    \end{split}
\end{equation}
In this interval, all Bell pairs available for cross-bipartition pairing are connected to trace projectors, and no contributing Bell pairs remain. For these bipartitions the TE vanishes and the bath can be treated as a purely Markovian perfect dephaser. Numerical results on these diagrams for generic dual-unitary Clifford circuits are presented in Appendix \ref{app:generic_DU_class_4}. 

The second bipartition interval is $2L-T\le t_i< L$. There, the contracted diagram has the shape of Eq.~\eqref{eq:diag_class4a}, and $n_{\text{cs}}=t_i/2 -L+T/2$. The phenomenology is the same as the one for the second interval of Regime 2. 
The third bipartition interval is $L\le t_i< T$. There, the contracted diagram has the shape of Eq.~\eqref{eq:diag_class3}, and $n_{\text{cs}}=(T-t_i)/2$. The phenomenology is the same as the one for the third interval of Regime 1 and 2. 

The analysis for the three intervals of Regime 3 matches the profile shown in Fig. \ref{fig:profile_T72L48SDKI}. The $t_i$ for maximal entanglement entropy is at $t_i=L$. The corresponding temporal entanglement follows as $S_T=2\cdot \text{max}_{t_i}(n_{\text{cs}})=T-L$.
\subsubsection{Regime 4: $L\ge T$} 
We now consider the final limit where the bath size is larger than the number of discrete time steps. In Regime 4, the contracted diagram always takes on the following shape:
\begin{equation}
    \begin{split}     \vcenter{\hbox{\includegraphics[width=0.25\columnwidth]{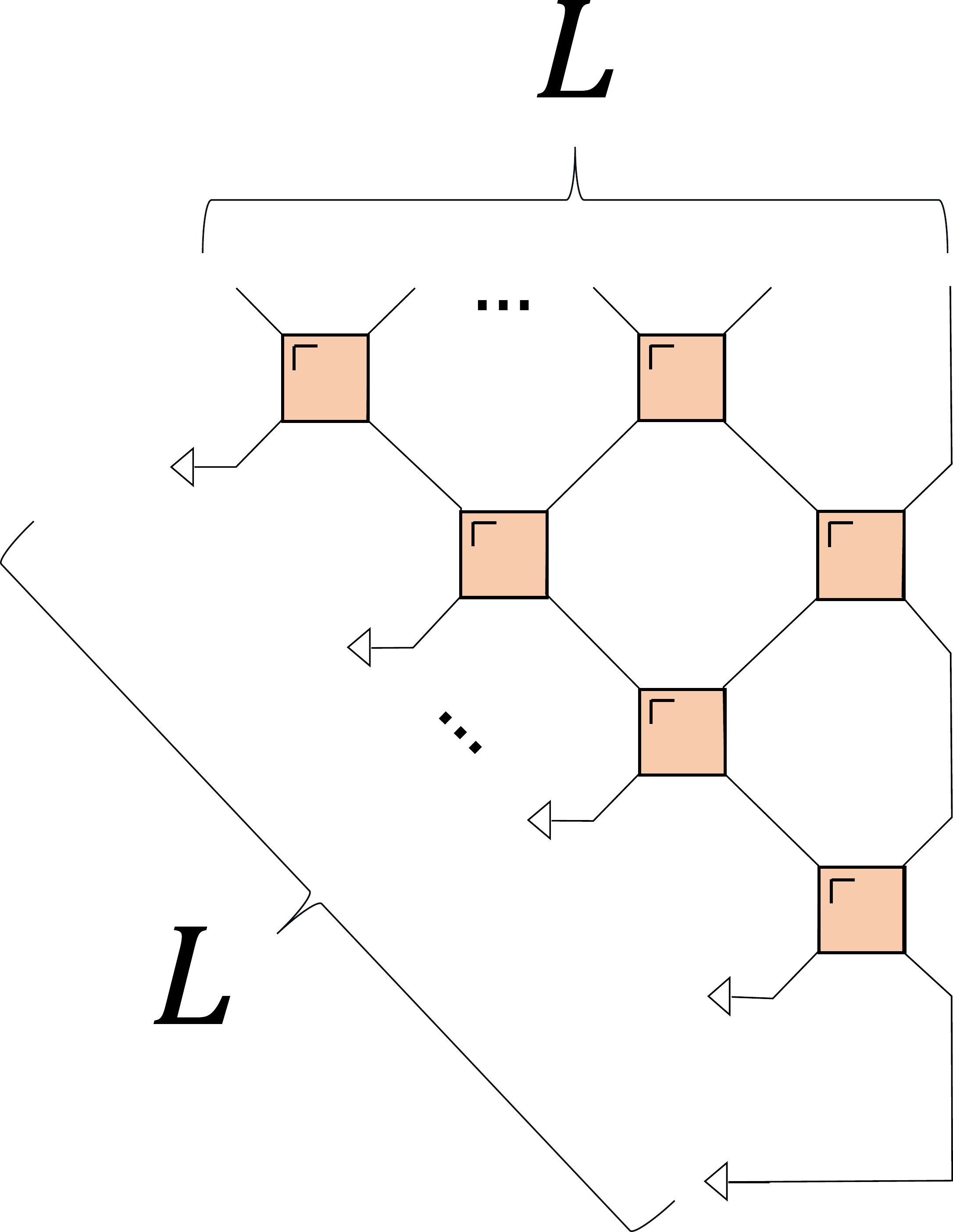}}}\quad=\quad\vcenter{\hbox{\includegraphics[width=0.25\columnwidth]{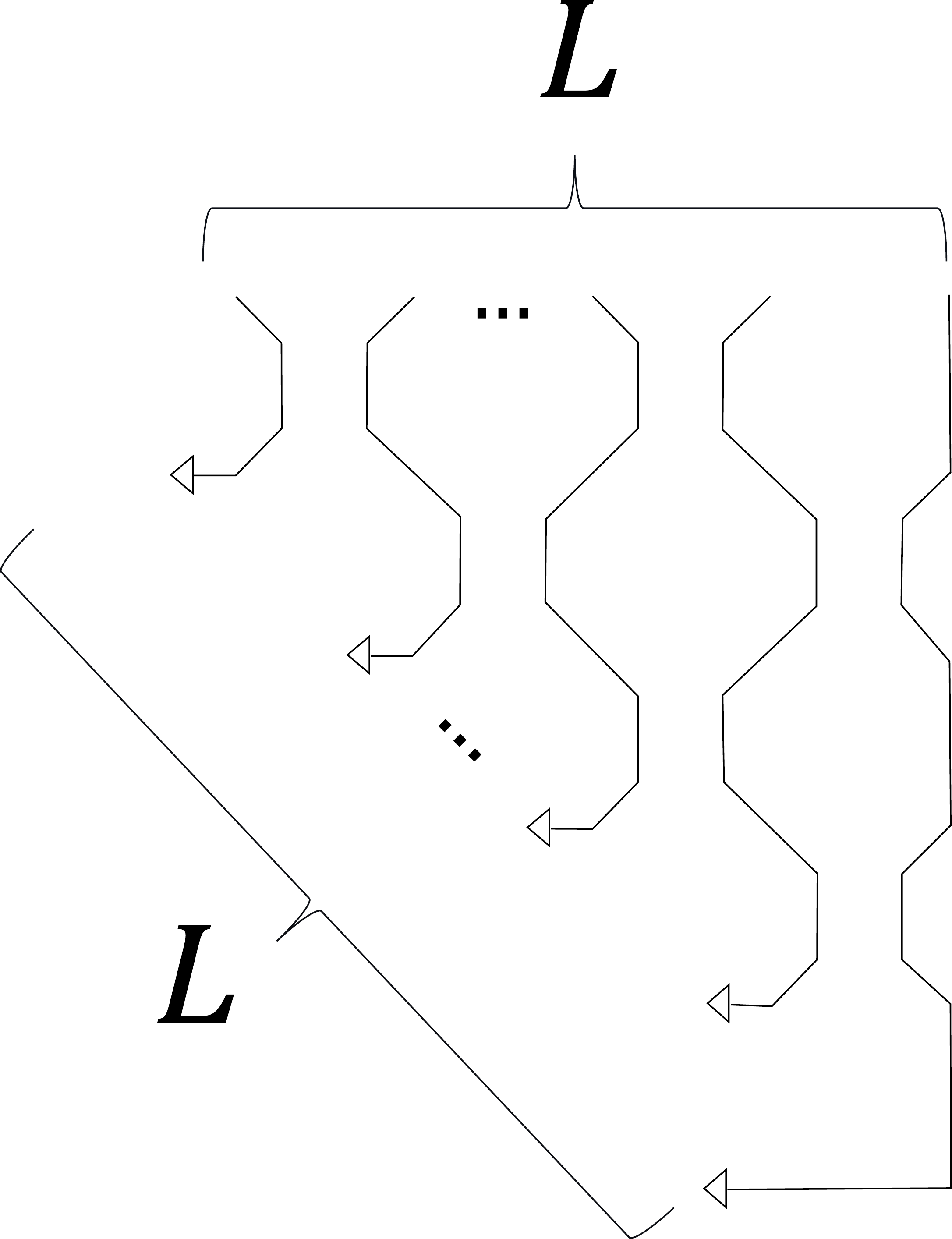}}}
    \end{split}
\end{equation}

In these diagrams $n_{\text{cs}}=0$ independent of $t_i$. All Bell pairs available for cross-bipartition pairing are connected to trace projectors. Therefore in this regime $S_T=0$ always, and the bath reduces to the expected perfect dephaser limit. 

These results exhaust the possible temporal entanglement profiles, and match with the piecewise linear form presented in Eq. \eqref{eqn:num_res_no-meas}. 
The linear profile can be intuitively understood through the ballistic dynamics of the (ends of the) Bell pairs. E.g., the initial growth of temporal entanglement with each update step can be understood by noting that the boundary Bell pairs spread ballistically under the action of the dual-unitary circuit, and the radius of each Bell pair grows linearly with each spatial update step. Since the update preserves the center of mass of each Bell pair, the number of Bell pairs crossing the optimal temporal bipartition site also grows linearly with each update step. In other regimes the obtained linear profile follows by additionally taking into account reflection at the spatial boundary and absorption due to the trace at the temporal boundary.

\section{Numerical Results on Circuits with Measurements}\label{sec:DU_with_meas}

Let us now consider the effect of projective measurements on the dynamics of the TE. In order to preserve the spatial unitarity and the Clifford nature of the dynamics, we restrict ourselves to measurements in the Bell-pair basis. 

\subsection{Measurements Preserving Spatial Unitarity}\label{sec:meas_formalim}

We focus on two-site measurements in the Bell-pair basis, also known as unitary-error-basis (UEB) measurements \cite{knill_non-binary_1996}.
There are four basis states for two qubits, denoted by:
\begin{align}
        |B_I\rangle&=(|00\rangle+|11\rangle)/\sqrt{2}, \quad
        |B_X\rangle=(|01\rangle+|10\rangle)/\sqrt{2},\\
        |B_Y\rangle&=(|01\rangle-|10\rangle)/\sqrt{2}, \quad
        |B_Z\rangle=(|00\rangle-|11\rangle)/\sqrt{2}.
\end{align}
The Bell-pair measurements break temporal unitarity while preserving spatial unitarity for any measurement outcome~\cite{piroli_exact_2020,claeys_exact_2022,claeys_emergent_2022,ippoliti_dynamical_2023,claeys_dual_2023}. Therefore, the spatial transfer matrix remains unitary upon adding measurements.
Upon space-time rotation, projectors onto the four basis states map to (unitary) products of single-site Pauli operators:
\begin{align}
        |B_I\rangle\langle B_I| &\rightarrow  \frac{\mathds{1}\otimes \mathds{1}}{2}, &&\qquad
        |B_X\rangle\langle B_X| \rightarrow  \frac{\sigma_x \otimes \sigma_x}{2}, \\
        |B_Y\rangle\langle B_Y| &\rightarrow  \frac{\sigma_y \otimes \sigma_y}{2}, &&\qquad
        |B_Z\rangle\langle B_Z| \rightarrow  \frac{\sigma_z \otimes \sigma_z}{2}.
\end{align}
Graphically, for $\alpha\in\{ I, X, Y, Z \}$ we can write that:
\begin{equation}
    \vcenter{\hbox{\includegraphics[width=0.15\linewidth]{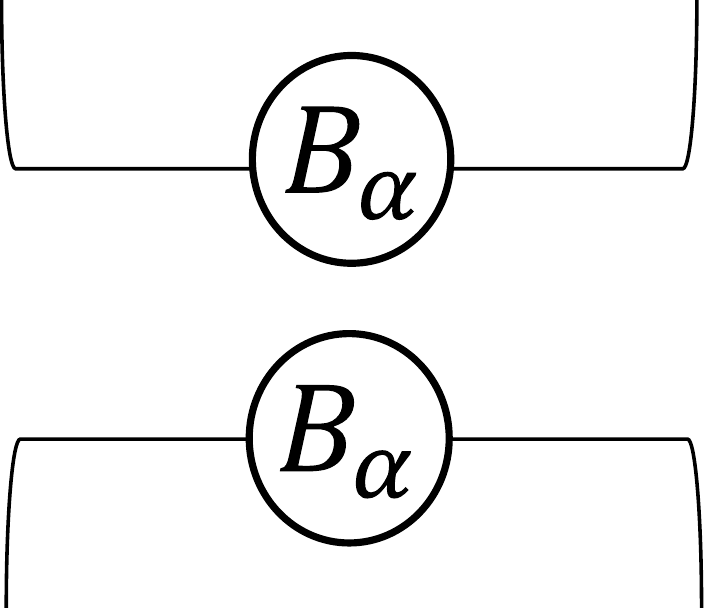}}}\quad\xrightarrow{\text{space-time rotation}}\quad \frac{1}{2}\;\times\quad \vcenter{\hbox{\includegraphics[width=0.15\linewidth]{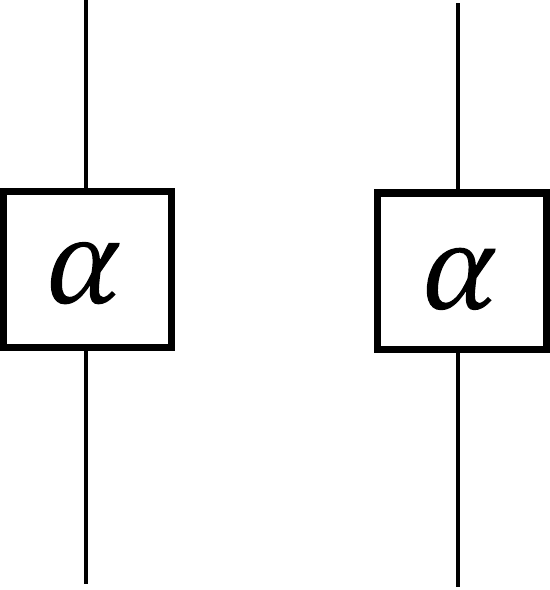}}}
\end{equation}
In a true measurement, the outcome of having one of the four basis states is probabilistic, with the probability given by the Born rule. Viewed spatially, the different outcomes correspond to applications of different Pauli operators. Since Pauli operators only incur a potential sign-change on the stabilizers, they do not change the entanglement structure of the state \cite{aaronson_improved_2004}. Therefore, without loss of generality, the present work replaces each true measurement by the projector $|B_I\rangle\langle B_I|$, such that the two qubits are forced to be in the $|B_I\rangle$ state after the measurement. This protocol is also known as the ``forced measurement" or ``post-selected measurement" protocol \cite{ippoliti_fractal_2022, nahum_measurement_2021, fidkowski_how_2021, lu_spacetime_2021}.

Each two-site unitary gate in the circuit has a probability $p$ of being replaced by a forced measurement. This random choice is made independently among all gates. 
Once the gate at sites $i$, $i+1$ and times $t_i$, $t_{i+1}$ along the forward time contour is chosen to be replaced by a measurement, the same choice must be made for the gate at the corresponding sites and times on the backward time contour, since the time evolution operator is identical between the two time contours for any chosen stochastic trajectory. In the folded representation, the measurement outcomes are referred to as being ``locked" among all layers. 

Graphically,
\begin{equation}\label{eqn:layer-locked}
    \begin{split}
        \vcenter{\hbox{\includegraphics[width=0.1\linewidth]{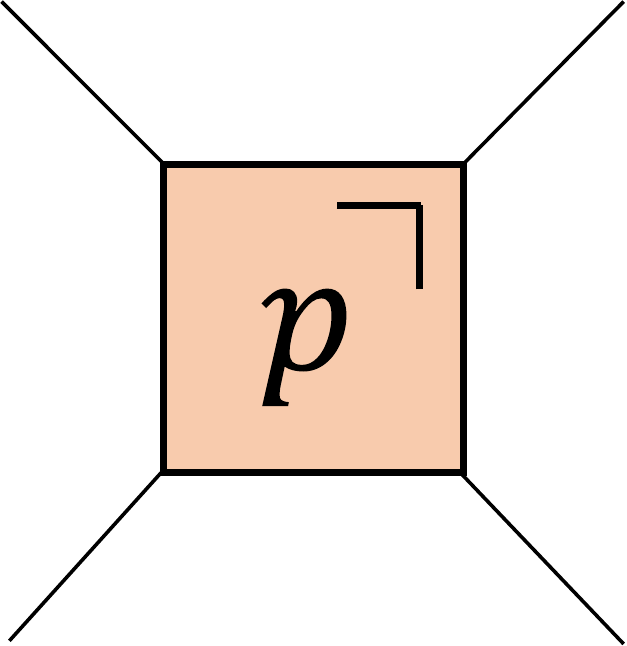}}} = \;\begin{cases}
        \quad \vcenter{\hbox{\includegraphics[width=0.1\linewidth]{merged_gate1_original.pdf}}} \quad &\text{with probability $1-p$,}\\
         &\text{}\quad\\
          \quad\vcenter{\hbox{\includegraphics[width=0.1\linewidth]{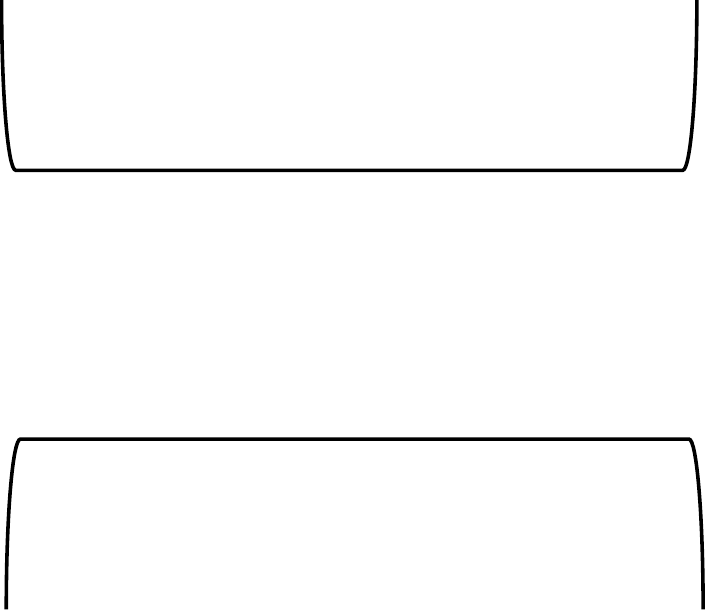}}}\;\ \quad &\text{with probability $p$,}
        \end{cases}
    \end{split}
\end{equation}
where the gates are already in the merged representation. With this choice of projector the evolved state remains normalized after each measurement, as will be discussed in Sec.~\ref{sec:analytical_decay}.

\subsection{Good Scrambler Circuits with Measurements: Numerical Results}
\begin{figure}[tb!]
\includegraphics[width=\columnwidth]{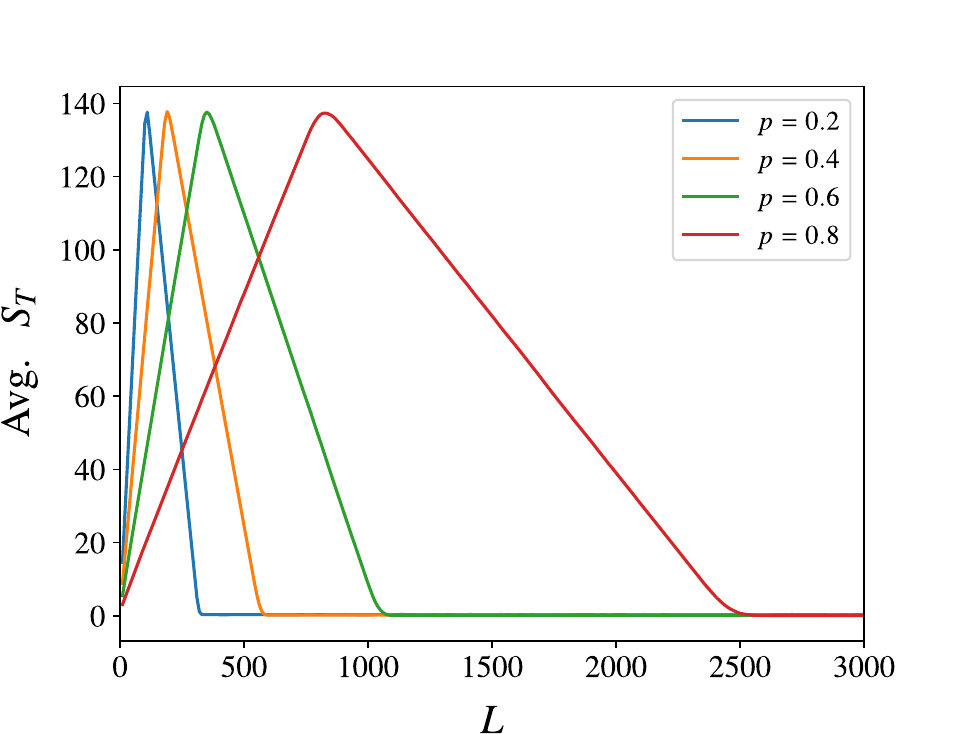}
\caption{Average temporal entanglement $S_T$ plotted against bath size $L$ for total evolution time $T=216$ and different measurement rates $p$. The dual-unitary gates are chosen to be the SDKI-r gates as defined in Eq. \eqref{eqn:SDKI-r}. The averaging is done with respect to both the choice of single-site Clifford gates and positions of the Bell measurements.}
\label{fig:TE_diff_meas_SDKI-r}
\end{figure}

\begin{figure}[tb!]
\includegraphics[width=\columnwidth]{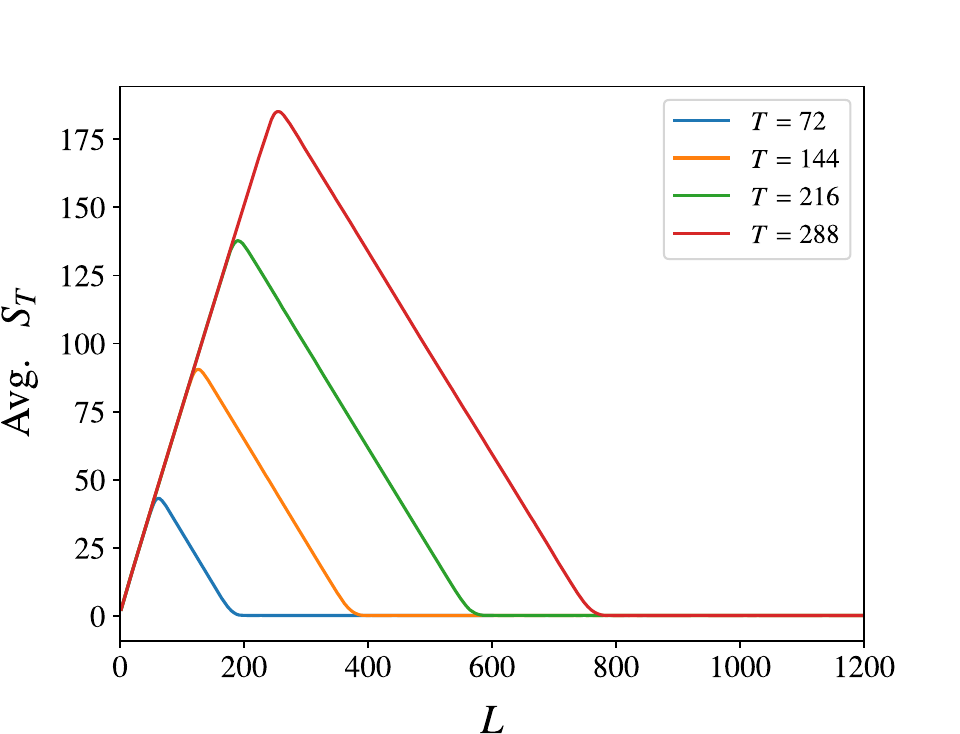}
\caption{Average temporal entanglement $S_T$ plotted against bath size $L$ for measurement rate $p=0.4$ and different total evolution times $T$. The dual-unitary gates are chosen to be the SDKI-r gates as defined in Eq. \eqref{eqn:SDKI-r}. The averaging is done with respect to both the choice of single-site Clifford gates and positions of the Bell measurements.}
\label{fig:TE_scale_T_p04_SDKI-r}
\end{figure}

Fig. \ref{fig:TE_diff_meas_SDKI-r} shows the trajectory-averaged TE as a function of the bath size for a fixed evolution time $T$ and varying measurement rates. The dual-unitary gates are chosen to be the SDKI-r gates defined in Eq.~\eqref{eqn:SDKI-r}. 
After introducing measurements, while the TE still grows and decays linearly with bath size, the slopes now depend continuously on the measurement rate $p$. For the growth regime, we find:
\begin{equation}
    S_T=2L\; \rightarrow \;S_T = 2v_{E}(p) L, \qquad 0<v_{E}(p)<1 
\end{equation}
where $v_{E}(p)$ is the entanglement velocity. For the decay regime we similarly find:
\begin{equation}
    S_T=T-L\;\rightarrow \; S_T=T-v_{E}(p)L,\qquad 0<v_{E}(p)<1\,.
\end{equation}
Fig.~\ref{fig:TE_scale_T_p04_SDKI-r} shows $S_T$ as a function of $L$ at $p=0.4$ and for various total evolution times $T$. The peak value of $S_T$ retains approximately the same value of $ \frac{2}{3} T$ as in the measurement-free case. The critical bath size at which this peak value is reached scales as $L_c\propto T$, qualitatively the same as in the measurement-free circuit. 

In summary, in circuits consisting of SDKI-r gates, or good scrambler gates in general, introducing measurements preserves the qualitative behavior of ballistic growth and decay of TE with bath size, albeit at a reduced entanglement velocity $0<v_E<1$ that depends continuously on the measurement rate $p$. The peak TE retains approximately the same value as in the measurement-free circuit, and the critical bath size scales linearly with $T$, just as in the measurement-free circuit. 

\subsection{Poor Scrambler Circuits with Measurements: Numerical Results}\label{sec:num_with_meas}

\begin{figure}[tb!]
\includegraphics[width=\columnwidth]{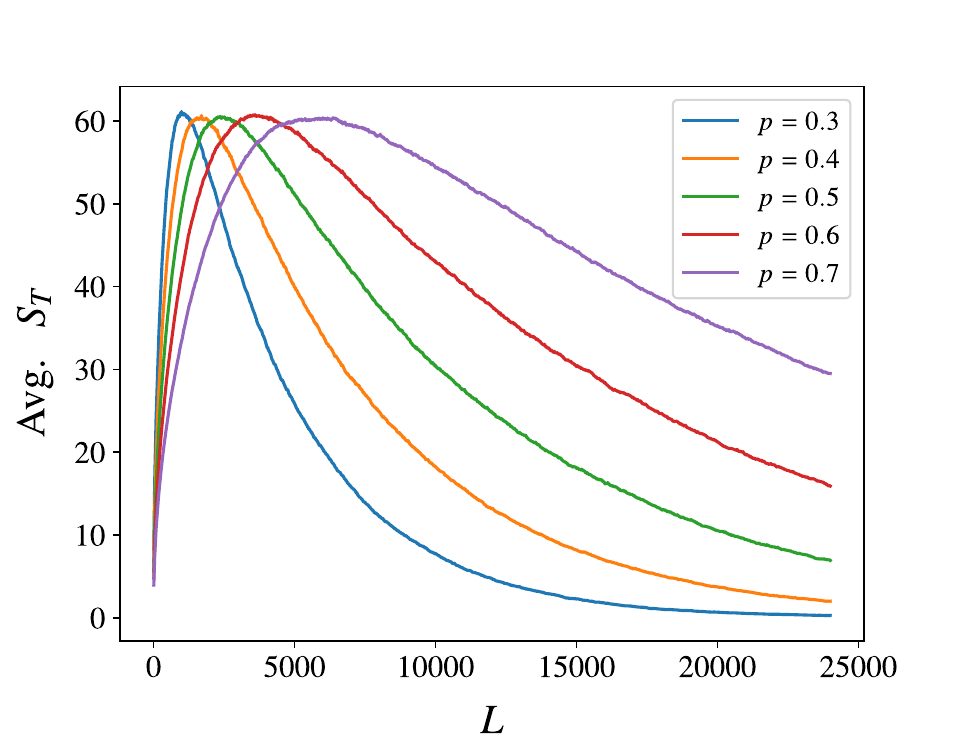}
\caption{Average temporal entanglement $S_T$ plotted against bath size $L$ for total evolution time $T=216$ and different measurement rates $p$. The dual-unitary gates are chosen to be the SDKI-f gates as defined in Eq. \eqref{eqn:SDKI-f}. The averaging is done with respect to positions of the Bell measurements.}
\label{fig:TE_diff_meas_SDKI-f}
\end{figure}

Fig. \ref{fig:TE_diff_meas_SDKI-f} shows the trajectory-averaged TE as a function of the bath size for a fixed evolution time $T$ and varying measurement rates. The dual-unitary gates are chosen to be the SDKI-f gates from Eq.~\eqref{eqn:SDKI-f}.
Introducing measurements now induces a few qualitative changes compared to the case of unitary circuits without measurements. For the growth regime we find that: 
\begin{equation}
    S_T=2 L\quad\rightarrow\quad S_T\propto \sqrt{L},
\end{equation}
whereas for the decay regime
\begin{equation}
    S_T=T-L,\quad\rightarrow\quad S_T\propto \exp{(-L/\xi)}\,,
\end{equation}
where the characteristic scale $\xi$ increases with $p$.  The linear growth and decay of TE is replaced by a diffusive growth and an exponential decay, respectively.

\begin{figure}[tb!]
\includegraphics[width=\columnwidth]{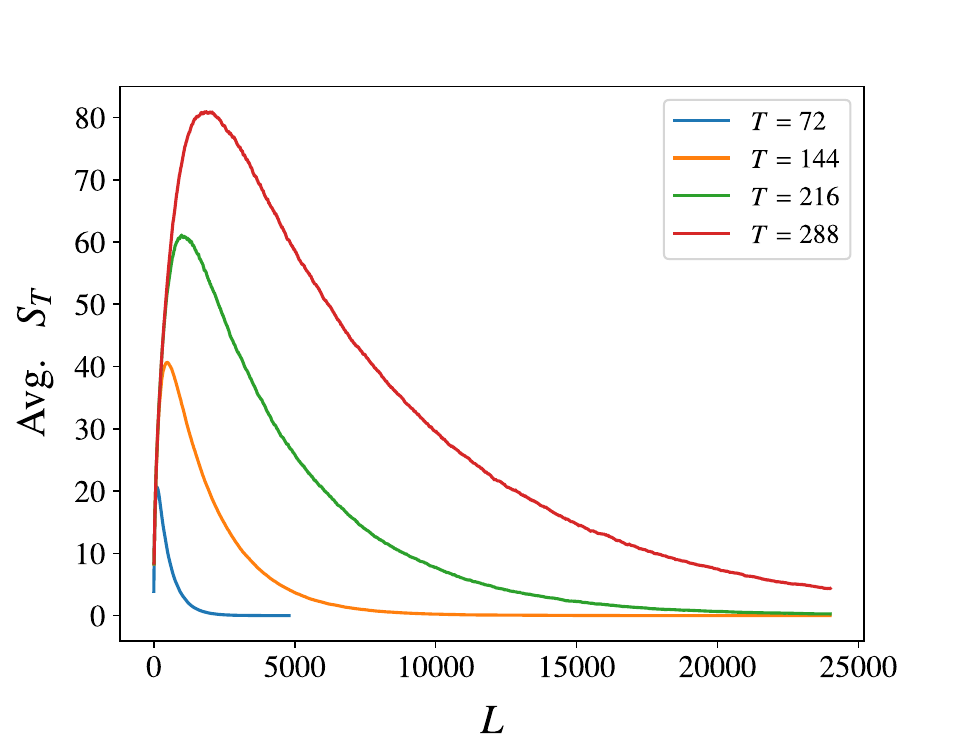}
\caption{Average temporal entanglement $S_T$ plotted against bath size $L$ for measurement rate $p=0.3$ and different total evolution times $T$. The dual-unitary gates are chosen to be the SDKI-f gates as defined in Eq. \eqref{eqn:SDKI-f}. The averaging is done with respect to positions of the Bell measurements.}
\label{fig:TE_scale_T_p03_SDKI-f}
\end{figure}

Fig. \ref{fig:TE_scale_T_p03_SDKI-f} shows $S_T$ as a function of $L$ at $p=0.3$ and for various total evolution times $T$. The peak value of $S_T$ and the critical bath size at which this peak value is reached scale as
\begin{equation}
    \begin{split}
        S_{T,\,\text{peak}}\propto T, \qquad L_c\propto T^2\,.
    \end{split}
\end{equation}

\section{Analytical Understanding of the Growth Regime} \label{sec:analytical-growth}

The different behaviors observed when introducing measurements can be directly related to the different operator dynamics in these circuits. For random unitary circuits operator dynamics can be mapped to a biased random walk, as discussed in Refs. \cite{nahum_quantum_2017, von_keyserlingk_operator_2018, nahum_operator_2018}, and we here obtain the corresponding random walk pictures for the different operator dynamics as the biased and unbiased persistent random walk for good and poor scramblers respectively.

Since we are working with stabilizer states, it is natural to examine the dynamics of operator strings under the stabilizer formalism~\cite{aaronson_improved_2004}. An operator $S$ is said to stabilize a state $\ket{\psi}$ if $S\ket{\psi} = \ket{\psi}$, and in the stabilizer formalism the state $\ket{\psi}$ is represented by the full set of operators that stabilize it. This stabilizer set $\mathcal{S}$ is hence defined as
\begin{equation}
    \mathcal{S}=\{S \,|\,S|\psi\rangle=|\psi\rangle\},
\end{equation}
and instead of evolving the state, one evolves the stabilizer set as
\begin{equation}
    \begin{split}
        \mathcal{S}\rightarrow U\mathcal{S}U^\dagger,
    \end{split}
\end{equation}
where for each $S_i\in \mathcal{S}$, $S_i\rightarrow US_i U^\dagger$. The action of an operator $U$ on a stabilizer string $S_i$ is denoted $U(S_i)=US_i U^\dagger$.

For $n$ sites, the basis consists of Pauli strings of the form $S^{(n)}=S_1 \otimes S_2\otimes\ldots\otimes S_n$, where the single-site Pauli operator is $S_i\in \{I,X,Y,Z\}$. 
By construction, Clifford circuits map one Pauli string to another Pauli string without generating superpositions of Pauli strings. One may factorize every Pauli string into an $X$-string and a $Z$-string, only containing the identity and $X$ and $Z$ operators respectively, and keep track of the stabilizer action on each string separately. For a 2-site gate $U$, its action on $\{IX,XI,IZ,ZI\}$ completely defines its action on all Pauli strings.


\subsection*{Measurement-free case}
We consider the stabilizer dynamics in the measurement-free case, first focusing on the poor scramblers before arguing how these results extend to good scramblers. 

The SWAP gate acts as
\begin{equation}
    \begin{split}
        &\text{SWAP}(ZI)=IZ,\quad\quad \text{SWAP}(IZ)=ZI,\\
        &\text{SWAP}(XI)=IX,\quad\quad \text{SWAP}(IX)=XI,
    \end{split}
\end{equation}
and the SDKI-f gate acts as
\begin{equation}
    \begin{split}
        &\text{SDKI-f}(ZI)=XZ,\quad\quad \text{SDKI-f}(IZ)=ZX,\\
        &\text{SDKI-f}(XI)=IX,\quad\quad \text{SDKI-f}(IX)=XI.
    \end{split}
\end{equation}
The stabilizer set for a Bell pair connecting sites $t_i$ and $t_{i+1}$,
\begin{equation}
        |\phi\rangle_{i,\,i+1}=\;\vcenter{\hbox{\includegraphics[width=0.18\linewidth]{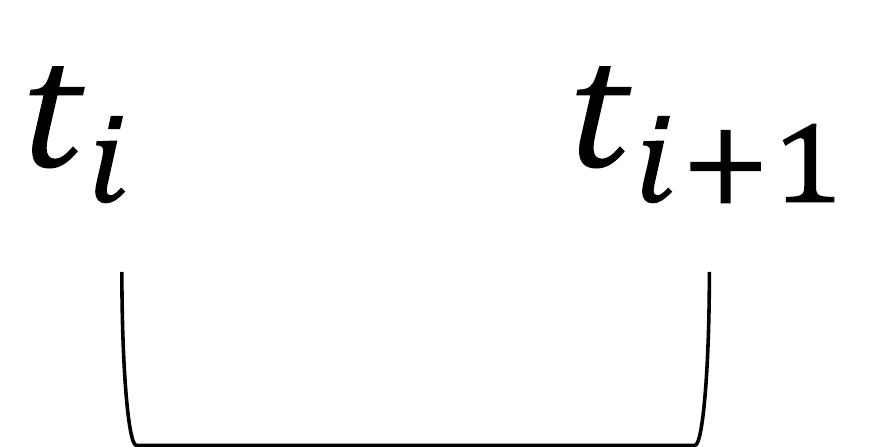}}},
\end{equation}
is given by
\begin{align}
        \mathcal{S}\left(|\phi\rangle\right)=\{\,&I_1 I_2\ldots I_{i-1}X_i X_{i+1}I_{i+2}\ldots I_T,\; \nonumber\\
        &I_1 I_2\ldots I_{i-1}Z_i Z_{i+1}I_{i+2}\ldots I_T\,\}\,.
\end{align}
Starting from the state $|\Phi\rangle=\bigotimes |\phi\rangle$, under the action of the SWAP gates the $IXXI$ and $IZZI$ stabilizers spread into $XIIX$ and $ZIIZ$, respectively:
\begin{equation}
\begin{split}
   (\text{SWAP}\otimes\text{SWAP})(IXXI)&=XIIX,\\
   (\text{SWAP}\otimes\text{SWAP})(IZZI)&=ZIIZ.
    \end{split}
\end{equation}
and for the SDKI-f circuit the $IXXI$ and $IZZI$ stabilizers spread as
\begin{equation}
\begin{split}
   (\text{SDKI-f}\otimes\text{SDKI-f})(IXXI)&=XIIX,\\
   (\text{SDKI-f}\otimes\text{SDKI-f})(IZZI)&=ZXXZ.
    \end{split}
\end{equation}
It can be directly verified that for initial Bell pair states updated by monitored poor scrambler circuits, stabilizers whose endpoints are $Z$ are either of the form $ZII\dots IIZ$ for the all-SWAP circuit or $ZXX \dots XXZ$ for the all-SDKI-f circuit, whereas operators whose endpoints are $X$ are all of the form $XII\dots IIX$. Crucially, these endpoints of the operator strings completely determine the entanglement structure of the state \cite{nahum_quantum_2017, li_measurement_2019}. As such, the precise Pauli operators in the bulk of the string are irrelevant (explaining why these different models give rise to the same TE entanglement profile). Under the action of the full brickwork circuit, the endpoints of the stabilizers strings spread ballistically with velocity $|v|=1$ while their overall structure remains unchanged. This result directly returns the maximal entanglement growth of the measurement-free case.

For the good scrambler case, considering e.g. the space-time rotated version of the SDKI-S gate as defined in Eq.~\eqref{eqn:SDKI-s}, the endpoints of the stabilizers strings again spread ballistically with maximum velocity (as implied by dual-unitarity~\cite{bertini_operator_2020a,claeys_maximum_2020}). Consider the right end point, for which the gate acts as
\begin{equation}
        \begin{split}
        &\text{SDKI-S}(ZI) = XZ,  \quad \text{SDKI-S}(XI) = XY,\\
        &\text{SDKI-S}(YI) = IX.
        \end{split}
\end{equation}
While the overall stabilizer strings no longer have the same simple structure, the endpoints move in the same way. This is because the each stabilizer operator front always ends on the first leg of the dual-unitary gate. By dual-unitarity, the operator front always grows, resulting in the same temporal entanglement dynamics for both classes of scramblers.

\subsection*{Monitored Poor Scrambler Circuits} The situation changes when introducing measurements. The Bell-pair measurements act as two-site identity gates upon space-time rotation, such that they do not grow the stabilizers. Furthermore, because of the even-odd staggering structure of the brickwork circuit, the stabilizers may also shrink. This is because instead of always ending on the first leg of the dual-unitary gate, each stabilizer operator front may now also end on the second leg of the gate. This scenario can be easily illustrated using a specific realization of an all-SWAP circuit with measurements (the argument directly extends to the SKDI-f gates):
\begin{align}\label{eqn:traj_prw}
    \vcenter{\hbox{\includegraphics[width=0.6\linewidth]{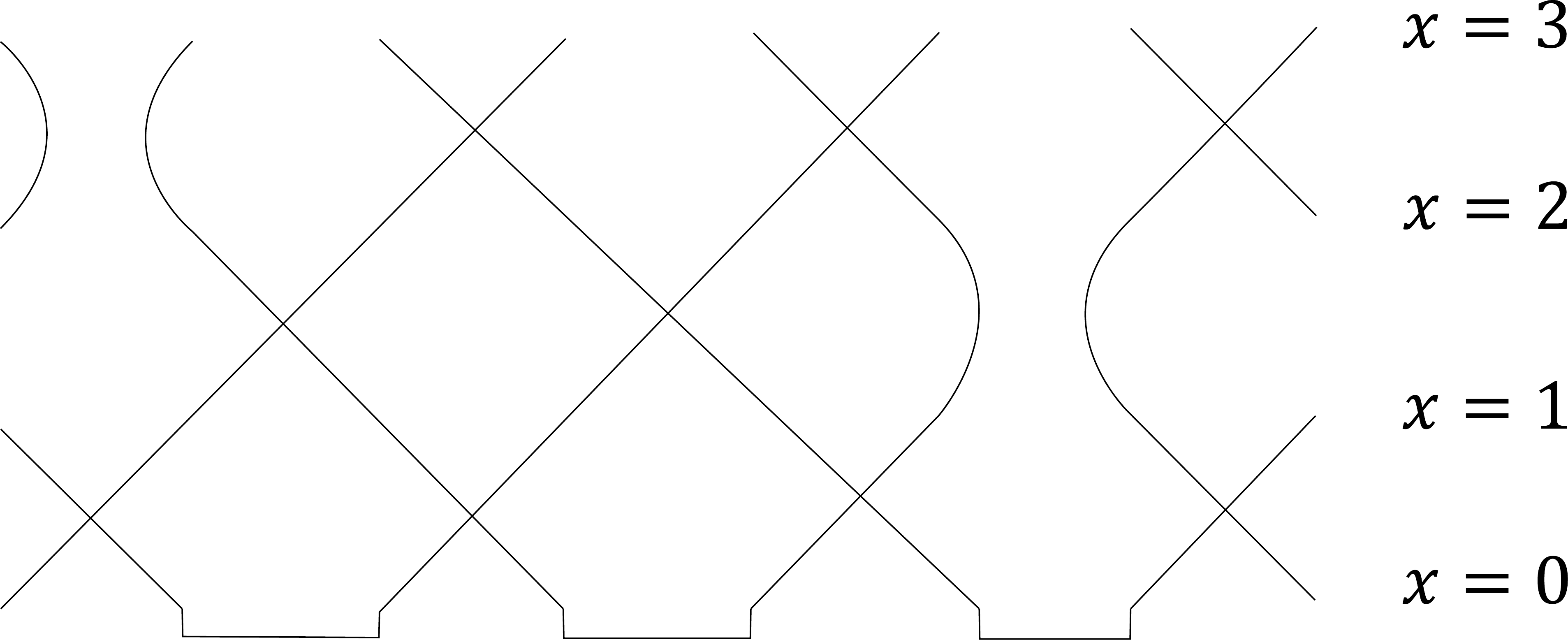}}}
\end{align}
In the first update step no measurement occurs, and both ends of the middle Bell pair propagate outwards. In the second update step a measurement occurs on the right operator front of the middle Bell pair, and the operator front stays put. In the third step, a SWAP gate acts on the right operator front of the middle Bell pair. Since the operator front ends on the second leg of the SWAP gate, it shrinks by one. 

Crucially, when acted on by a poor scrambler gate, the operator front always shrinks if it ends on the second leg of the gate, whereas it always spreads if it ends on the first leg of the gate. As a result, there is no overall preferred direction of movement, and the operator front of each Bell pair undergoes an unbiased \emph{persistent random walk } (PRW)~\cite{kac_stochastic_1974, pottier_analytic_1996}. 

The PRW is a variant of the random walk analogous to the run-and-tumble model in the context of biophysics and active matter \cite{berg_e_2004}. 
Consider a random walk model in discrete time on a discrete one-dimensional lattice, where the displacement after $n$ update steps is given by $\sum_{i=1}^{n} \tau_i $, where $\tau_i=\pm 1$ is the displacement at update step $i$. A movement to the right corresponds to $\tau_i=+1$, and a movement to the left corresponds to $\tau_i=-1$. Initially the particle starts in one direction with velocity $v_0=+1$ or $v_0=-1$. In the unbiased case, at each step $i$, the particle may either continue with the same velocity, with probability $(1-\alpha)$, or reverse its direction of traveling, with probability of reflection $\alpha$. Due to the lack of a preferred direction, the particle experiences no drift but instead diffuses, with diffusion constant \cite{kac_stochastic_1974, pottier_analytic_1996}: 
\begin{equation}\label{eqn:D_of_p}
    D(\alpha)=\frac{1-\alpha}{\alpha}\,.
\end{equation}
In our scenario we recover diffusive motion for the temporal displacement of the Bell pairs ends, where the average displacement $\sqrt{\langle \Delta t ^2\rangle}$ after $n$ updates follows as 
\begin{equation}
        \sqrt{\langle \Delta t ^2\rangle}= \sqrt{2D(p)\,n}\,.
\end{equation}
Here $D(p)$ is the diffusion constant as a function of the measurement rate $p$. From Eq.~\eqref{eqn:traj_prw}, it can be seen that the reflection probability coincides with the measurement probability. The diffusion constant follows as $D(p)=(1-p)/p$. 

The operator dynamics of the stabilizer strings hence corresponds to the outer edges undergoing a persistent random walk, while the (internal) structure of the stabilizer strings remain unchanged. We identify the number of Bell pairs crossing the time point $t_i$ as the ``local Bell pair density", denoted $m_{t_i}$. Since the Bell pairs are initially uniformly placed, and the update preserves the center of mass on average, the local Bell pair density remains proportional to the average radius of each Bell pair, 
\begin{equation}
m_{t_i}=\kappa \sqrt{\langle \Delta t^2 \rangle},
\end{equation}
where we have introduced a proportionality constant $\kappa$ that can be interpreted as an effective packing factor. Since $S_T=2\cdot \max_{t_i} (m_{t_i})$, temporal entanglement is hence expected to grow diffusively with each update step as: 

\begin{equation}\label{eqn:S_diffusion}
    \begin{split}
        S_T&=2\cdot \max_{t_i}(m_{t_i})=2\kappa \sqrt{\langle \Delta t^2 \rangle}\\&=2\kappa \sqrt{2 D(p)\,n} = 2\sqrt{2}\kappa\sqrt{ \frac{1-p}{p}\,n} \,,
    \end{split}
\end{equation}
where $n$ is again the number of spatial update steps. 

\begin{figure}[tb!]
\includegraphics[width=\columnwidth]{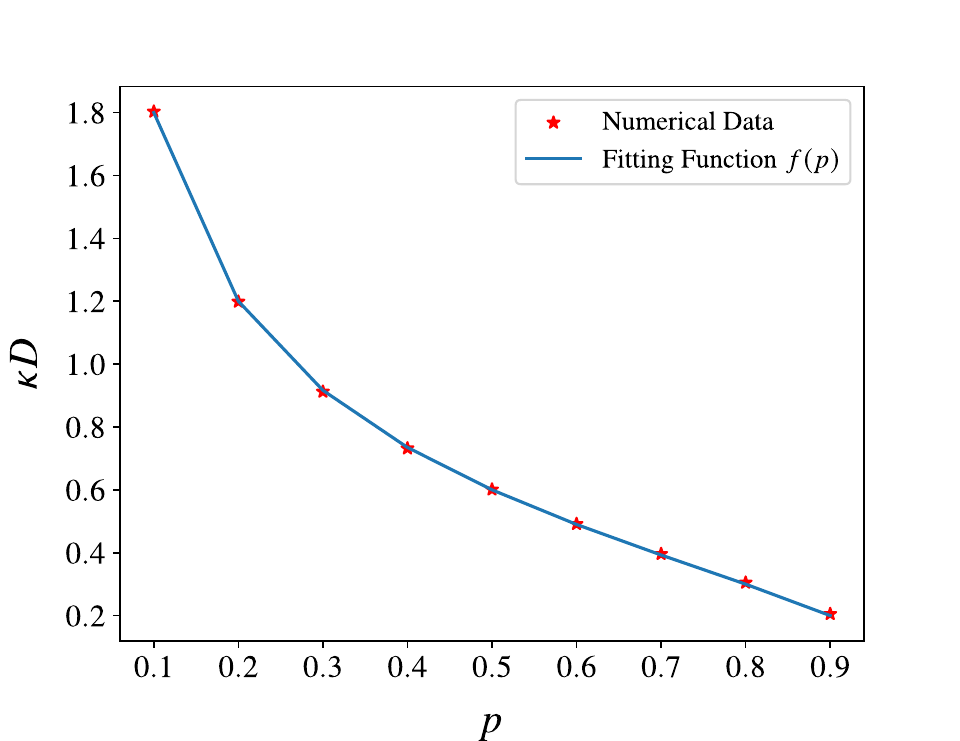}
\caption{Numerically extracted effective diffusion constant, $\kappa D$, for total evolution time $T=432$. A one-parameter fit using $f(p)=\kappa \sqrt{1-p}/\sqrt{p}$ yields excellent agreement, with $\kappa=0.36$ as the fitting parameter. }
\label{fig:diffusion_const}
\end{figure}

A one-parameter fit is performed on the numerical data with $\kappa$ as the fitting parameter, and the results are shown in Fig. \ref{fig:diffusion_const}. The fitting curve agrees well with numerical data, confirming the proposed functional dependence on $p$, and the optimal fitting value is found as $\kappa\approx 0.36$. These results confirm that the operator front dynamics in the monitored poor scrambler corresponds to an unbiased persistent random walk.

 While the presented picture shows a PRW of the ends of the Bell pairs, the exact same PRW also describes the ends of operator strings associated with each Bell pair. This equivalence between operator fronts and entanglement fronts is due to the fact that the generating stabilizer set always remains in the so-called clipped gauge \cite{nahum_quantum_2017, li_measurement_2019} when updated by the the monitored poor scrambler circuits.

\subsection*{Monitored Good Scrambler Circuits} In monitored good scramblers, this equivalence between operator fronts and entanglement fronts is lost, and the butterfly velocity $v_B$ at which operator fronts travel is generally different from the entanglement velocity $v_E$ at which entanglement spreads. Only in the measurement-free dual-unitary case do these coincide, leading to a maximal butterfly velocity $v_B=1$~\cite{claeys_maximum_2020} and entanglement velocity $v_E=1$~\cite{zhou_maximal_2022}.

For these circuits, the action on the Pauli strings is more involved since (products of) $X$ and $Z$ matrices can be mapped to $Y$ matrices, where for completeness we give the remaining transformations of the space-time rotated SDKI-S gate below:

\begin{align}
    &\text{SDKI-S}(IZ) = ZX, & \quad &\text{SDKI-S}(IX) = YX,\\
    &\text{SDKI-S}(IY) = XI, & \quad &\text{SDKI-S}(ZZ) = YY,\\
    &\text{SDKI-S}(XZ) = YZ, & \quad &\text{SDKI-S}(YZ) = ZI,\\  
    &\text{SDKI-S}(ZX) = ZY, & \quad &\text{SDKI-S}(XX) = ZZ,\\ 
    &\text{SDKI-S}(YX) = YI, & \quad &\text{SDKI-S}(ZY) = IZ,\\ 
    &\text{SDKI-S}(XY) = IY, & \quad &\text{SDKI-S}(YY) = XX,
\end{align}
The even-odd staggering of the circuit again results in a persistent random walk, which is now \emph{biased} since the operator fronts now have finite probability of \emph{not} shrinking when ending on the second leg of the good scrambler gate. In the biased PRW, the reflection probability at step $i$ differs depending on whether the particle's velocity at step $i-1$ is $v_{i-1}=+1$ or $v_{i+1}=-1$, and the corresponding reflection probabilities are denoted $\alpha_+$ and $\alpha_-$, respectively. The particle experiences both drift and diffusion, with the drift term dominating for a large enough number of update steps. The particle then travels ballistically with a drift velocity $v_{\text{drift}}$~\cite{pottier_analytic_1996}
\begin{equation}\label{eqn:v_drift}
    v_{\text{drift}}=\frac{\alpha_- - \alpha_+}{\alpha_- + \alpha_+}\,.
\end{equation}

These probabilities now depend on the specific choice of gate.
When these gates act on an operator front ending on the first leg of the gate, the operator string always grows by one site due to the dual-unitarity of the gate. Conversely, when acted on by the two-site identity the operator front stays put, corresponding to a ``reflection" event. Therefore, $\alpha_+ = p$. 

When these gates act on an operator front ending on the second leg of the gate, it is possible for the operator front to shrink.
 However, when acted on by a rotated good scrambler gate, the operator string has some probability $s$ of staying put and probability $(1-s)$ of shrinking by one site. This is in contrast to the monitored poor scrambler circuits, where the operator string always shrinks when acted on by the dual-unitary gate in this way, i.e. $s=0$. For the SDKI-r case the rate $s$ can be computed by averaging over the single-site Clifford gates and assigning equal probability to each allowed transition. The outmost Pauli can be only one of $X$, $Y$, or $Z$, whereas the second outmost Pauli can be $I$, $X$, $Y$, or $Z$, so there are $3\times 4=12$ possible configurations. Out of these 12, 3 lead to shrinking of the operator string, whereas 9 lead to the operator string staying put, such that $s=9/12$. When acted on by the two-site identity, the operator front still trivially stays put, such that we find $\alpha_-=p+s(1-p) = p/4+3/4$. 

Using Eq.~\eqref{eqn:v_drift}, the butterfly velocity at which the operator front travels can be computed as:
\begin{equation}\label{eqn:vb}
    v_B= \frac{\alpha_- -\alpha_+}{\alpha_- +\alpha_+} = \left(1+\frac{2p}{s(1-p)} \right)^{-1} \,,
\end{equation}
where $s=3/4$ for monitored SDKI-r circuit.

It can be directly checked that Eq.~\eqref{eqn:vb} yields the correct limiting behaviors: $\lim_{p\rightarrow 0} v_B=1$, $\lim_{p\rightarrow 1} v_B=0$, and, in the absence of bias, $\lim_{s\rightarrow 0} v_B=0$.
\begin{figure}[tb!]
\includegraphics[width=\columnwidth]{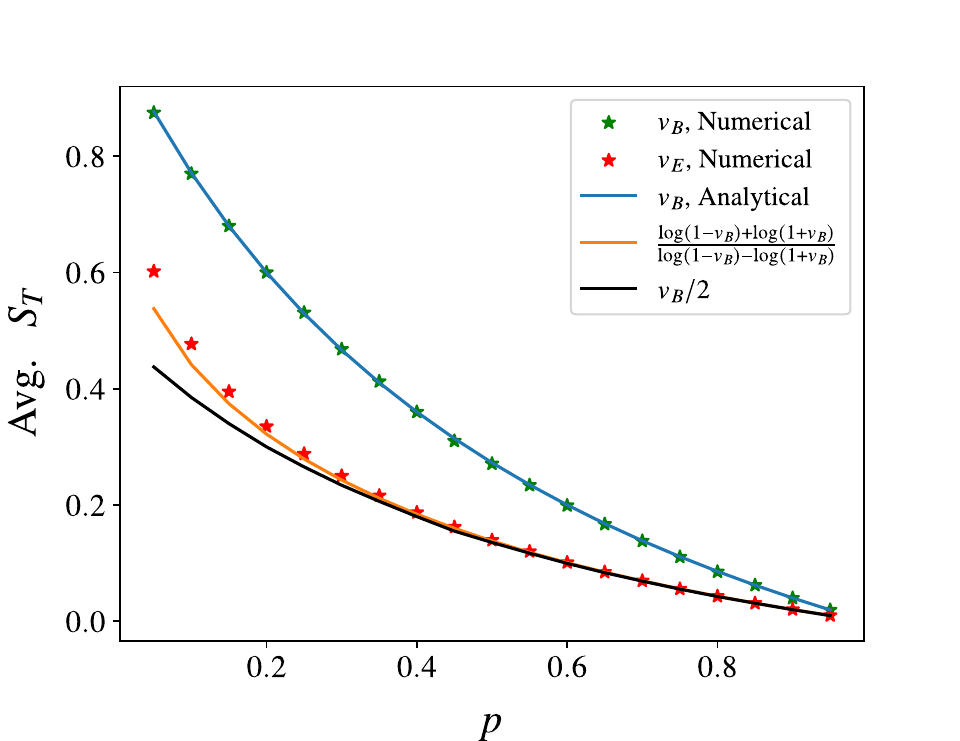}
\caption{Numerical data for the butterfly velocity $v_B$ and entanglement velocity $v_E$ for various measurement rates $p$ at $T=128$. The analytically obtained $v_B$ is also plotted, along with two relations expressing $v_E$ as a function of $v_B$.}
\label{fig:vb_ve}
\end{figure}

Due to the fact that the generating stabilizer set does not always remain in the clipped gauge, not all stabilizers ending at given site $t$ are independent~\cite{nahum_quantum_2017, li_measurement_2019}. Therefore, the entanglement velocity is generally smaller than the butterfly velocity, $v_E<v_B$. The former velocity is the relevant one for our study of temporal entanglement. Fig.~\ref{fig:vb_ve} shows the numerically obtained butterfly velocity $v_B$ and entanglement velocity $v_E$ for various measurement rates $p$ at $T=128$. The analytically computed $v_B$ shows excellent agreement with the numerical results. While there is no universal relation between $v_E$ and $v_B$, various functional dependencies of $v_E$ on $v_B$ have appeared in the literature. For the non-integrable Ising model it has been shown that $v_E\approx v_B /2$ \cite{mezei_entanglement_2017,jonay_coarse_2018}, whereas for Haar random circuits a more refined relation is given by (see Eq.~(27) of Ref.~\cite{von_keyserlingk_operator_2018}):
\begin{equation}\label{eq:v_E_v_B_2}
    v_E= \frac{\log(1+v_B)+\log(1-v_B)}{\log(1+v_B)-\log(1-v_B)}\,.
\end{equation}
Both relations are plotted in Fig.~\ref{fig:vb_ve}, and it can be observed that both agree with the numerically obtained $v_E$ for sufficiently large measurement rates $v_B\lesssim 0.4$.
In the case of no measurements the correct limiting value for $v_E$ is $\lim_{p\rightarrow 0}v_E=1$, which is only matched by Eq.~\eqref{eq:v_E_v_B_2}.  While no analytic relation between the entanglement velocity and the butterfly velocity was found in our setup, we see that Eq.~\eqref{eq:v_E_v_B_2} presents an accurate approximation to the entanglement velocity starting from the analytically obtained butterfly velocity. 

In this way we can conclude that the dynamics of the monitored good scrambler circuits corresponds to a biased persistent random walk. The ballistic drift term originating from the bias dominates the dynamics, resulting in the linear growth already observed in the measurement-free case. The maximal entanglement velocity $v_E=1$ in the measurement-free case however needs to be modified due to the measurements, and can here be indirectly obtained by expressing the entanglement velocity in terms of the butterfly velocity, which admits an exact expression in terms of the measurement rate.

\section{Analytical Understanding of the Decay Regime}\label{sec:analytical_decay}
The linear decay of TE in monitored good scrambler circuits can be understood in an analogous way as in the measurement-free circuits, using a similar underlying microscopic picture with a modified entanglement velocity $v_E<1$. The slope for decay is again always half the value of the slope for growth.

For this reason we will now focus on monitored poor scrambler circuits. Here the TE decays exponentially with the bath size, reaching the perfect dephaser limit only as $L\rightarrow \infty$, as opposed to the measurement-free case where this value is obtained at a finite bath size. In order to understand the exponential decay and extract the decay scale, we identify a ``mixed" spatial transfer matrix, $\mathcal{T}_p$, where the averaged TE can be directly calculated by absorbing the measurements into $\mathcal{T}$. This transfer matrix is given by:
\begin{equation}
    \begin{split}
        \mathcal{T}_p = \quad \vcenter{\hbox{\includegraphics[width=0.3\linewidth]{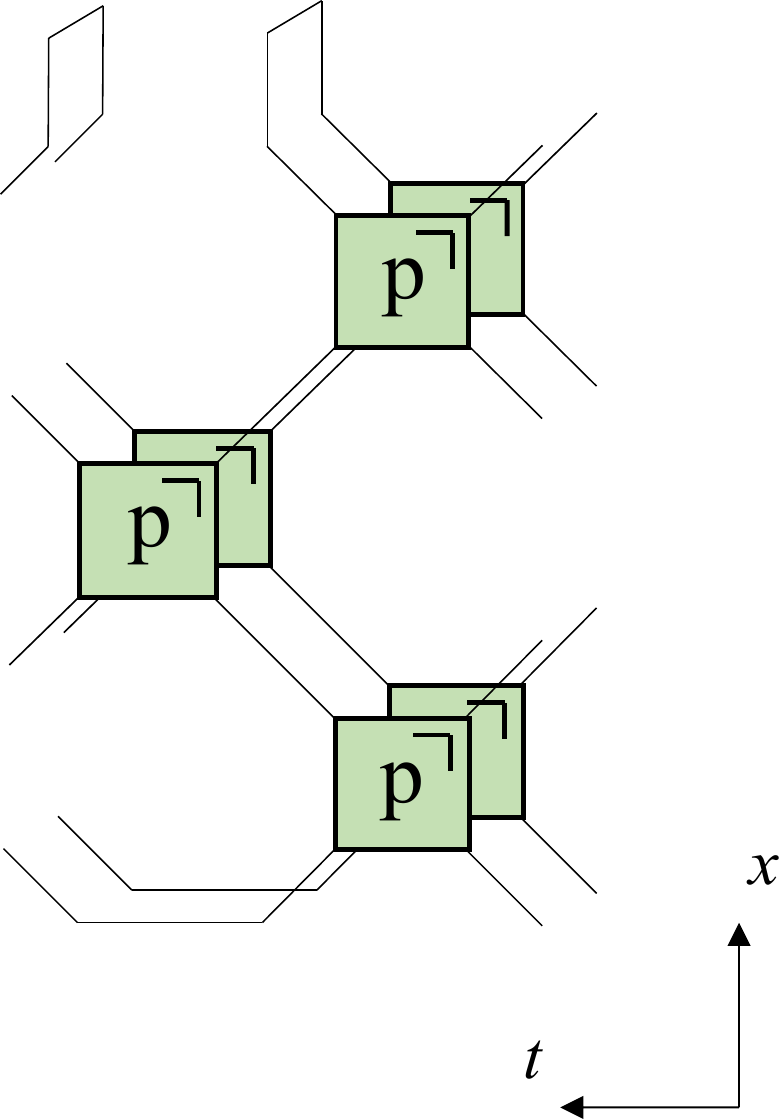}}}
    \end{split}
\end{equation}
where
\begin{equation}\label{eq:mixed-gate}
    \begin{split}
        \vcenter{\hbox{\includegraphics[width=0.15\linewidth]{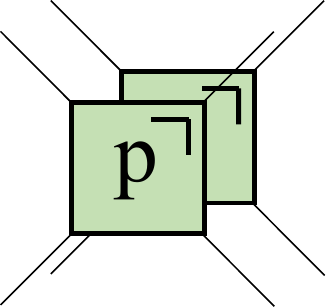}}}\equiv (1-p)\cdot \; \vcenter{\hbox{\includegraphics[width=0.15\linewidth]{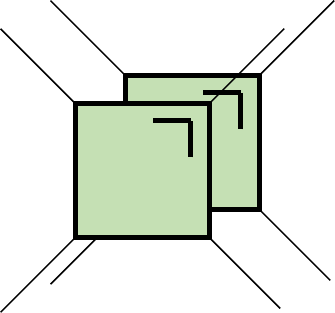}}}+p \cdot \vcenter{\hbox{\includegraphics[width=0.15\linewidth]{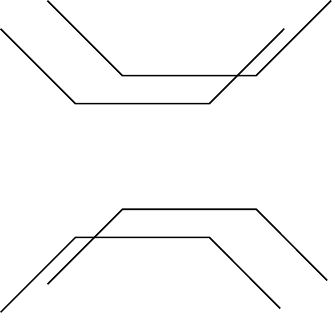}}}
    \end{split}
\end{equation}

In order to calculate the averaged TE, numerical tensor contraction is performed to construct $\mathcal{T}_p$ with SWAP gates and measurements. Since the spectrum of $\mathcal{T}_p$ will determine the exponential decay, the tensor contractions are done without bond-dimension truncation. 

That this averaged transfer matrix exactly determines the dynamics of the TE is specific to our setup, and depends on the specific choice of gates and measurements. 
Under real time evolution, the norm of the state may decrease after a measurement, and the state needs to be rescaled to ensure a norm of 1 after each measurement. With numerical simulation using the stabilizer formalism, this is always accounted for, and the state remains normalized at all times. When constructing the averaged transfer matrix analytically, however, care needs to be taken in choosing the coefficient in front of the projector to account for normalization of the state. This explicitly shows up in Eq. \eqref{eq:mixed-gate}, where the coefficient in front of the projector is chosen to be $p\times 1/4 \times 4 = p$. The factor of $p$ accounts for the probability for a measurement to occur, whereas the factor of factor of $1/4$ accounts for the normalization of the Bell-pair states in the projector. Finally, the factor of $4$ ensures that the norm of the state remains 1 after each measurement. That the final normalization constant is always 4 is specific to the choice of Bell-pair initial state and all the dual-unitary gates in the circuit being the SWAP gate. It can be easily verified that under such circuit dynamics, the state remains a tensor product of Bell pairs at all times, with the SWAP gates merely spreading the Bell pairs around. The measurement in the Bell-pair basis always merges two pairs while generating a new Bell pair -- in doing so, it conserves the total number of Bell pairs, while introducing a factor of $1/4$ to the norm of the state, reflecting the amount of overlap between the state and the measurement projector. 

The only potential counter-example to the above-described action of the measurement is if the Bell-pair projector acts on two legs of the \textit{same} Bell pair, in which case there would be an additional factor of 4 generated after the measurement. 

However, such a situation never occurs in our setup, since, due to causality, in order for the measurement to act on the two legs of the same Bell pair, the two legs of the Bell pair must have reflected off the two spatial boundaries. In our setup the right spatial boundary is always left open with open degrees of freedom representing that of the influence matrix, such that no Bell pair can have both of its legs reflecting off both spatial boundaries, and the above-described counter-example can never occur in our setup.

For extracting the decay scale for purity, one should generally consider constructing the 4-replica $\mathcal{T}_p$ rather than the 1-replica $\mathcal{T}_p$. However, this is unnecessary for analyzing the SWAP circuit with measurements. Any specific circuit realization will consist out of SWAP gates and identity matrices along the spatial direction, and the action on any initial product state will lead to a ``reshuffling" of the initial state, and this reshuffling is independent of the choice of local basis. Furthermore, since the gate choices are locked among replicas, the 4-replica $\mathcal{T}_p$ has the same set of eigenvalues as the 1-replica $\mathcal{T}_p$, albeit with different degeneracies. 
This argument directly extends to any calculation of the purity and R\'enyi entropies (see also Ref.~\cite{claeys_exact_2022}). 
Within the 1-replica $\mathcal{T}_p$, one may further reduce the local Hilbert space dimension from $q=4$ to $q=3$, since the gate choices are also locked between the forward and backward time contours. The detailed construction is presented in Appendix \ref{app:num_exact}, and significantly reduces the computational complexity of constructing the transfer matrix, allowing for numerical simulations for longer evolution times. 

\begin{figure}[tb!]
\includegraphics[width=0.9\columnwidth]{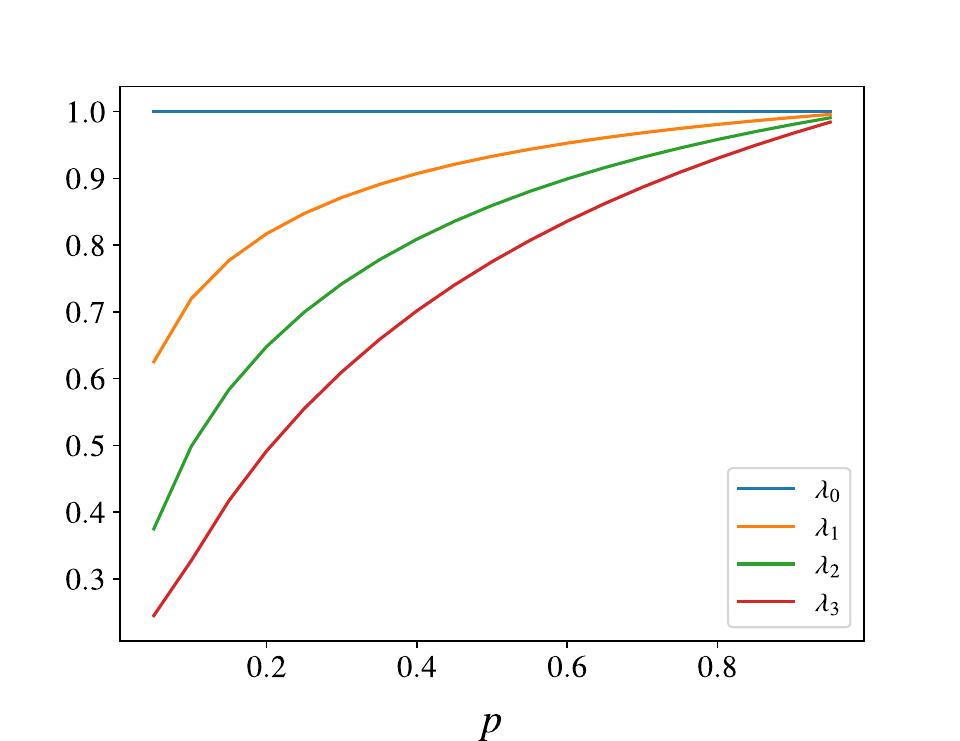}
\caption{First four leading eigenvalues of the mixed transfer matrix $\mathcal{T}_p$ plotted against the measurement rate, $p$, for total evolution time $T=6$.}
\label{fig:spec1}
\end{figure}

The leading eigenvalue of $\mathcal{T}_p$ satisfy $\lambda_0=1$ and corresponds to the perfect-dephaser steady state in the $L\rightarrow\infty$ limit, as can be directly checked. Furthermore, $|\lambda_i|<1$ for $i>0$ such that all sub-leading eigenmodes decay exponentially, qualitatively explaining the observed exponential decay of the TE with bath size $L$.
These leading eigenvalues of $\mathcal{T}_p$ with $T=6$ are plotted against $p$ in Fig. \ref{fig:spec1}. The leading eigenvalues $\lambda_i$ with $i=0,1,2,3$ are real and change smoothly with $p$.  

\begin{figure}[tb!]
\includegraphics[width=0.9\columnwidth]{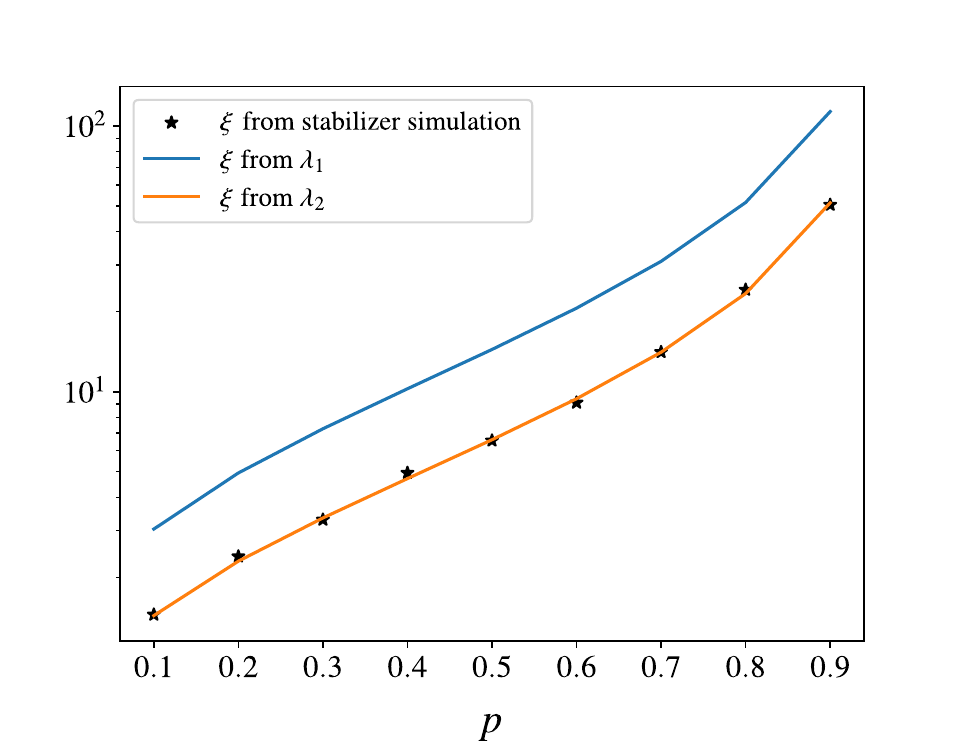}
\caption{Decay scales for $S_T$ as extracted from numerical data and as predicted from $\lambda_1$ and $\lambda_2$.}
\label{fig:scale1}
\end{figure}

Fig. \ref{fig:scale1} compares the decay scale $\xi$ for $S_T \propto \exp{(-L/\xi)}$ as extracted from numerical data to the decay scales predicted from $\lambda_1$ and $\lambda_2$. 
In generic cases, one expects the decay scale to be set by the leading nontrivial eigenvalue $\lambda_1$, $\xi=-(\log\lambda_1)^{-1}$. Nevertheless, numerical results indicate that  $\xi=-(\log\lambda_2)^{-1}$. This difference is indicative of a symmetry present in $\mathcal{T}_p$.


In order to highlight this symmetry it is convenient to define a local basis for two qubits at the same time point $t_i$ on the forward and backward time contours as:
\begin{equation}
    \begin{split}
    |B_\alpha\rangle_i\;\equiv\quad\vcenter{\hbox{\includegraphics[width=0.16\linewidth]{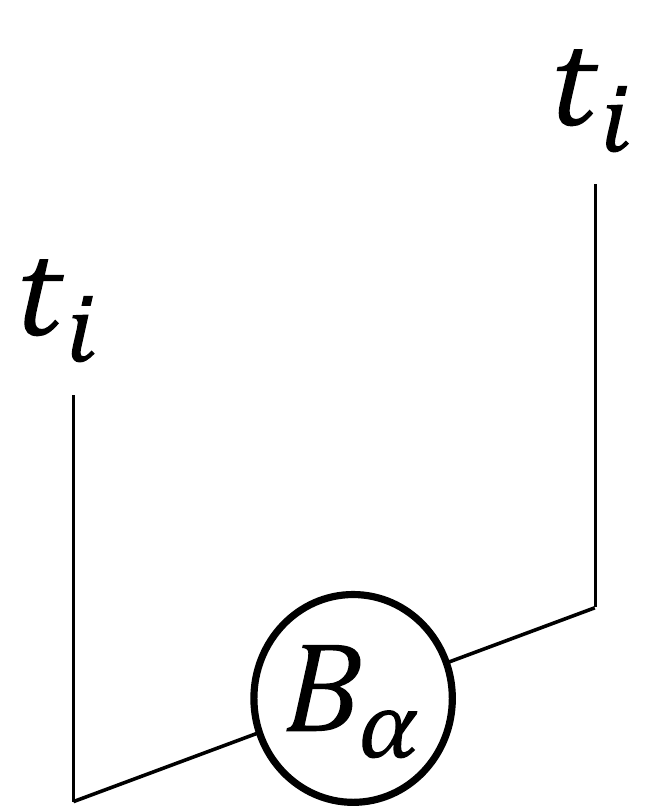}}}\,,
    \end{split}
\end{equation}
where $\alpha\in\{I,\, X,\,Y,\,Z\}$ again label the four Bell-pair states.
Graphically, 
\begin{align}\label{eq:def_Bs}
    |B_I\rangle\;\equiv\;\vcenter{\hbox{\includegraphics[width=0.03\linewidth]{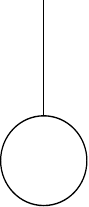}}}\;,\quad |B_X\rangle\;\equiv\;\vcenter{\hbox{\includegraphics[width=0.04\linewidth]{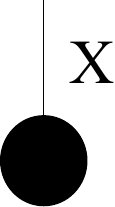}}}\;,\quad |B_Y\rangle\;\equiv\;\vcenter{\hbox{\includegraphics[width=0.04\linewidth]{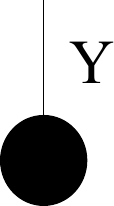}}}\;,\quad |B_Z\rangle\;\equiv\;\vcenter{\hbox{\includegraphics[width=0.04\linewidth]{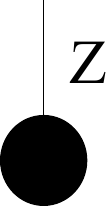}}}\,.
\end{align}
The trace projector at the temporal boundary $t=T$ projects onto the $|B_I\rangle$ mode and annihilates the X, Y, and Z states. The product state of $|B_I\rangle$ modes $|\circ\circ \dots \circ \circ \rangle$ corresponds to the perfect dephaser state and is an eigenstate of $\mathcal{T}_p$. We can denote the $|B_I\rangle$ mode as a reference state and the $|B_X\rangle$, $|B_Y\rangle$, $|B_Z\rangle$ modes as X-, Y- and Z-particles.

Consider the SWAP circuit with measurements. One trajectory realization may then, for example, look like the following: 
\begin{equation}
\vcenter{\hbox{\includegraphics[width=0.7\linewidth]{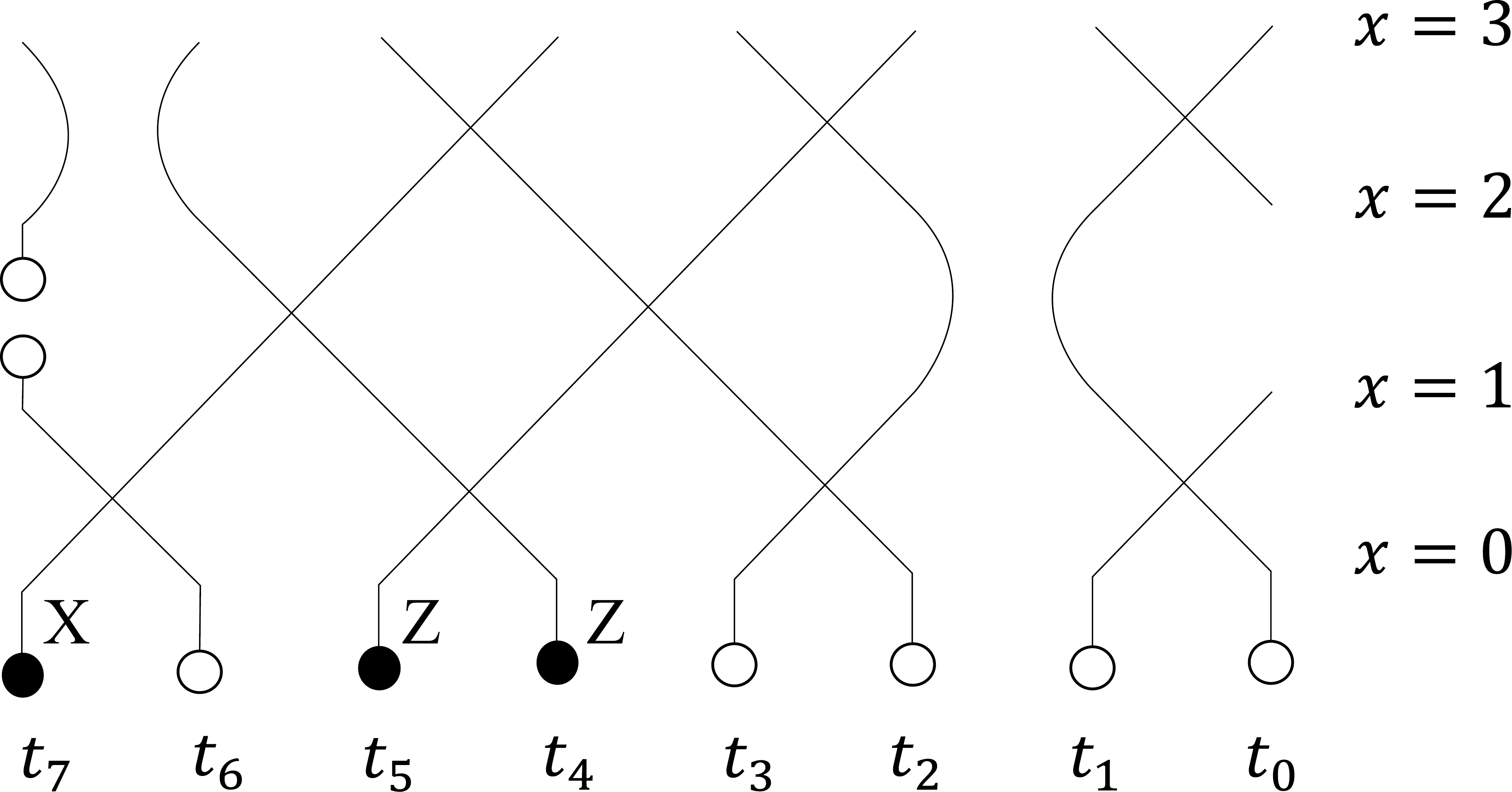}}}
\end{equation}

The actions of the SWAP and the identity gates both preserve the number and flavors of the particles (i.e. whether these are X, Y or Z Bell pairs), and the particles merely get shuffled around. The only operator that does not conserve the particle number is the projector, which however annihilates the state if it acts on a particle, such that $\mathcal{T}_p$ does not couple sectors with different number and flavors of particles.
Consequently, $\mathcal{T}_p$ decomposes into symmetry sectors corresponding to fixed numbers of particles: 

\begin{align}
    \mathcal{T}_p=\mathcal{T}^{(0)}_p \oplus \mathcal{T}^{(1)}_p \oplus \ldots \oplus\mathcal{T}^{(T-1)}_p\,,
\end{align}
where the superscript denotes the number of particles in the sector.

For simplicity, we assume that all particles are of the same flavor; predictions obtained under such simplification already match the numerical results. 
The first few leading eigenvalues, which are all real, correspond to the leading eigenvalues of the lowest particle-number sectors, i.e. $\lambda_0\leftrightarrow \mathcal{T}_p^{(0)}$, $\lambda_1\leftrightarrow \mathcal{T}_p^{(1)}$, $\lambda_2\leftrightarrow \mathcal{T}_p^{(2)}$, etc. In order to have a finite entanglement in the system, one needs at least one Bell pair connecting some $t_i$ to some $t_j$ on the same time contour. Such a Bell pair however requires at least two particles in the system. E.g., two replicas of a Bell pair connecting different times in two time contours can be expressed as a linear combination of four Bell pairs connecting the same times between different time contours:
\begin{equation}
    \begin{split}
    \vcenter{\hbox{\includegraphics[width=0.2\linewidth]{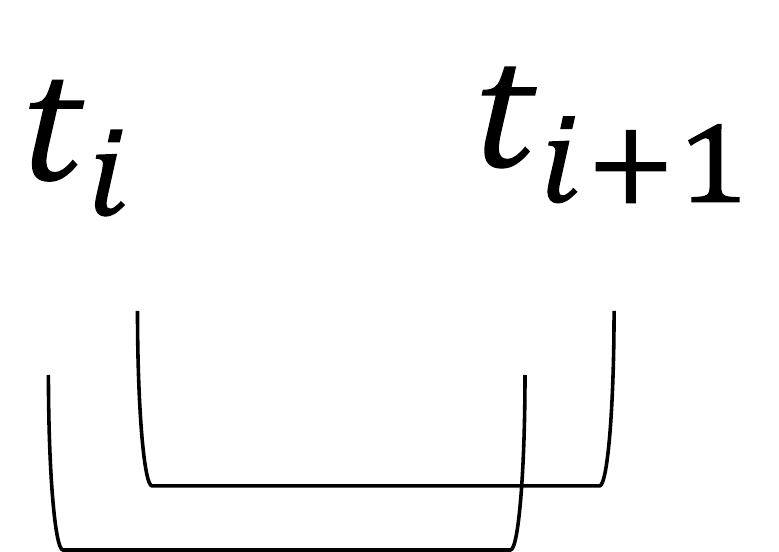}}}\quad =\quad \frac{1}{2}(& |B_I\rangle_i |B_I\rangle_{i+1}+|B_X\rangle_i |B_X\rangle_{i+1}\\& +|B_Y\rangle_i |B_Y\rangle_{i+1}+|B_Z\rangle_i |B_Z\rangle_{i+1}).
    \end{split}
\end{equation}

Therefore, neither the zero-particle perfect dephaser sector nor the one-particle sector contribute to the overall temporal entanglement of the system. The two-particle sector, $\mathcal{T}_2$, is the leading sector that contributes to the total temporal entanglement. Moreover, $\mathcal{T}_2$ can contribute an extensive amount of entanglement in the form of superposition of states. The eigenvalue that sets the decay scale for $S_T$ is therefore $\lambda_2$. Since $\lambda_2$ changes smoothly with $p$, the decaying part of $S_T$ also changes smoothly with $p$.

It is now worth comparing the structure of $\mathcal{T}_p$ with measurements to the structure of $\mathcal{T}$ in the measurement-free case. 
$\mathcal{T}$ for the purely unitary circuit consists of a $1\times 1$ block $\mathcal{T}^{(0)}$ corresponding to the steady state with eigenvalue $\lambda_0=1$ and Jordan blocks of various sizes, 
\begin{equation}
    \mathcal{T}=\mathcal{T}^{(0)}\oplus Q_2\oplus Q_3\oplus \ldots \oplus Q_{T-1}\,.
\end{equation}
Each Jordan block $Q_i$ has its corresponding eigenvalue zero since these are necessarily nilpotent. This is necessary to exactly reach the perfect dephaser limit after a finite number of update steps. The largest Jordan block $Q_{T-1}$ is of size $(T-1)\times (T-1)$, which vanishes after being raised to the $T$-th power, $(Q_{T-1})^T=0$, such that temporal entanglement decays linearly with bath size rather than exponentially and an exact steady state is reached at bath size $L=T$. 

This structure is in stark contrast to the structure of $\mathcal{T}_p$ for circuits with measurements. In $\mathcal{T}_p$, the Jordan blocks are no longer nilpotent. Instead of having all eigenvalues being zero, the Jordan blocks now have nontrivial diagonal elements. 
As such, the decay of temporal entanglement is again exponential in bath size. Any infinitesimal measurement rate immediately induces this structural change in the transfer matrix, since the appearance of nilpotent Jordan blocks generally requires fine-tuning.

\section{Peak Value of Temporal Entanglement and the Critical Bath Size}\label{sec:peak}

In between the growth and decay regimes, the TE reaches a peak value that sets the height of the temporal entanglement barrier.
In dual-unitary circuits without measurements, the peak value can be trivially computed as the intersection of the growth and the decay lines, yielding a peak value $S_{T,\text{peak}}=2T/3$, which occurs at the critical bath size $L_c=T/3$. In monitored good scrambler circuits, $S_{T,\text{peak}}$ and $L_c$ can be analogously computed, yielding $S_{T,\text{peak}}=2T/3$ and $L_c=T/(3v_{E})$.

\begin{figure}[tb!]
\includegraphics[width=0.95\columnwidth]{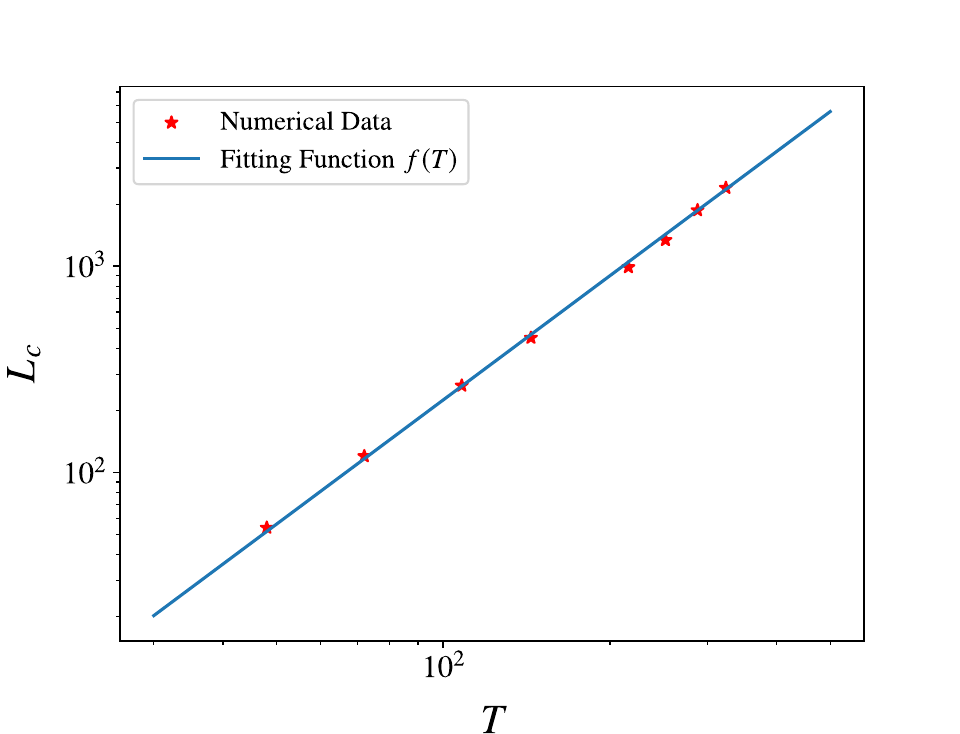}
\caption{Scaling of $L_c$ with $T$. The data is fitted with a one-parameter function of the form $f(T) \propto T^2$.}
\label{fig:Lc_scale_T}
\end{figure}

\begin{figure}[tb!]
\includegraphics[width=0.95\columnwidth]{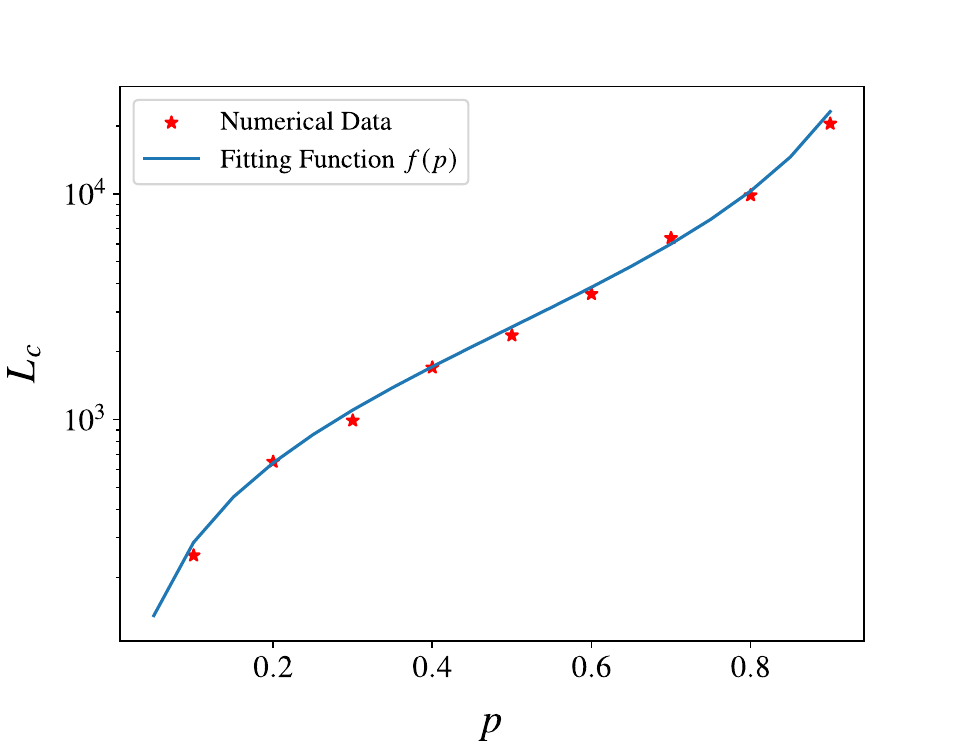}
\caption{Scaling of $L_c$ with $p$. The data is fitted with a one-parameter function of the form $f(p)\propto p/(1-p)$.}
\label{fig:Lc_scale_p}
\end{figure}

In monitored poor scrambler circuits, we observe numerically that $S_{T,\text{peak}}=T/3$. Furthermore, the critical bath size $L_c$ scales as:
\begin{equation}\label{eqn:num_L_c_scale}
    L_c (T,p) \propto \frac{T^2}{D(p)},
\end{equation}
as shown in Figs. \ref{fig:Lc_scale_T} and \ref{fig:Lc_scale_p}.


To understand the values of $S_{T,\text{peak}}=T/3$ and $L_c$, we examine the condition under which $L_c$ is reached. In both circuits with and without measurements, $L_c$ is reached when all the temporal Bell pairs initially in the interval $2T/3\le t <T$ have their left ends hitting the temporal boundary. This condition can be understood in the measurement-free setup as follows: for $L<L_c$, entanglement builds up in the interval $0<t<2T/3$ and decays in the interval $2T/3<t<T$. At $L=L_c$, entanglement in the interval $0<t<2T/3$ is ``saturated", with all $T/3$ Bell pairs forming a rainbow state:
\begin{equation}\label{eqn:rainbow}
    \begin{split}
        |\Psi_{\text{RB}}\rangle =\quad&\vcenter{\hbox{\includegraphics[width=0.6\linewidth]{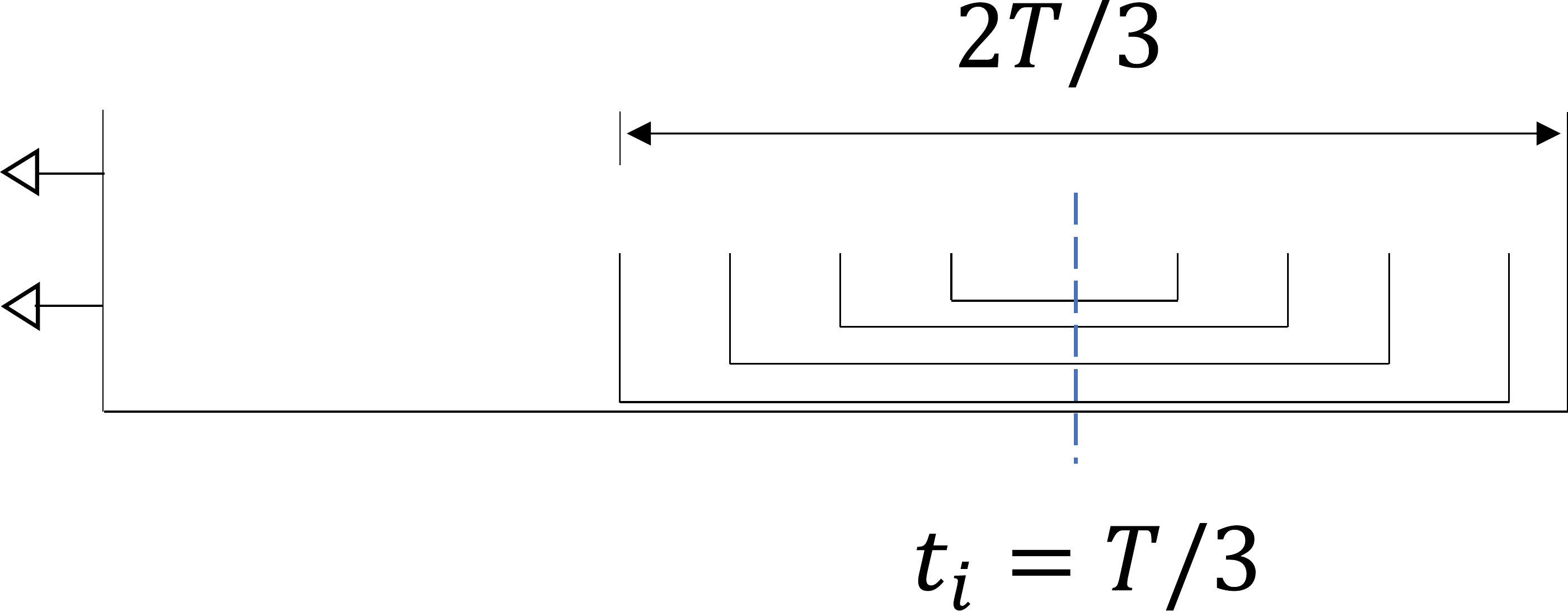}}}
    \end{split}
\end{equation}
The optimal bipartition is at $t_i=T/3$, and the peak entanglement is given by the number of Bell pairs crossing the bipartition on both the forward and backward time contours: $S_{T,\,\text{peak}}=2\cdot T/3=2T/3$. For $L>L_c$, ends of Bell pairs start reflecting at $t=0$, and the rainbow state is destroyed. The maximal entanglement thus starts decreasing. 

In monitored poor scrambler circuits the ends of Bell pairs were already argued to spread diffusively, resulting in
\begin{equation}
\begin{split}
    \sqrt{2D L_c}=\Delta t  = \frac{T}{3}\quad\Rightarrow L_c = \frac{(\Delta t)^2}{2D}\,,
    =\frac{T^2}{18D},
    \end{split}
\end{equation}
which matches the numerically observed scaling of Eq. \eqref{eqn:num_L_c_scale}. 
At $L=L_c$, entanglement in the interval $0<t<2T/3$ is saturated on average, with the only difference with the case without measurements being that the Bell pairs in this interval no longer form a rainbow state. 

Adding random measurements results in random distributions of Bell pairs, and the resulting different distribution the TE ranges from a minimal value of zero to a maximal value given by that of the rainbow state. The averaged value then reaches a steady state of approximately half the maximum value, independent of $p$, yielding a peak value of $S_{T,\,\text{peak}}\approx T/3$. 

The situation is different in monitored good scrambler circuits, since the distribution of TE due to random distribution of Bell pairs fluctuate only within a narrow window of the maximal value given by that of the rainbow state. Consequently, upon trajectory averaging, $S_{T,\text{peak}}$ remains approximately the rainbow-state value, $S_{T,\text{peak}}\approx 2T/3$. 
\section{Non-Hermitian Phase Transition in the Mixed Transfer Matrix Spectrum of Monitored Poor Scrambler Circuits}\label{sec:nonH}

In monitored poor scrambler circuits, while the exponential decay of the TE for large bath sizes is fully set by the leading eigenvalue of the transfer matrix, the intermediate dynamics generally requires knowledge of the full eigenspectrum. The leading eigenvalue was already observed to be real, resulting in purely exponential decay, but in general it is not guaranteed that the eigenvalues of the transfer matrix are real. In this Section we show that both the eigenspectrum and the eigenstates of the transfer matrix  -- except for the leading eigenvalue -- change qualitatively as the measurement rate is varied, indicating nonanalytic transitions in the ``dynamics" of the averaged TE as the measurement rate is varied.

The spatial transfer matrix is non-Hermitian and hence not guaranteed to be diagonalizable or have real eigenvalues. The eigenspectrum is however constrained because the transfer matrix generally possesses PT (parity-time) symmetry: it is invariant under the combined action of a unitary parity operator, here the exchange of the forward and backward time contour, and an anti-unitary time reversal operator, here complex conjugation. While this symmetry is readily apparent for our choice of gates, resulting in purely real transfer matrices that are invariant under complex conjugation, this symmetry holds more generally. 

Eigenvalues of PT-symmetric matrices are constrained to be either real of part of a complex conjugate pair. As pointed out in Refs.~\cite{bender_real_1998, bender_making_2007}, certain non-Hermitian matrices with PT symmetry possess spectra that are entirely real, in which case the system is said to be in a PT-symmetric phase. The spectrum of PT-symmetric Hamiltonian can change nonanalytically as some underlying parameter is tuned, and the PT-symmetry can be spontaneously broken when the spectrum changes from purely real to being a combination of complex-conjugate pairs of eigenvalues and real eigenvalues \cite{bender_real_1998}. The PT-broken phase is known to host a proliferation of exceptional points (EPs) at which the model is not diagonalizable but rather exhibits Jordan blocks~\cite{heiss_physics_2012, ashida_non_2020, bergholtz_exceptional_2021}. 

\begin{figure}[tb!]
    \centering
    \includegraphics[width=0.85\columnwidth]{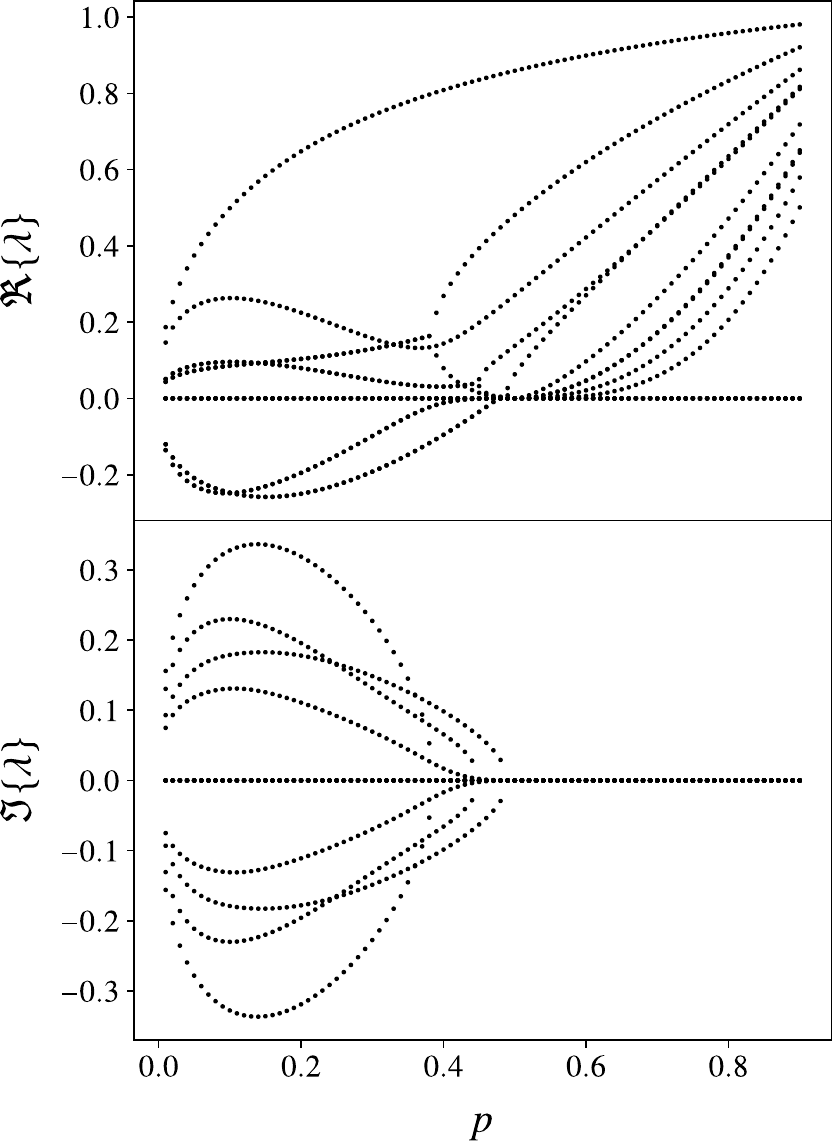}
    \caption{Real and imaginary parts of the spectra of $\mathcal{T}_p^{(2)}$ for total evolution time $T=6$ and varying measurement rates $p$.}
    \label{fig:ep}
\end{figure}

For the spatial transfer matrix~\eqref{eqn:spatial_transfer_matrix} such a transition is observed at a critical measurement rate $p_c = 1/2$.
For $1/2<p<1$ all eigenvalues are purely real, whereas for $0<p<1/2$ the PT symmetry is spontaneously broken in all sectors $\mathcal{T}_p^{(i)}$ with $i>1$ and the spectrum contains pairs of complex conjugate eigenvalues. This is illustrated in Fig.~\ref{fig:ep}, showing the spectra of the two-particle sector, $\mathcal{T}_p^{(2)}$, for total evolution time $T=6$.

\begin{figure}[h!]
 \centering
 \includegraphics[width=0.9\columnwidth]{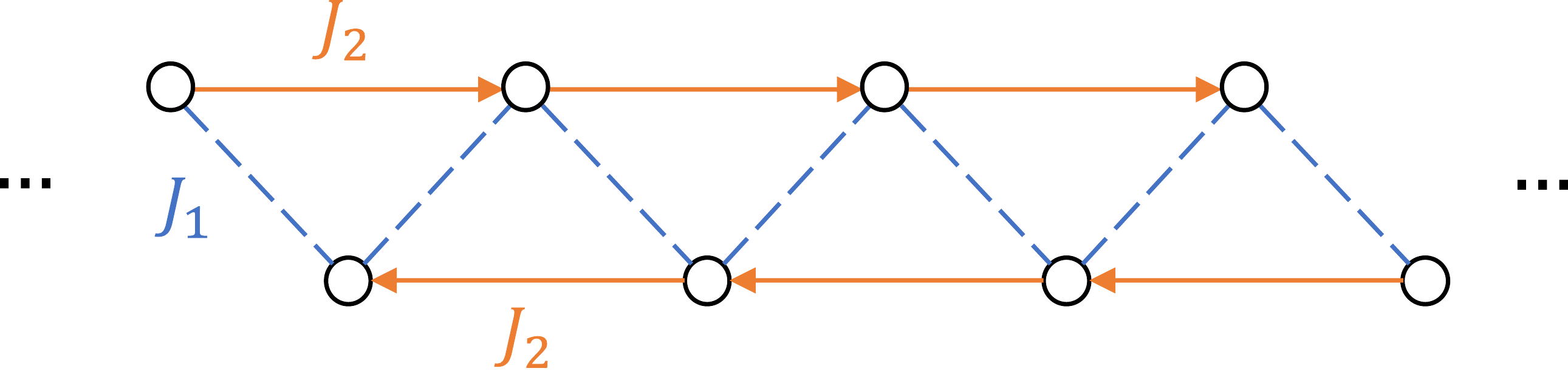}
 \caption{Hopping model for the spatial transfer matrix restricted to a single particle. The lattice is divided into odd and even sub-lattices to account for the even-odd dependence in hopping amplitudes. $J_1$ is the symmetric nearest-neighbor hopping amplitude; $J_2$ is the uni-directional next-nearest-neighbor hopping amplitude.}
 \label{fig:t-s_chain}
\end{figure}

In order to understand the critical measurement rate $p_c = 1/2$, it is instructive to rewrite the transfer matrix as a non-Hermitian Hamiltonian. We consider the simplest example of the one-particle sector, which has a natural choice of basis, e.g. for $T=6$:
\begin{equation}
    \begin{split}
        |1\rangle &\equiv |\bullet\circ\circ\circ\circ\circ\rangle\\
        |2\rangle &\equiv |\circ\bullet\circ\circ\circ\circ\rangle\\
        |3\rangle &\equiv |\circ\circ\bullet\circ\circ\circ\rangle\\
        |4\rangle &\equiv |\circ\circ\circ\bullet\circ\circ\rangle\\
        |5\rangle &\equiv |\circ\circ\circ\circ\bullet\circ\rangle\\
        |6\rangle &\equiv |\circ\circ\circ\circ\circ\bullet\rangle
    \end{split}
\end{equation}
Here $\ket{\circ}$ and $\ket{\bullet}$ correspond to the particle notation of Eq.~\eqref{eq:def_Bs}. The matrix elements of the spatial transfer matrix can be analytically obtained in closed form, resulting in a hopping Hamiltonian with symmetric nearest-neighbor hopping and uni-directional next-nearest neighbor hopping, with the hopping direction being different depending on whether the lattice site is odd or even. For a single particle the transfer matrix $\mathcal{T}_p^{(1)}$ can be written as a non-Hermitian Hamiltonian:
\begin{align}\label{eq:H_hopping}
  H =&\,  p^2 \sum_{j=2}^{T-1} |j\rangle \langle j | + J_1  \sum_{j=2}^{T-2}\left(|j\rangle\langle j-1|+|j \rangle \langle j+1|\right) \nonumber\\
  &+J_2 \sum_{\substack{j=2 \\ j\,\textrm{even}}}^{T-2} |j\rangle \langle j+2|+J_2 \sum_{\substack{j=3 \\ j\,\textrm{odd}}}^{T-3} |j\rangle \langle j-2| \nonumber\\
  &+ p |1\rangle \langle 1| + (1-p)|1\rangle \langle 2|\,,
\end{align}
with nearest-neighbor hopping amplitude $J_1 = p(1-p)$ and next-nearest-neighbor hopping amplitude $J_2 = (1-p)^2$. This model is illustrated in Fig. \ref{fig:t-s_chain}, where the lattice is divided into odd and even sub-lattices. Both the boundary terms and the next-to-nearest-neighbor interaction are explicitly non-Hermitian. In higher-particle sectors the corresponding Hamiltonian has the same hopping amplitudes, as well as an additional hard-core constraint on the particles.

This model can be explicitly solved in the one-particle sector, with the solution being representative of the physics in the higher-particle sector. It is instructive to first consider the model with periodic boundary conditions, i.e.
\begin{align}
  H_{\textrm{PBC}} =&\,  p^2 \sum_{j=1}^T |j\rangle \langle j | + J_1  \sum_{j=1}^T\left(|j\rangle\langle j-1|+|j \rangle \langle j+1|\right) \nonumber\\
  &+J_2 \sum_{j\,\textrm{even}} |j\rangle \langle j+2|+J_2 \sum_{j\,\textrm{odd}} |j\rangle \langle j-2|,
\end{align}
where we identify $j+T = j$. The periodic boundary conditions allow this model to be solved by going to Fourier space, writing an eigenstate $\ket{\psi}$ with components 
\begin{align}\label{eq:psi_k}
    \psi_j = 
    \begin{cases}
        \alpha_+ e^{ikj} + \alpha_- e^{-ikj} , \qquad &\textrm{for}\, j\,\textrm{even},\\
        \beta_+ e^{ikj} + \beta_- e^{-ikj} , \qquad &\textrm{for}\, j\,\textrm{odd}.
    \end{cases}
\end{align}
The coefficients $\alpha_{\pm}$ and $\beta_{\pm}$ can be obtained by solving the eigenvalue equation in Fourier space:
\begin{align}
 \begin{bmatrix} p^2+ J_2 e^{-2ik} & 2J_1 \cos{k}\\ 2J_1\cos{k} & p^2+ J_2 e^{2ik} \end{bmatrix} 
 \begin{bmatrix}
     \alpha_{\pm} \\
     \beta_{\pm}
 \end{bmatrix} = 
\lambda
 \begin{bmatrix}
     \alpha_{\pm} \\
     \beta_{\pm}
 \end{bmatrix},
\end{align}
which additionally returns the dispersion relation, giving a pair of eigenvalues $\lambda_{\pm}$ as a function of the momentum $k$:
\begin{equation}\label{eqn:eigvals}
    \lambda_\pm(k) =1-2J_1 -2J_2 \sin^2 k \pm 2 \cos{k}\sqrt{J_1^2-J_2^2\sin^2 k }\,.
\end{equation}
The periodic boundary conditions quantize $k = 2\pi n /T,n=0,1 \dots T-1$. 

The non-Hermitian Hamiltonian with periodic boundary conditions already highlights how the structure of the eigenspectrum is determined by the relative magnitudes of the (positive) hopping amplitudes $J_1$ and $J_2$. In the regime $J_1 > J_2$ and hence $p>1/2$, all eigenvalues \eqref{eqn:eigvals} are purely real for all $k$-modes. For $J_1 < J_2$, a transition from complex to real eigenvalues occurs for the $k$-mode satisfying $\sin{k}=J_1/J_2=p/(1-p)$. The first complex eigenvalues are possible at $J_1=J_2$ and hence $p=1/2$, with a proliferation of exceptional points occurring in the $J_1<J_2$ regime corresponding to $0<p<1/2$. In the former regime the Hermitian nearest-neighbor hopping is the dominant contribution in the Hamiltonian and the eigenvalues are purely real, whereas in the latter regime the unidirectional hopping dominates the dynamics and results in strongly non-Hermitian dynamics. However, while the eigenvalues exhibit a qualitative change as $p$ is varied, the eigenstates remain plane waves at any measurement rate.
\begin{figure}[h!]
 \centering
 \includegraphics[width=0.9\columnwidth]{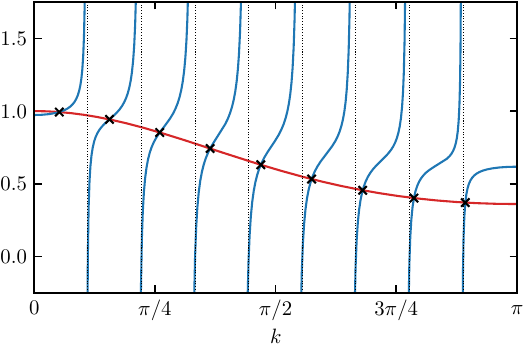}
 \caption{Graphical illustration of both sides of the quantization condition [Eq.~\eqref{eq:selfconsistent}] for the momentum with $\lambda(k) = \lambda_+(k)$ for total evolution time $T=10$ and measurement rate $p=0.8$.  Vertical dotted lines indicate the approximate poles  $k = n \pi/(T-1), n=1 \dots T-2$, and crosses indicate the solutions for the quantized momentum $k$ at the intersections between the left-hand side (red) and right-hand side (blue line).
 \label{fig:selfconsistent}}
\end{figure}

Taking into account the exact boundary conditions from Eq.~\eqref{eq:H_hopping}, the ansatz~\eqref{eq:psi_k} still returns the exact eigenstates, but the momentum $k$ now needs to be determined in a self-consistent way. As shown in Appendix \ref{app:selfconsistent}, for the boundary conditions in the spatial transfer matrix the eigenvalue $\lambda(k)$ needs to satisfy
\begin{align}\label{eq:selfconsistent}
    \lambda(k) =p \frac{\lambda(k) - p^2 + J_2}{\lambda(k)-p^2-J_2 \sin(k(T-3))/\sin(k(T-1))},
\end{align}
where $\lambda(k) = \lambda_{\pm}(k)$ [both choices of the sign lead to the same solutions since $\lambda_+(k) = \lambda_-(\pi-k)$]. This equation is graphically illustrated in Fig.~\ref{fig:selfconsistent} in the regime where $p>1/2$ and all eigenvalues are real.

The left-hand side is a smooth function of $k$, whereas the right-hand side exhibits a series of vertical asymptotes in between which this function is monotonically increasing for $0 < k < \pi$. The corresponding poles are located at the values of $k$ for which
\begin{align}
    \frac{\lambda(k)-p^2}{J_2} = \frac{\sin(k(T-3))}{\sin(k(T-1))}.
\end{align}
The locations of these poles can be approximately determined when $p\approx 1$. In this limit we have that $(\lambda(k)-p^2)/J_2 \gg 1, \forall k,$ such that the poles need to satisfy $\sin(k(T-1)) \approx 0$, leading to poles at the quantized momentum values $k = n \pi/(T-1), n=1 \dots T-2$. In between any pair of neighboring poles a solution to the self-consistent equation can be found, returning the expected $T-1$ nontrivial eigenvalues, with $\lambda=0$ the remaining trivial eigenvalue. The number of poles remains fixed for $p>1/2$ and in this way all eigenstates in the regime $p > 1/2$ can be obtained. We find that in this regime all eigenvalues are real and the eigenstates resemble the plane waves also observed for periodic boundary conditions. 
\begin{figure}[h!]
 \centering
 \includegraphics[width=0.85\columnwidth]{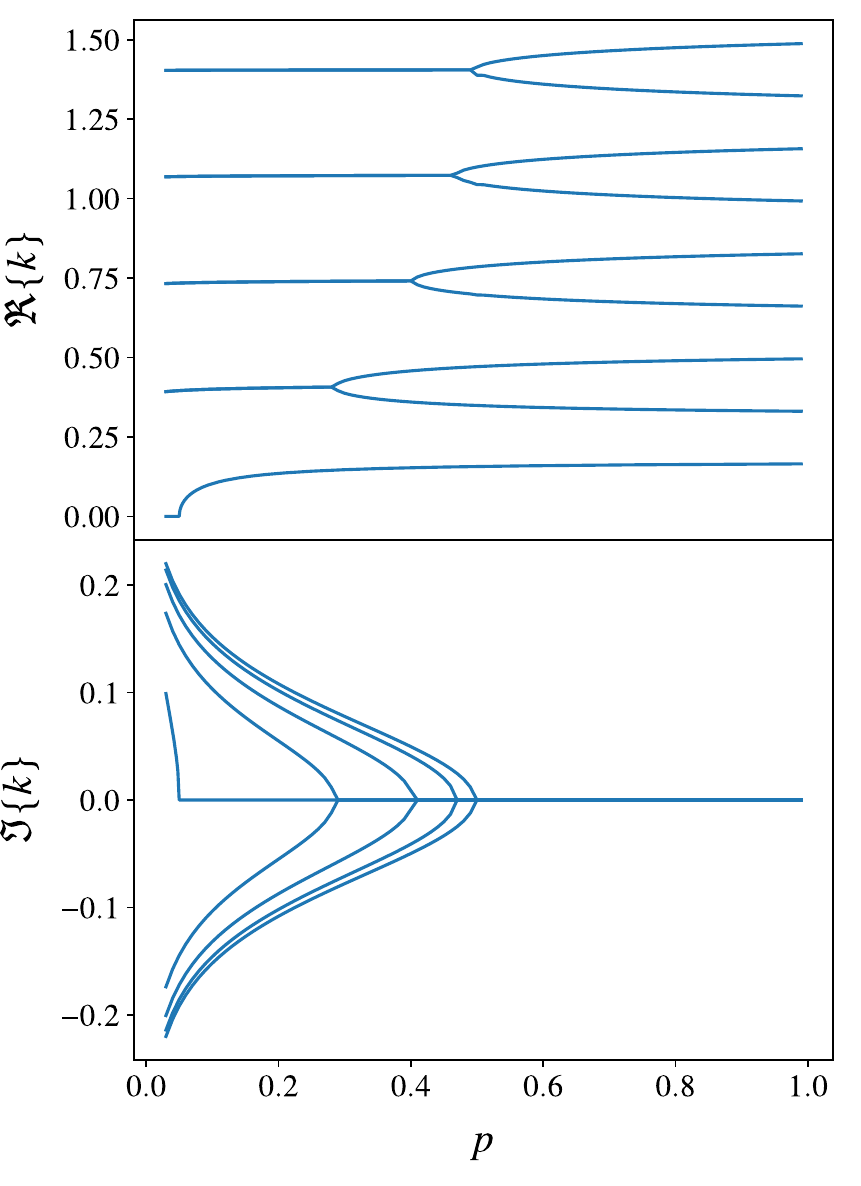}
 \caption{Quantized values of the momentum $k$ solving the quantization condition [Eq.~\eqref{eq:selfconsistent}] for total evolution time $T=10$ as function of the measurement rate $p$. For $p>1/2$ all momenta and the corresponding eigenvalues are real, whereas for $p<1/2$ a series of exceptional points occur at which two real momenta and corresponding eigenvalues coalesce and continue as a pair of complex conjugate eigenvalues. At $p=1/(2T)$ the value of $k$ associated with the leading eigenvalue changes from purely real to purely imaginary.
 \label{fig:kvals}}
\end{figure}

For $p < 1/2$ all eigenvalues can still be found as solutions to the self-consistent equation~\eqref{eq:selfconsistent}, but now both the eigenvalue and the corresponding momentum can be complex. The poles move into the complex plane as $p$ is decreased, requiring complex values of $k$ in order to satisfy Eq.~\eqref{eq:selfconsistent}. In Fig.~\ref{fig:kvals} we show the resulting values of the momentum as function of $p$ for a small system size of $T=10$. As opposed to the case with periodic boundary conditions, the corresponding eigenstates are no longer given by plane waves but rather by states localized at the (temporal) boundaries. Any nonzero imaginary part in $k$ results in an exponential decay in the eigenstates away from the boundaries $t=0$ and $t=T$, inducing localization at the boundaries reminiscent of the non-Hermitian skin effect~\cite{lu_spacetime_2021,yao_edge_2018,zhang_review_2022}. 

However, as can also be observed in Fig.~\ref{fig:spec1}, the leading eigenvalue is always real. The corresponding momentum $k$ changes from purely real to purely imaginary at $p=1/(2T)$, at which point the eigenstates again decay exponentially away from the temporal boundaries -- with a localization length that is however on the order of $T$, such that the wave function is still supported on the full system of $T$ sites. Exactly at $p=1/(2T)$ the eigenstates have a linear profile (see Appendix \ref{app:selfconsistent}). That the leading eigenvalue is real can be understood from the PT-symmetry: for an even number of eigenvalues it is not possible for all eigenvalues to be part of a complex conjugate pair since $\lambda=0$ is always a real eigenvalue, requiring an additional real eigenvalue in the spectrum.

This argument directly extends to the 2-particle sector governing the decay of temporal entanglement. While the above derivation focused on the single-particle sector, results in the two-particle sector are qualitatively similar, indicating that for large $T$ the interaction between the two particles can be treated perturbatively. 

The non-Hermitian transition observed in the spectra of $\mathcal{T}_p$ implies a corresponding nonanalyticity in the decay of physical observables and entanglement upon crossing any exceptional point. In particular, when going from the PT-symmetric to the PT-broken phase, oscillatory (spatial) dynamics is expected to show up due to the appearance of non-zero imaginary parts in the eigenvalues of the transfer matrix. However, this is a subleading effect: The leading eigenvalue in the 2-particle sector is always real and smooth for all values of $p$, and it is this leading eigenvector that determines the leading large-bath decay dynamics of physical observables and entanglement.

One potential way to observe such nonanalyticity at exceptional points is by choosing spatial boundary states that are orthogonal to the leading eigenstate in the 2-particle sector of the transfer matrix. Since such choice of boundary state does not show up naturally in our setup of interest, we defer detailed investigation along this direction to future works.

\section{Conclusion}\label{sec:conclude}
In this work, we characterized the shape of the temporal entanglement barrier in dual-unitary Clifford circuits with and without measurements. By leveraging the spatial unitarity of the circuit, we are able to efficiently simulate the evolution of the influence matrix with bath size and obtain the temporal entanglement profile in both space and time. 

In dual-unitary circuits without measurements, the observed linear growth and decay of TE are explained through exact tensor network contractions. In the presence of measurements, the precise shape of the temporal entanglement barrier depends on the scrambling properties of the individual gates. For ``good scramblers" the linear growth and decay regime are preserved, albeit with a reduced slope corresponding to an entanglement velocity $v_E < 1$ (to be contrasted with $v_E=1$ for measurement-free dual-unitary circuits). The functional dependence of this entanglement velocity on the measurement rate was obtained by mapping the stabilizer dynamics to a biased persistent random walk. 

For ``poor scramblers'', the linear growth underlied by ballistic spreading of temporal Bell pairs is modified to a diffusive growth. The functional dependence of the diffusion constant on the measurement rate is explained in terms of an unbiased random walk motion for the ends of temporal Bell pairs and stabilizer strings. This diffusion picture for temporal Bell pairs can also be used to predict the peak value of TE and the critical bath size at which this peak is reached. 
Rather than exactly reaching the perfect dephaser limit with vanishing TE at a finite bath size, for these poor scrambler circuits the decay of the TE to this limit becomes exponential in the presence of measurements. The corresponding characteristic decay scales are explained by constructing a mixed spatial transfer matrix, identifying a symmetry, and examining its eigenspectrum in different symmetry sectors. By tuning the measurement rate, the decay rate can be made arbitrarily slow, vanishing in the limit of purely measurement dynamics.

In all considered cases the temporal entanglement barrier scales linearly with the number of time steps, similar to volume law entanglement, such that there is no measurement-induced phase transition in the current setup. However, although no nonanalyticity shows up in the TE as the measurement rate is tuned, there is a PT phase transition in the eigenspectrum of the mixed spatial transfer matrix for the poor scrambler at $p=1/2$. This transition can be understood in a specific symmetry sector, where we find that both the eigenvalues and eigenstates exhibit a quantitative change, with the latter localizing at the temporal boundaries. It would be interesting to explore conditions for which the PT transitions manifest itself in the dynamics of TE, for instance by restricting to particular sets boundary states. Going beyond the current work, it would be worth investigating the interplay between measurements and temporal entanglement in more generic setups. The mapping of the spatial dynamics to a non-Hermitian hopping model furthermore suggests using space-time duality as a way of realizing non-Hermitian dynamics.
It would also be interesting to further investigate the relation between the entanglement velocity and the butterfly velocity in this space-time dual picture using e.g. entanglement membrane theory \cite{jonay_coarse_2018, nahum_dynamics_2018,zhou_entanglement_2020}.

\section*{Acknowledgements}
We are grateful for helpful discussions with Gerald Fux, Alessio Lerose, Yujie Liu, Sen Mu, Lorenzo Piroli, Michael A. Rampp, Yuan Wan, Heran Wang, Zhong Wang, Hongzheng Zhao, and Marko Znidaric. We thank Grace M. Sommers for helpful comments on the manuscript. 

\appendix

\section{Temporal Entanglement Profile of Generic Dual-Unitary Clifford Circuits without Measurements}
\label{app:generic_DU_class_4}

This section presents numerical results on the temporal entanglement profile for dual-unitary Clifford circuits where Eq.~\eqref{eqn:corner} does not hold. This situation can occur either because the gates fail to be both self-dual and real or because the circuit fails to be translationally invariant in space and time, or a combination of both. In this case the diagrams~\eqref{eq:diag_class4a} and \eqref{eq:diag_class4b} cannot be analytically evaluated. We will first consider circuits of random dual-unitary Clifford circuits that are inhomogeneous in both time and space, before consider homogeneous circuits that do not satisfy Eq.~\eqref{eqn:corner}.
\begin{figure}[h!]
    \centering
    \includegraphics[width=0.9\columnwidth]{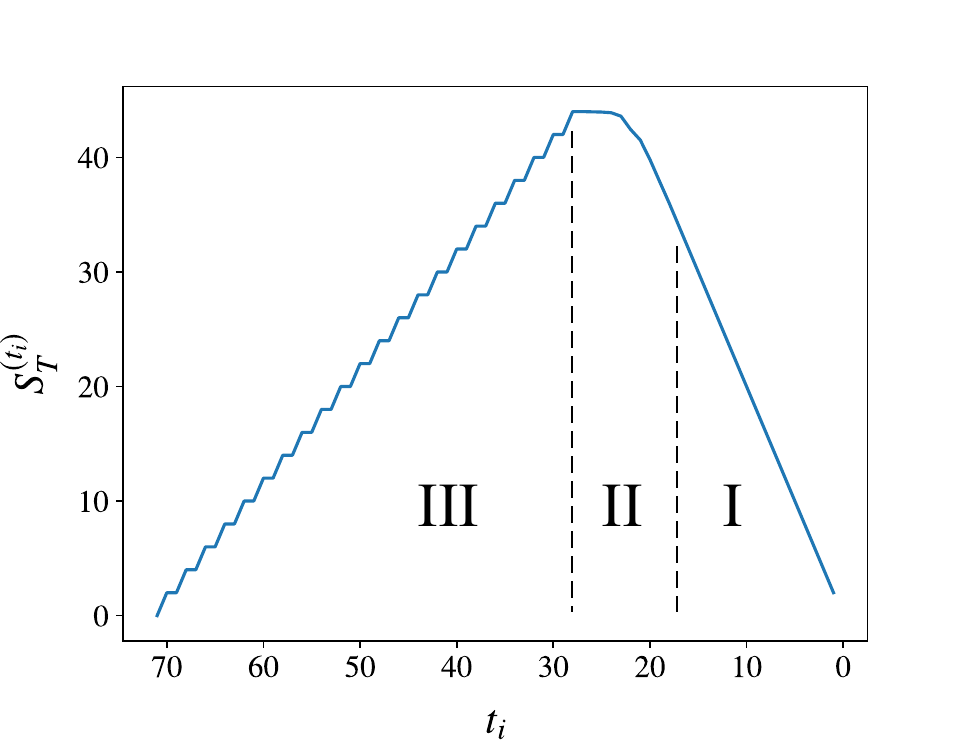}
    \caption{Temporal entanglement plotted against temporal bipartition point $t_i$, with total evolution time $T=72$ and bath size $L=28$. The circuit consists of random dual-unitary Clifford gates, and the results are averaged over $N=100$ realizations. There are 3 bipartition intervals, each showing different phenomenologies. Interval I is $0<t_i<T-2L$, Interval II is $T-2L\le t_i<L$, and Interval III is $L\le t_i<T$.}
    \label{fig:profile_T72L28rand100}
\end{figure}

Fig. \ref{fig:profile_T72L28rand100} shows $S_T^{(t_i)}$ as function of $t_i$ for $T=72$ and $L=28$ with random dual-unitary Clifford gates. The parameter choice corresponds to Regime 2.
While qualitatively similar to Fig. \ref{fig:profile_T72L28SDKI}, there is a quantitative difference: Instead of nonanalytically changing the slope to half of that in Interval I, $S_T^{(t_i)}$ maintains the same slope as in Interval I upon entering Interval II, before flattening out.
\begin{figure}[h!]
    \centering
    \includegraphics[width=0.9\columnwidth]{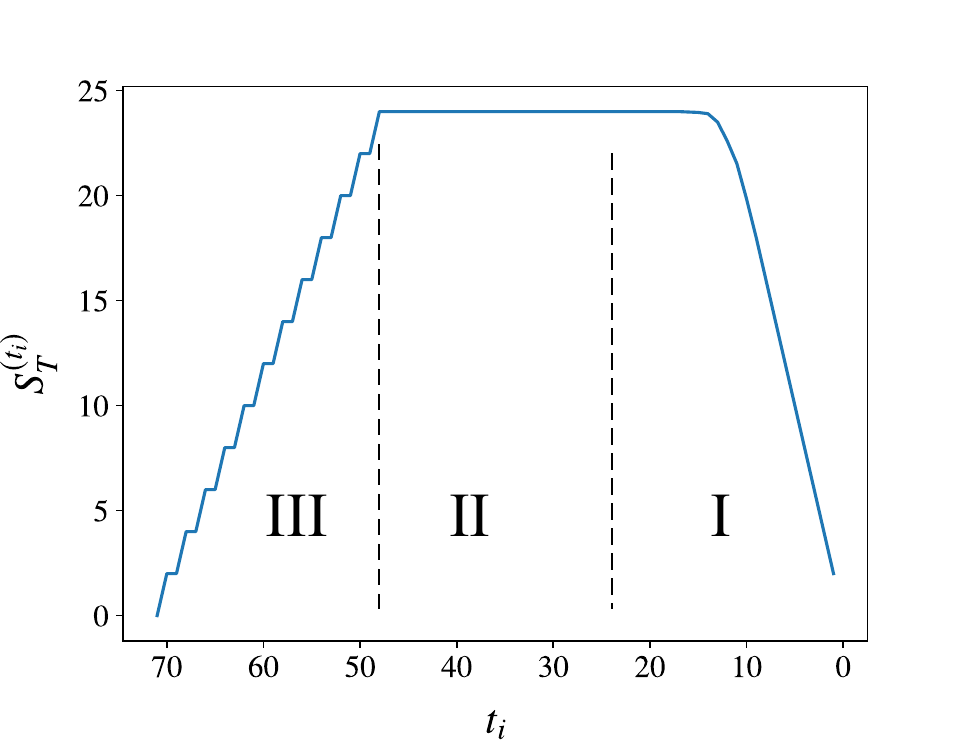}
    \caption{Temporal entanglement plotted against temporal bipartition point $t_i$, with total evolution time $T=72$ and bath size $L=48$. The circuit consists of random dual-unitary Clifford gates, and the results are averaged over $N=100$ trajectories. There are 3 bipartition intervals, each showing different phenomenologies. Interval I is $0<t_i<2L-T$, Interval II is $2L-T\le t_i<L$, and Interval III is $L\le t_i<T$.}
    \label{fig:profile_T72L48rand100}
\end{figure}

Fig. \ref{fig:profile_T72L48rand100} shows $S_T^{(t_i)}$ as function of $t_i$ for $T=72$, $L=48$ with random dual-unitary Clifford gates. The parameter choice corresponds to Regime 3. The contracted diagram correspond to Eq.~\eqref{eq:diag_class4b} in Interval I and Eq.~\eqref{eq:diag_class4a} in Interval II, respectively. Contrasting with Fig. \ref{fig:profile_T72L48SDKI}, $S_T^{(t_i)}$ behaves differently in these intervals: In Interval I, instead of having always $S_T^{(t_i)}=0$, $S_T^{(t_i)}$ grows linearly with $t_i$ and saturates to the peak value somewhere in the middle of Interval I. In Interval II, $S_T^{(t_i)}$ remains constant at the peak value instead  of growing linearly with $t_i$.  

\begin{figure}[h!]
    \centering
    \includegraphics[width=0.9\columnwidth]{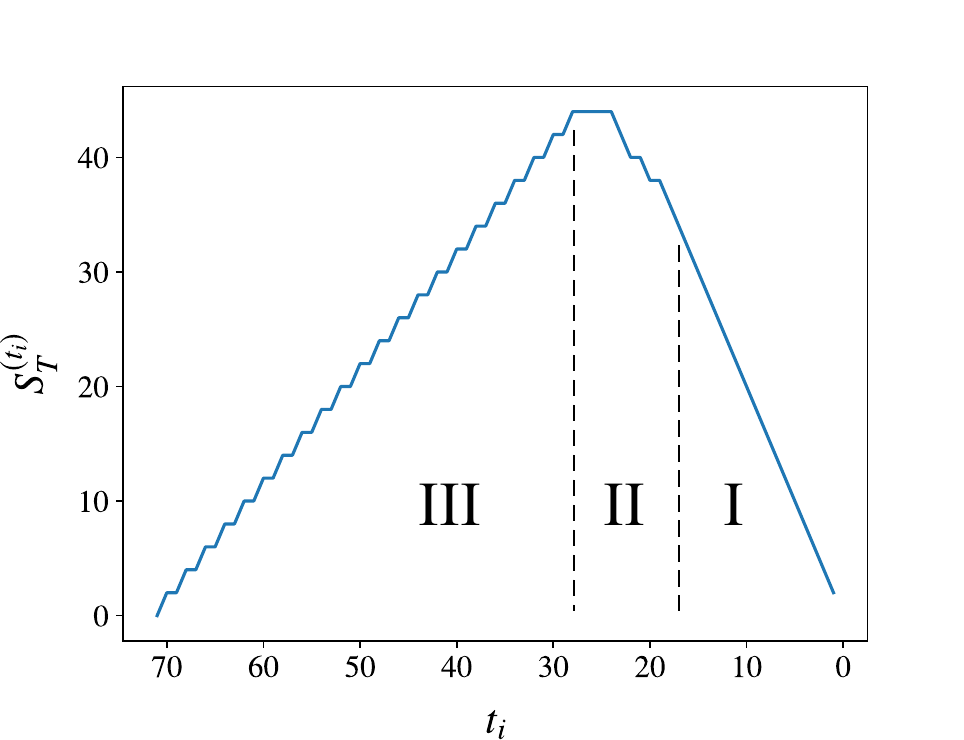}
    \caption{Temporal entanglement plotted against temporal bipartition point $t_i$, with total evolution time $T=72$ and bath size $L=28$. The circuit consists of only SDKI-H gates of the form~\eqref{eq:SDKI_LH}. There are 3 bipartition intervals, each showing different phenomenologies. Interval I is $0<t_i<T-2L$, Interval II is $T-2L\le t_i<L$, and Interval III is $L\le t_i<T$.}
    \label{fig:profile_T72L28SDKIleftH}
\end{figure}

We next consider the following gate obtained by applying Hadamard gates to the left input and out legs of the SDKI-f gate:
\begin{equation}\label{eq:SDKI_LH}
   \text{SDKI-H}\equiv(H\otimes I)\,\text{SDKI-f}\,(H\otimes I)
\end{equation}
This gate does not satisfy the condition given in Eq. \eqref{eqn:corner}.
Fig. \ref{fig:profile_T72L28SDKIleftH} shows the resulting $S_T^{(t_i)}$ as function of $t_i$ for $T=72$, $L=28$. The parameter choice corresponds to Regime 2. In Interval II, the contracted diagram is of the form~\eqref{eq:diag_class4a}. Contrasting with Fig. \ref{fig:profile_T72L28SDKI}, the temporal entanglement now behaves similarly as in the case of inhomogeneous dual-unitary Clifford gates.  

\begin{figure}[h!]
    \centering
    \includegraphics[width=0.9\columnwidth]{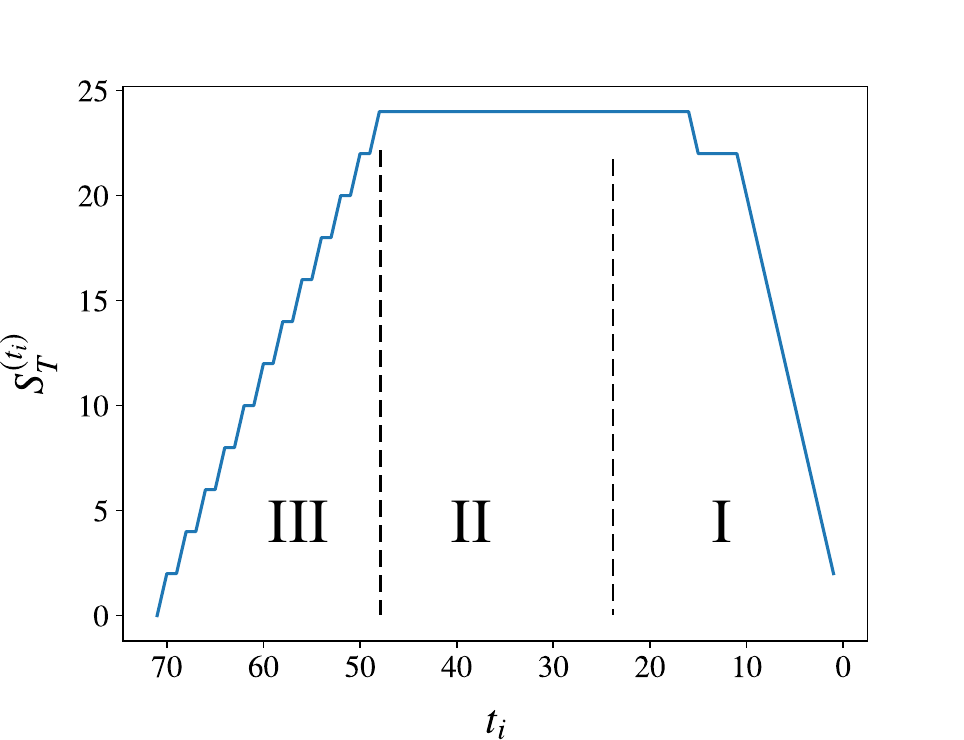}
    \caption{Temporal entanglement plotted against temporal bipartition point $t_i$, with total evolution time $T=72$ and bath size $L=48$. The circuit consists of only SDKI-H gates. There are 3 bipartition intervals, each showing different phenomenologies. Interval I is $0<t_i<2L-T$, Interval II is $2L-T\le t_i<L$, and Interval III is $L\le t_i<T$.}
    \label{fig:profile_T72L48SDKIleftH}
\end{figure}

Fig.~\ref{fig:profile_T72L48SDKIleftH} shows $S_T^{(t_i)}$ as function of $t_i$ for $T=72$, $L=48$ with gates of the form~\eqref{eq:SDKI_LH} in Regime III. The contracted diagram is of the form Eq.~\eqref{eq:diag_class4b} and Eq.~\eqref{eq:diag_class4a} in Interval I and II, respectively. We again observe the same behavior as for inhomogeneous dual-unitary Clifford gates.
Overall, failure to fulfill Eq.~\eqref{eqn:corner} leads to a drastic change in the behavior of $S_T^{(t_i)}$ in intervals where the simplified diagram is of the form Eq.~\eqref{eq:diag_class4a} or Eq.~\eqref{eq:diag_class4b}.
Nonetheless, the maximal value and the corresponding TE remains the same regardless of whether Eq. \eqref{eqn:corner} is fulfilled or not. Therefore, the temporal entanglement profile given by Eq. \eqref{eqn:num_res_no-meas} remains valid for all dual-unitary Clifford circuits and is insensitive to the condition of Eq. \eqref{eqn:corner}.


\section{Numerically Exact Results on Temporal Entanglement with Measurements: Tensor Contraction of the Mixed Transfer Matrix}\label{app:num_exact}

This section details the procedure for computing $S_T$ through tensor contraction of $\mathcal{T}_p$. The procedure does not involve averaging over stochastic trajectories and thus avoids stochastic noise. 
It is convenient to temporarily fix the normalization factor $\tr\left( (\rho^{\langle I_{\text{left}}|})^2  \right)=1$. Although the purity $D_{t_i}^{\langle \text{left}|}$ is nonlinear in $\left|\rho^{\langle \text{left}|}\right\rangle$, it is linear in $\left|\rho^{\langle \text{left}|}\right\rangle^{\otimes 4}$. Graphically, one may define:
\begin{equation}
    \begin{split}
        \vcenter{\hbox{\includegraphics[width=0.12\linewidth]{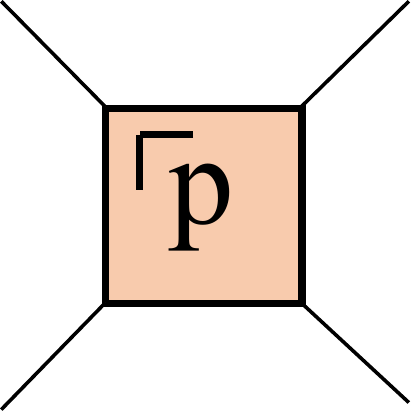}}}\; \equiv \; (1-p)\cdot \vcenter{\hbox{\includegraphics[width=0.14\linewidth]{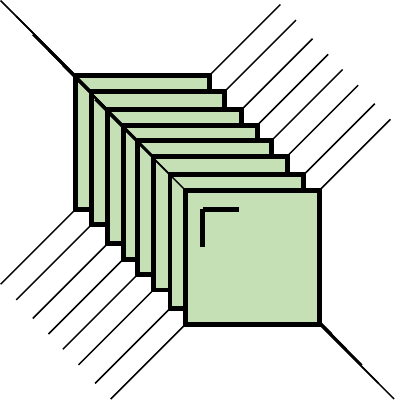}}}\; +\; p\cdot \vcenter{\hbox{\includegraphics[width=0.14\linewidth]{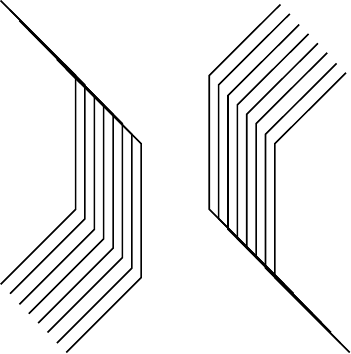}}}
    \end{split}
\end{equation}
The resulting tensor network is given by, e.g. for $T=8$, $L=4$, 
\begin{equation}
    \begin{split}
        \left|\rho^{\langle \text{left}|}\right\rangle^{\otimes 4}=\quad\vcenter{\hbox{\includegraphics[width=0.7\linewidth]{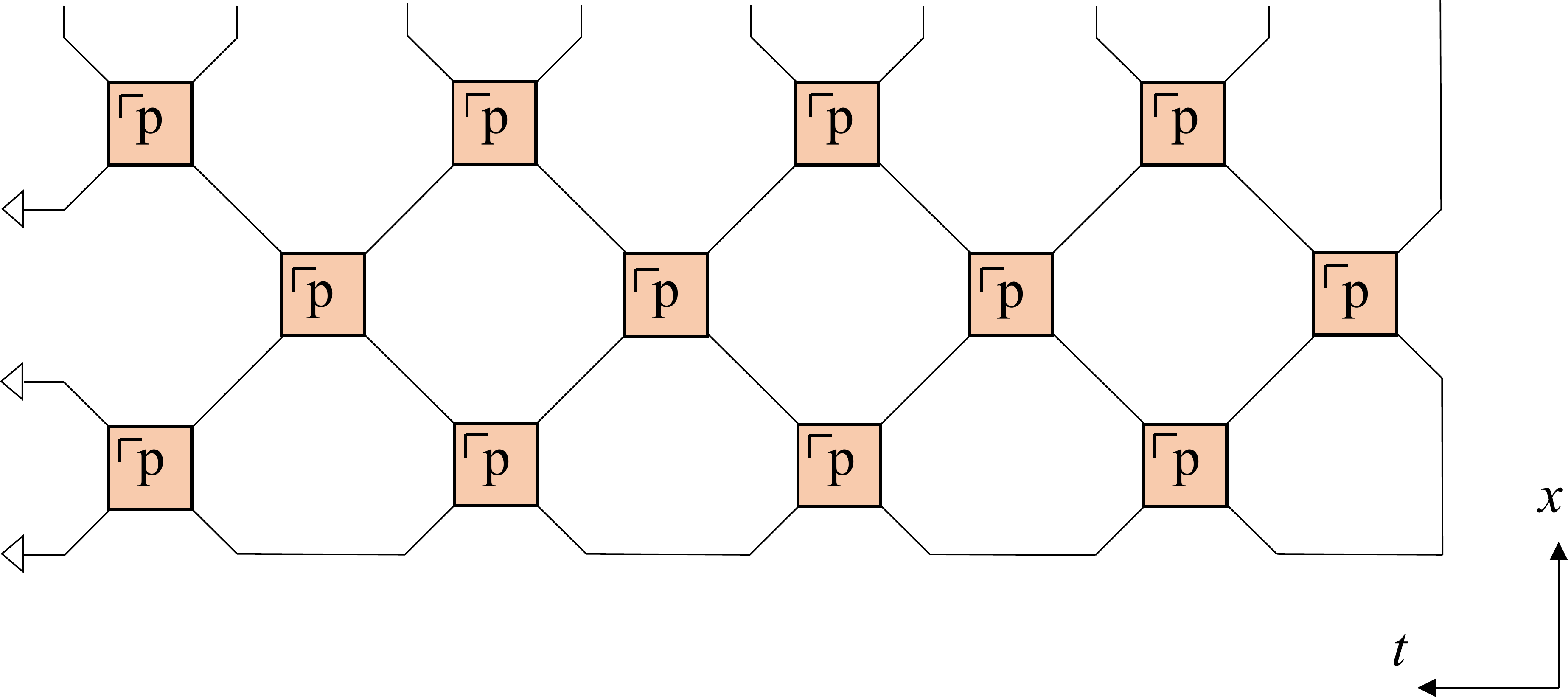}}}
    \end{split}
\end{equation}

Consider for example the contraction order with bipartition at $t_i=4$, in which case the purity follows as the contraction of the above circuit with
\begin{equation}
    \begin{split}
        \langle \text{CO}_{4,8}|=\;\vcenter{\hbox{\includegraphics[width=0.3\linewidth]{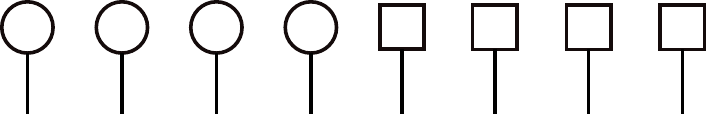}}}
    \end{split}
\end{equation}
as
\begin{align}
    &D_{t_i=4}^{\langle \text{left}|}=\langle \text{CO}_{4,8}|\left(\left|\rho^{\langle \text{left}|}\right\rangle^{\otimes 4}\right)\quad \nonumber\\
    &= \quad \vcenter{\hbox{\includegraphics[width=0.7\linewidth]{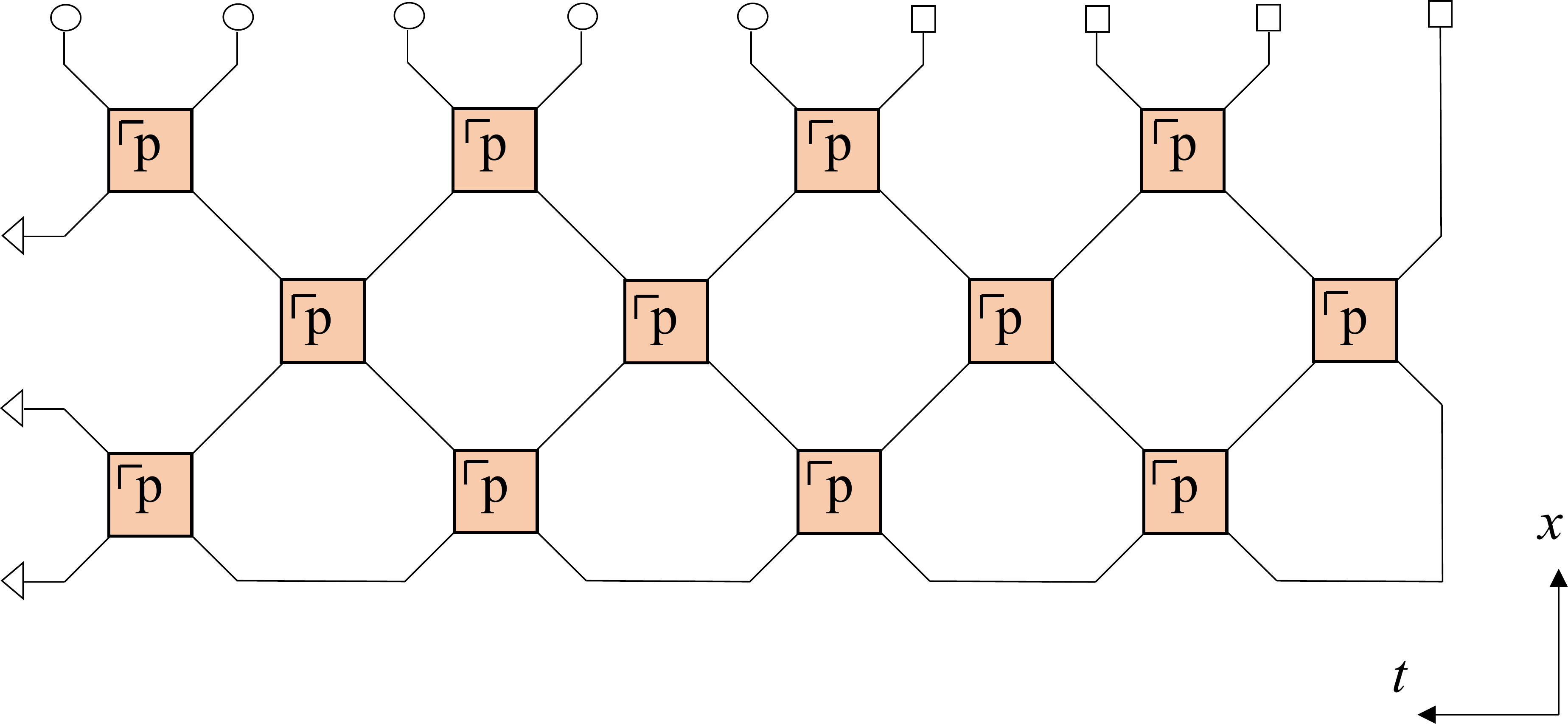}}}
\end{align}

For the all-SWAP circuit with measurements, $\mathcal{T}_p$ merely moves the initial temporal Bell pairs around. Analogous to the measurement-free case, for each trajectory realization, the purity is determined by the number of circle-square pairs:
\begin{equation}
        \left(D_{t_i}^{\langle \text{left}|}\right)_j=\left(\frac{1}{4}\right)^{(n_{\text{cs}})_j}\;,
\end{equation}
with  $(n_{\text{cs}})_j$ the number of contractions $\vcenter{\hbox{\includegraphics[width=0.06\linewidth]{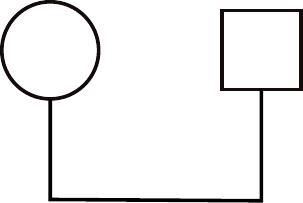}}}$ in trajectory $j$. One may deduce the inner product values between different boundary states as follows:
\begin{equation}\label{eqn:inn_prod_annealed}
    \begin{split}
        &\bra{\circ}(\ket{\square})^* = \frac{1}{4}\;,\quad \langle \circ \ket{\triangle}=\langle \square\ket{\triangle}=1,\\
        & \bra{\circ}(\ket{\circ})^*=\bra{\square}(\ket{\square})^*=\langle\triangle\ket{\triangle}=1
    \end{split}
\end{equation}

With the definition of the inner products in Eq. \eqref{eqn:inn_prod_annealed}, one may construct explicit vector representations for each square, circle and triangle boundary state in a reduced Hilbert space. One needs 3-component vectors to satisfy all the numerical constraints, leading to a minimal local Hilbert space dimension of $q=3$. The three basis states are denoted as $\ket{\uparrow}$, $\ket{0}$, and $\ket{\downarrow}$. One may also construct explicit vector representations for the Bell-pair state, the SWAP gate and the identity gate. The chosen vector representations are as follows:
\begin{equation}\label{eqn:vector_rep_annealed}
    \begin{split}
        \bra{\circ}&=\bra{\uparrow}-\sqrt{\frac{3}{8}}\bra{0}+i \sqrt{\frac{3}{8}}\bra{\downarrow},\;\\
        \bra{\square}&=\bra{\uparrow}+\sqrt{\frac{3}{8}}\bra{0}+i \sqrt{\frac{3}{8}}\bra{\downarrow},\;\\
        \ket{\triangle}&=\ket{\uparrow}, \qquad
    \ket{\sqcup}=\ket{\uparrow\uparrow}+\ket{00}+\ket{\downarrow\downarrow}.
    \end{split}
\end{equation}
where $\ket{\sqcup}$ denotes the Bell-pair state.


As for the gates, the identity gate is trivially defined in the $q=3$ representation, and the SWAP gate is now defined as a $3\times3\times3\times3$ tensor:
\begin{equation}
    \vcenter{\hbox{\includegraphics[width=0.1\linewidth]{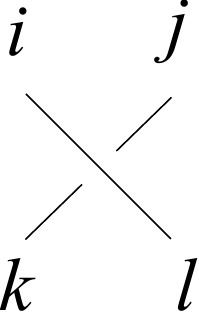}}}=(\text{SWAP})_{ijkl}=\delta_{il} \delta_{jk}
\end{equation}

It is worth remarking on several aspects of the choice of the numerical representation. Firstly, due to the existence of the projector from spacetime rotation of the trace operation, the norm of the state can decrease upon successive action by the projector. Without re-normalizing the state, factors coming from the norm of the state are multiplied with the purity, and this can result in erroneous calculation of the purity. In the stabilizer formalism, the state is always re-normalized after each measurement, and such problem does not occur. 

Therefore, for exact computation using the transfer matrix approach, the norm of the state is ignored. This is reflected in two key aspects: 1) the initial state is defined to be product of Bell pairs, where each Bell pair has norm 3; 2) the only relevant numerical values are the inner products defined in Eq. \eqref{eqn:inn_prod_annealed}, while the value of a ``loop" is not explicitly defined, since it does not appear in any contraction diagrams. This way, each trajectory is only weighted by the probability of the gate configuration and not by the norm of the state.  

The second remark is that the components of the circle and the square covectors are necessarily complex. This can be intuitively seen as follows: if only real entries are used, the numerical conditions specified in Eq. \eqref{eqn:inn_prod_annealed} translate to having three normalized vectors with circle aligned with triangle, and square also aligned with triangle, but circle not aligned with square, which is impossible. Complex entries, on the other hand, relax the normalization constraint, such that circle and square are no longer normalized to $1$, and the conditions specified in Eq. \eqref{eqn:inn_prod_annealed} can then be satisfied. 

The above prescription yields the average purity, which in turn yields the ``annealed" average entanglement. The annealed average is defined as first averaging over the purity then taking the logarithm, whereas the quenched average is defined as taking the average of the entanglement entropy itself:
\begin{equation}
    S^{\text{anneal}}_A\equiv -\log_2 \tr{\overline{\rho^2_A}},\quad S^{\text{quench}}_A\equiv -\overline{\log_2 \tr{\rho^2_A}}
\end{equation}

It is possible to compute the quenched average from the mixed transfer matrix by introducing a perturbative parameter $\epsilon$ into the vector representation of different boundary states. Redefining the overlaps such that:
\begin{align}
    \vcenter{\hbox{\includegraphics[width=0.08\linewidth]{circle-square-pair.pdf}}} \rightarrow 1-\epsilon \log_2 \vcenter{\hbox{\includegraphics[width=0.08\linewidth]{circle-square-pair.pdf}}},
\end{align}
we have that the calculation for the purity results in
\begin{align}
    D^{\langle \text{left}|}_{t_i}=\left(\;\vcenter{\hbox{\includegraphics[width=0.08\linewidth]{circle-square-pair.pdf}}}\;\right)^{n_{\text{cs}}}&\rightarrow \left(1-\epsilon \log_2 \;\vcenter{\hbox{\includegraphics[width=0.08\linewidth]{circle-square-pair.pdf}}}\;\right)^{n_{\text{cs}}} \nonumber\\
    &=1-\epsilon n_{\text{cs}}\log_2 \; \vcenter{\hbox{\includegraphics[width=0.08\linewidth]{circle-square-pair.pdf}}}\;+O(\epsilon^2).
\end{align}
Crucially, the entanglement entropy for any given trajectory is proportional to $n_{\text{cs}}$, which can be directly obtained from the above overlap.

Calculating the averaged overlap, which can be done by again absorbing the averaging into the gates, hence returns the averaged entanglement entropy.
The numerical values of the inner products are:
\begin{equation}\label{eqn:inn_prod_quenched}
    \begin{split}
        &\bra{\circ}(\ket{\square})^* = 1+2\epsilon\;,\quad 
        \langle\circ\ket{\triangle}=\langle\square\ket{\triangle}=1\;,\\
        & \bra{\circ}(\ket{\circ})^*=\bra{\square}(\ket{\square})^*=\langle\triangle\ket{\triangle}=1\,,
    \end{split}
\end{equation}
with chosen vector representations:
\begin{equation}\label{eqn:vector_rep_quenched}
    \begin{split}
        \bra{\circ}&=\bra{\uparrow}+\epsilon\bra{0}+i \epsilon\bra{\downarrow}, \quad
        \bra{\square}=\bra{\uparrow}+\bra{0}-i\bra{\downarrow},\\
        \ket{\sqcup}&=\ket{\uparrow\uparrow}+\ket{00}+\ket{\downarrow\downarrow}, \quad
        \ket{\triangle}=\ket{\uparrow}.
    \end{split}
\end{equation}
Eqs. \eqref{eqn:inn_prod_quenched} and \eqref{eqn:vector_rep_quenched} are the direct counterparts to Eqs. \eqref{eqn:inn_prod_annealed} and \eqref{eqn:vector_rep_annealed}. The representation of the SWAP and identity gates remain unchanged in the quenched representation. Therefore, properties of $\mathcal{T}_p$ directly affects properties of $S_T$. The decay scales extracted from the spectrum of $\mathcal{T}_p$ match the decay scales of $S_T$ from stabilizer simulations, as should be the case.

\section{Alternative averaging procedures for computing decay scale of temporal entanglement from exact tensor contraction of the mixed transfer matrix}\label{app:alternate_avg}

Besides the choice of taking either the annealed or the quenched average, one may also choose to average either before or after maximizing $S_T^{(t_i)}$ over different $t_i$, henceforth dubbed ``average-first (AF)" and ``extreme-value-first (EF)", respectively. This order is nontrivial because taking the extreme value is a nonlinear operation and in general does not commute with trajectory averaging.

\begin{figure}
        \centering
        \includegraphics[width=0.9\columnwidth]{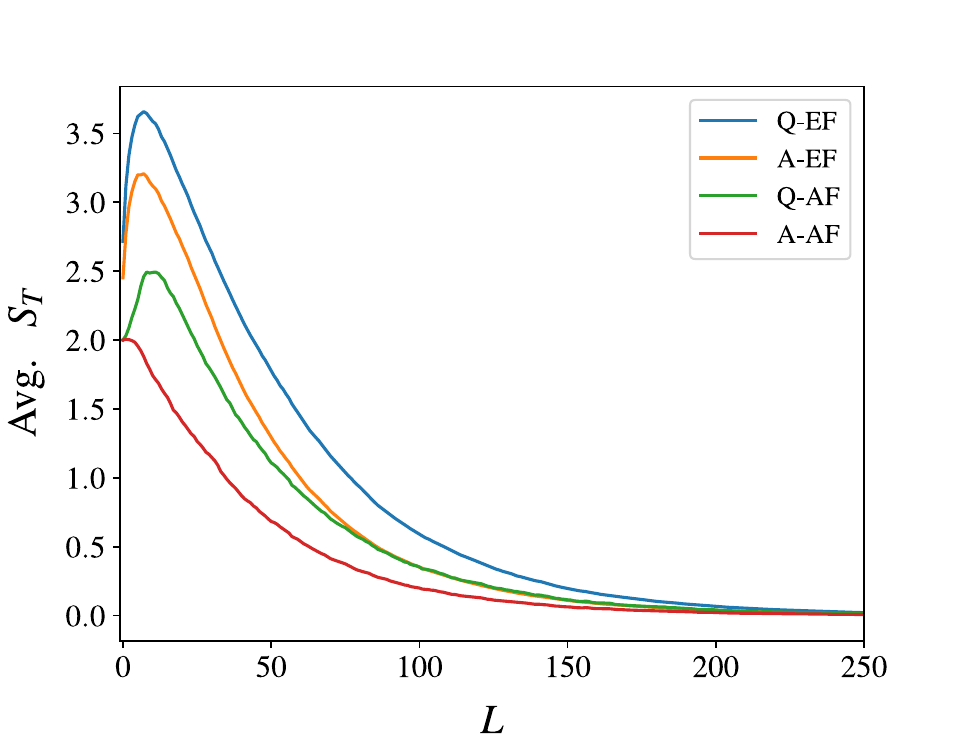}
        \caption{Four types of averages from stabilizer simulations of the all-SWAP circuit with measurements, for $T=8$, $p=0.8$. The first letter being either Q or A indicates whether the average is quenched or annealed, respectively. The last two letters being either EF or AF indicates whether the extreme value is computed first or the average is taken first, respectively.}
        \label{fig:T8p08_comp}
\end{figure}

Fig. \ref{fig:T8p08_comp} shows the average $S_T$ obtained from stabilizer simulation of the all-SWAP circuit with measurements, at measurement rate $p=0.8$, with four different ways of averaging. The labeling convention is detailed in the captions. By construction, fixing either the annealed (A) or the quenched (Q) average, the extreme-value-first (EF) curve always has higher value than the average-first (AF) curve. The results from exact $\mathcal{T}_p$ evolution yields either the annealed average-first (A-AF) or the quenched average-first (Q-AF) $S_T$. Although the precise $S_T$ values are different for the four types of averages, the resulting decay scale is the same regardless of which average is taken. 

\section{Temporal Entanglement Profile as Bath Size Increases}\label{app:profile}
\begin{figure}[h!]
 \centering
 \includegraphics[width=\columnwidth]{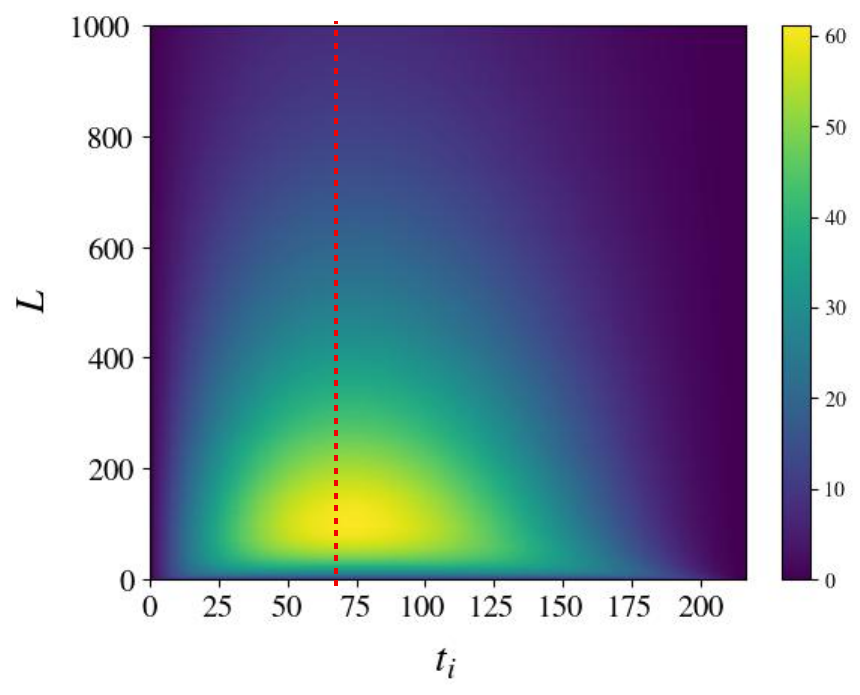}
 \caption{Average temporal entanglement profile in space and time of the monitored all-SDKI-f circuit, for total evolution time $T=216$ and measurement rate $p=0.3$. The numerical label on the color bar indicates value of the averaged temporal entanglement. The red dashed line marks the temporal cut that yields maximal average TE, which occurs on average at $t_i=T/3$, just like in the measurement-free case. }
 \label{fig:temporal_profile}
\end{figure}

Fig. \ref{fig:temporal_profile} shows the average $S_T^{(t_i)}$ profile for monitored ``poor scrambler" (all-SDKI-f) circuits as a 2D plot in both space and time with $T=216$ and $p=0.3$. On average, the $t_i$ that yields maximal $S_T^{(t_i)}$ occurs at $t_i=T/3$, consistent with the arguments given in the main text. 

\section{Derivation of the self-consistent equation}\label{app:selfconsistent}
In this Appendix we explicitly derive the self-consistent equation~\eqref{eq:selfconsistent} determining the eigenvalues and momenta of the spatial transfer matrix in the single-particle sector~\eqref{eq:H_hopping}. Consider the parametrization of the eigenstate $\ket{\psi}$ as 
\begin{align}\label{eq:psi_k_app}
    \psi_j = 
    \begin{cases}
        \alpha_+ e^{ikj} + \alpha_- e^{-ikj} , \qquad &\textrm{for}\, j\,\textrm{even},\\
        \beta_+ e^{ikj} + \beta_- e^{-ikj} , \qquad &\textrm{for}\, j\,\textrm{odd}.
    \end{cases}
\end{align}
where we take $j=0,1 \dots T-1$. For $j=1, \dots, T-2$ the corresponding components of the eigenvalue equation are satisfied provided
\begin{align}
 \begin{bmatrix} p^2 + J_2 e^{-2ik} & 2J_1 \cos{k}\\ 2J_1\cos{k} & p^2+J_2 e^{2ik} \end{bmatrix} 
 \begin{bmatrix}
     \alpha_{+} \\
     \beta_{+}
 \end{bmatrix} = 
\lambda
 \begin{bmatrix}
     \alpha_{+} \\
     \beta_{+}
 \end{bmatrix},
\end{align}
and similarly
\begin{align}
 \begin{bmatrix} p^2 + J_2 e^{2ik} & 2J_1 \cos{k}\\ 2J_1\cos{k} & p^2+ J_2 e^{-2ik} \end{bmatrix} 
 \begin{bmatrix}
     \alpha_{-} \\
     \beta_{-}
 \end{bmatrix} = 
\lambda
 \begin{bmatrix}
     \alpha_{-} \\
     \beta_{-}
 \end{bmatrix},
\end{align}
returning the eigenvalue $\lambda(k)$~\eqref{eqn:eigvals}. For $j=T-1$, taking $T$ even for convience, the boundary condition reads
\begin{align}
    \beta_+ e^{ik(T-1)}+\beta_- e^{-ik(T-1)} = 0,
\end{align}
whereas for $j=0$ we have that
\begin{align}\label{eq:eigenvec_0}
    p (\alpha_++\alpha_-)+(1-p)(\beta_+e^{ik }+\beta_-e^{-ik}) = \lambda(k)(\alpha_++\alpha_-).
\end{align}
Without loss of generality we can fix $\beta_+=e^{-ik(T-1)}$, such that the boundary condition at $j=T$ returns $\beta_+ = -e^{ik(T-1)}$. The bulk eigenvalue equation fixes $\alpha_{\pm}$ in terms of $\beta_{\pm}$, leading to
\begin{align}
    \alpha_+ &= - e^{-ik(T-1)} \frac{p^2-\lambda(k) + J_2 e^{2ik}}{2J_1 \cos(k)}, \nonumber\\
    \alpha_- &= e^{ik(T-1)} \frac{p^2-\lambda(k) + J_2 e^{-2ik}}{2J_1 \cos(k)}.
\end{align}
Rewriting the boundary condition~\eqref{eq:eigenvec_0} as
\begin{align}
    \lambda(k) = p + (1-p)\frac{ \beta_+ +  \beta_-}{\alpha_+ + \alpha_-},
\end{align}
and plugging in the above expressions returns the result from the main text~\eqref{eq:selfconsistent} after some straightforward manipulations. If these equations are satisfied the full state returns an eigenstate with eigenvalue $\lambda(k)$.

At $p=1/(2T)$ the value of $k$ for the leading eigenvalue changes from purely imaginary to purely real, with $k=0$ exactly at $p=1/(2T)$. For $k=0$ the eigenvalue \eqref{eqn:eigvals} evaluates to $\lambda = (1-2p)^2$, and plugging this expression in the self-consistent equation~\eqref{eq:selfconsistent} fixes $T=1/(2p)$. At this point the wave function of the leading eigenvalue is exactly linear, which follows as a linearization of Eq.~\eqref{eq:psi_k_app} in the limit $k\to 0$. The corresponding unnormalized eigenstates are given by
\begin{align}
    \psi_j = 
    \begin{cases}
        T+j, \qquad &\textrm{for}\, j\,\textrm{even},\\
        T-j-1 , \qquad &\textrm{for}\, j\,\textrm{odd}.
    \end{cases}
\end{align}
as can be verified by direct calculation.

\bibliography{main}

\end{document}